\documentclass[authoryear,preprint,11pt,3p]{elsarticle}
\journal{European Journal of Operational Research}

\usepackage{graphicx}
\usepackage{xcolor}
\usepackage{physics}
\usepackage{amsmath}
\usepackage{amssymb}
\usepackage{amsthm}
\theoremstyle{plain}%
\newtheorem{theorem}{Theorem}
\newtheorem{corollary}{Corollary}
\newtheorem{lemma}{Lemma}
\theoremstyle{definition}
\newtheorem{example}{Example}
\newtheorem{definition}{Definition}
\newtheorem{remark}{Remark}
\usepackage{ulem}

\let\today\relax
\makeatletter
\def\ps@pprintTitle{%
    \let\@oddhead\@empty
    \let\@evenhead\@empty
    \def\@oddfoot{\footnotesize\itshape
         {} \hfill\today}%
    \let\@evenfoot\@oddfoot
    }
\makeatother

\makeatletter
\def\els@aparagraph[#1]#2{\elsparagraph[#1]{#2\@addpunct{.}}}
\def\els@bparagraph#1{\elsparagraph*{#1\@addpunct{.}}}
\makeatother

\usepackage{bm}
\usepackage{amsfonts}
\usepackage{hyperref}
\hypersetup{colorlinks=True}
\usepackage{booktabs}
\usepackage{multirow}
\usepackage{mathtools}
\usepackage[bb=boondox]{mathalfa}
\usepackage{cleveref}
\crefformat{section}{§#2#1#3}
\usepackage{nicefrac}
\usepackage{enumitem}
\usepackage{chngcntr}
\usepackage{tikz,tikz-3dplot}
\usepackage{pgfplots,pgfplotstable}
\usetikzlibrary{arrows,automata,calc,shapes,plotmarks, positioning,shapes.geometric,decorations.pathmorphing,patterns,fillbetween,pgfplots.groupplots}
\usepackage{circuitikz}
\pgfplotsset{compat=1.15}
\pgfplotsset{
    table/search path={plotdata},
}
\usepgfplotslibrary{colorbrewer,units}
\pgfplotsset{cycle list/Set3}
\definecolor{maincolor}{HTML}{032F99} %blue
\definecolor{secondcolor}{HTML}{ff5722} % dark orange
\definecolor{thirdcolor}{HTML}{c7d3d7}  %bluish gray
\definecolor{newtextcolor}{HTML}{bf360c}
\newcommand{\nt}[1]{\textcolor{black}{#1}}

\newcommand{\minimize}[1]{\underset{{#1}}{\text{min}}}
\newcommand{\maximize}[1]{\underset{{#1}}{\text{max}}}

\newcommand{\st}{\text{s.t.}}

\usepackage{mathtools}

\usepackage{esvect}

\makeatletter
\newcommand{\oset}[3][0ex]{%
  \mathrel{\mathop{#3}\limits^{
    \vbox to#1{\kern-2\ex@
    \hbox{$\scriptstyle#2$}\vss}}}}
\makeatother

\def\storage#1#2{
\begin{scope}[shift={#1}, rotate=#2]
    \node[anchor=north west,inner sep=0] at (0,0) {
            \includegraphics[height=1.1cm]{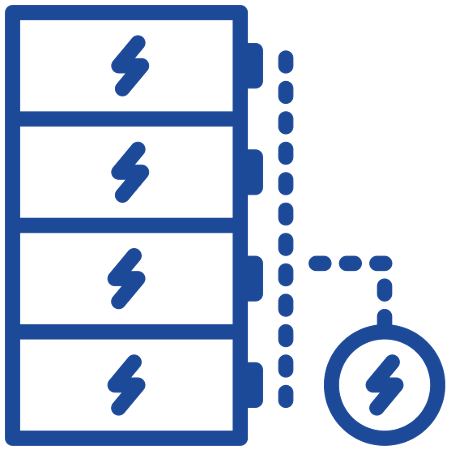}};
\end{scope}
}

\begin{document}

\begin{frontmatter}
\title{Moving from Linear to Conic Markets for Electricity}
\author[inst1,inst2]{Anubhav Ratha}\corref{cor1}\ead{arath@dtu.dk}%\fnref{fn1}
\author[inst3,inst1]{Pierre Pinson}\ead{p.pinson@imperial.ac.uk}
\author[inst4]{Hélène Le Cadre}\ead{helene.le-cadre@inria.fr}
\author[inst2]{Ana Virag}\ead{ana.virag@vito.be}
\author[inst1]{Jalal Kazempour}\ead{jalal@dtu.dk}

\affiliation[inst1]{organization={Technical University of Denmark},
                    city={Kgs. Lyngby},
                    Denmark}
\affiliation[inst2]{organization={Flemish Institute for Technological Research (VITO)}, city={Mol}, country={Belgium}}
\affiliation[inst3]{organization={Imperial College London}, city={London}, country={United Kingdom}}
\affiliation[inst4]{organization={Univ. Lille, Inria, CNRS, Centrale Lille, UMR 9189 CRIStAL}, city={Lille}, country={France}}
\cortext[cor1]{Corresponding author.}

\begin{abstract}
We propose a new forward electricity market framework that admits heterogeneous market participants with second-order cone strategy sets, who accurately express the nonlinearities in their costs and constraints through conic bids, and a network operator facing conic operational constraints. In contrast to the prevalent linear-programming-based electricity markets, we highlight how the inclusion of second-order cone constraints \nt{improves} uncertainty-, asset-, and network-awareness of the market, which is key to the successful transition towards an electricity system based on weather-dependent renewable energy sources. We analyze our general market-clearing proposal using conic duality theory to derive efficient spatially-differentiated prices for the multiple commodities, comprised of energy and flexibility services. Under the assumption of perfect competition, we prove the equivalence of the centrally-solved market-clearing optimization problem to a competitive spatial price equilibrium involving a set of rational and self-interested participants and a price setter. Finally, under common assumptions, we prove that moving towards conic markets does not incur the loss of desirable economic properties of markets, namely market efficiency, cost recovery, and revenue adequacy. Our numerical studies focus on the specific use case of uncertainty-aware market design and demonstrate that the proposed conic market brings advantages over existing alternatives within the linear programming market framework.
\end{abstract}

\begin{keyword}
OR in energy \sep spatial equilibrium \sep mechanism design \sep electricity markets \sep conic economics
\end{keyword}

\end{frontmatter}

\section{Introduction}\label{sec:1_introduction}
The spatial price equilibrium problem, as first analyzed by \cite{Enke1951} and \cite{Samuelson1952}, computes commodity prices and trade flows that satisfy partial equilibrium conditions over a network. In a two-sided auction framework, this problem involves price-quantity supply offers and demand bids matched by an auctioneer to maximize the social welfare, contingent on the spatial constraints. Historically, spatial price equilibrium problems rely on linear programming (LP) theory \citep{Kantorovich1960} to derive the market equilibrium prices from \textit{marginal equalities}. Despite the success of LP in achieving optimal market-clearing outcomes and efficient prices with satisfactory computational effort, it is potentially limiting in physical systems as it may fail to accurately represent the \textit{nonlinear} operational characteristics of assets and the network. Examples of such physical systems include electricity \citep{Bohn1984}, natural gas \citep{deWolf2000}, water \citep{Cai2001}, heat \citep{Mitridati2020}, telecommunication networks \citep{Courcoubetis2003} and supply chains \citep{Snyder2014}. Even in non-physical systems facing uncertainty, the use of LP compels a linear modeling of uncertainty and potential risk preferences, which is limiting at times.

A natural question then arises, why should the market-clearing problem be confined to the LP framework? This question is pertinent since LP is the simplest mathematical framework in the convex optimization theory, while more general convex frameworks such as conic programming are available. If such frameworks enable a more accurate representation of physical assets and networks as well as uncertainty, while retaining the advantages of LP in terms of optimality, pricing, and computational ease, it is then, indeed, appealing to adopt them. Leveraging the recent mathematical and computational advances in conic programming \citep{Alizadeh2003, MosekApS2021}, in this work, we introduce and analyze spatial price equilibrium in a market-clearing context based on the second-order cone programming (SOCP) framework, and demonstrate how it outperforms the LP-based markets. Although our theoretical results are generalizable to any market-clearing context, we choose the domain of electricity markets for exposition as it is rich in examples that worsen the adverse impacts of the limitations imposed by LP-based markets.

\subsection{Limitations of LP-based Electricity Markets}\label{subsec:1_1_lp_limitations}
In electricity markets, the nonlinearities arise from the costs (utilities) and constraints of producers (consumers) and the physics of power flow in the network. Currently, real-world electricity markets follow the original proposal by \cite{Bohn1984} to solve the spatial price equilibrium problem as an LP to obtain optimal production and consumption quantities and the spatially-differentiated nodal electricity prices, commonly referred to as \textit{locational marginal prices} (LMPs) in the industry \citep{Kirschen2018}. To reach climate change mitigation goals, electricity systems are transitioning towards a more sustainable future \citep{Chu2012}, by integrating larger shares of weather-dependent renewable energy sources such as wind and solar. This trend challenges LP-based markets on three accounts.

First, with large shares of variable and unpredictable renewable energy, electricity markets are exposed to significant uncertainty, which needs to be accounted for. Classical methods within the LP framework, such as scenario-based stochastic programs \citep{Pritchard2010,Zavala2017} and robust optimization techniques \citep{Bertsimas2013} are unsuitable in practical settings as they suffer from computational intractability and solution conservativism, respectively. Going beyond LP-based markets is beneficial from an uncertainty modeling perspective. For instance, chance-constrained programming \citep{Kuang2018}, which admits nonlinear yet convex, computationally tractable and analytically expressable uncertainty models, paves the way for \textit{uncertainty-aware} electricity markets in practice. 

Second, accommodating the uncertainty requires resources that provide \textit{flexibility services} by adapting their operational status. Such flexible resources include, among others, energy storage \citep{Kim2011}, flexible consumers \citep{Anjos2017}, \nt{upcoming power-to-x units \citep{Varela2021}} as well as the coordination with natural gas \citep{Thompson2013} and district heating \citep{Mitridati2020} sectors. \nt{Operational characteristics of such flexible resources are typically nonlinear, i.e, they incur quadratic costs and admit nonlinear feasibility sets. LP-based markets require linear or piecewise-linear approximation of such cost functions, which fail to accurately characterize the costs and deter flexible resources from offering flexibility. Cost functions aside, adopting the inner (outer) polyhedral approximation required by LP-based markets leads to the undermining (overestimation) of the amount of flexibility available.} At times, ignoring the characteristics induces operational and reliability risks for the flexibility provider and the electricity system, as witnessed during the 2014 polar vortex event in Northeastern United States \citep{PJM2014} and more recently, during the cold weather event in Texas \citep{Bushnell2021}. The need for an \textit{asset-aware} electricity market remains unfulfilled within the LP framework. \nt{Apart from nonlinearities, the feasibility sets of various assets may entail non-convexity arising from integrality constraints, e.g., the commitment status of power producers \citep{Hobbs2001book}. This non-convexity is resolved in practice by solving a unit-commitment problem as a mixed-integer linear programming (MILP) problem before the actual market clearing from which optimal prices and quantities are derived \citep{ONeill2005}. While considering non-convexities is beyond the scope of this work, Remark \ref{remark:non_convexities} further discusses them in the context of the market design proposed.}

Third, accurate operational modeling of flexible resources is valuable only if the network constraints are represented in sufficient detail, which is crucial as flexible resources are dispersed across the electricity network. Conventionally, a linear approximation of electricity network constraints is adopted to retain an LP-based market-clearing problem. \nt{With this approximation, the market-clearing problem typically results in physical flows that are infeasible in reality \citep{Baker2021}. Consequently, flexibility services are procured in a way that the network constraints may not allow the flexibility to be delivered when needed.} This is critical in the case of electricity markets as the loss of \textit{real-time balance} between the production and consumption of electricity in the system may lead to minor, localized supply disruptions at best and a large-scale cascading blackout at worst \citep{Daqing2014}. A \textit{network-aware} procurement of flexibility services is vital for maintaining the real-time balance.

On the above three accounts, the LP framework falls short in meeting the challenges of electricity markets of the future, necessitating a more advanced yet practical alternative.

\subsection{Towards Conic Economics}
\nt{Augmenting the LP-based market-clearing problem with conic constraints alleviates these limitations to a great extent. A market-clearing problem that involves conic constraints leads to the emergence of \textit{conic economics}. Coined by \cite{Raissi2016}, conic economics was introduced in the context of general equilibrium theory, focusing on the mitigation of financial risk. In contrast, we argue that the inclusion of second-order cone (SOC) constraints based on the Lorentz cones into the spatial price equilibrium problem leads to a convexity-preserving modeling of nonlinearities within the market framework. The resulting SOCP-based market-clearing problems are efficiently solved in polynomial time using interior-point methods \citep{Alizadeh2003} by several off-the-shelf commercial solvers such as \texttt{MOSEK}, \texttt{Gurobi}, and \texttt{CPLEX}.}

\nt{In the context of electricity markets, SOC constraints have recently gained interest in market proposals based on chance-constrained programming \citep{Dvorkin2020,Mieth2020a}. The single-period stochastic market clearing proposed in \cite{Dvorkin2020} discusses the internalization of uncertainty in the price formation process, highlighting the advantages of chance-constrained electricity markets over scenario-based stochastic markets in terms of potential acceptability in a real-world implementation. The work in \cite{Mieth2020a} showed that SOC reformulations of chance constraints also enable an analytical characterization of the risk faced by electricity markets in mitigating the uncertainty from renewable energy sources, leading to risk- and variance-aware electricity prices. In regards to asset-awareness, \cite{Kuang2019} study quadratic costs of deliverability in unit commitment problems, essential to modeling ramping costs while providing flexibility. Finally, towards improving network-awareness, considering the SOC relaxation of power flows in distribution systems, pricing schemes based on conic duality were proposed in a deterministic setting in \cite{Papavasiliou2018}, and extended in \cite{Mieth2020b} to include uncertainty modeled via chance constraints.}

\nt{Previous works have primarily focused on only one of the three aspects, i.e., uncertainty-, asset-, or network-awareness of the markets, whereas integration of large shares of renewable energy in electricity markets requires a combined approach. Our work generalizes the prior works, such that heterogeneous market participants with nonlinear (and potentially inter-temporal constraints) and quadratic costs could participate in multiple commodity trades in an electricity market aimed at harnessing flexibility in a network-aware, cost-efficient manner. Consequently, flexibility providers take a central role in the proposed market design, transforming electricity markets from energy-centric to flexibility-centric.}

\nt{\begin{remark}[Addressing non-convexities in electricity markets]\label{remark:non_convexities}
Non-convexity of market participants' strategy sets implies that the optimal quantity allocations and prices do not necessarily support social welfare-maximizing equilibrium, e.g., producers may not recover their costs \citep{Liberopoulos2016} and therefore, have incentives to deviate from the equilibrium. In practice, this is resolved by out-of-market side payments made by system operators to ensure cost recovery for producers and to deter such deviations, see \cite{Gribik2007, Schiro2016} for details. Neglecting cost recovery of participants, similar to the two-step unit commitment approach using MILP \citep{ONeill2005}, these non-convexities can be potentially incorporated as an extension to our proposed conic market framework, rendering the problem as a mixed-integer second-order cone program (MISOCP). Commercial nonlinear programming solvers can already solve MISOCP problems, adopting a variety of algorithms \citep{Benson2013}. However, further computational advances and analysis of prices are needed prior to adoption in real-world electricity markets. 
\end{remark}}

\subsection{Contributions}
As a broad contribution, our work generalizes the prevalent LP-based market-clearing problem to the SOCP framework and applies conic duality to analyze the market equilibrium prices and the economic properties of the underlying market-clearing problem\footnote{Beyond the SOCP framework, semidefinite programming (SDP) which operates on the cone of semi-definite matrices instead of the Lorentz cone, allows for further broadening of the scope of the market-clearing, albeit at the cost of a higher computational burden. Our theoretical results and their proofs build on Lagrangian duality involving \textit{generalized inequalities}, which lay the foundation for an SDP-based market framework in future.}. In the following, we discuss the specific contributions of this work from three perspectives, i.e., from a market design perspective, theoretical perspective, and finally from a practitioner's perspective.

From a market design perspective, our primary contribution is an original proposal for a \textit{general} conic electricity market. By general, we imply a market framework that \nt{improves uncertainty-, asset- and network-awareness of electricity markets beyond the prevalent LP-based markets within the convex optimization realm.} Our proposed market framework is uncertainty-aware by design, as it admits a chance-constrained market-clearing formulation. Towards an asset-aware electricity market, we enable heterogeneous market participants to accurately express SOC-representable nonlinearities in their cost (or utility) functions and constraints. Nonlinear network flow models underlying the physical delivery associated with the trades are also included in our proposal, leading to network-aware electricity markets.

From a theoretical perspective, we first formulate the market-clearing problem as a centrally-solved optimization problem and address the challenge of \textit{robust solvability} of SOCP problems. Theorem \ref{thm:strongduality} provides the necessary and sufficient conditions for optimality and robustness of the market-clearing outcomes, while Theorem \ref{thm:theorem_priceformation} gives an analytical expression for conic spatial prices of the traded commodities.  Connecting the centrally-solved optimization problem to a spatial equilibrium problem involving rational and self-interested actors, Theorem \ref{thm:comp_spatial_price_eqm} leverages conic duality to prove the equivalence of the optimization to a competitive equilibrium. Under common assumptions, Theorem \ref{thm:theorem_economic_properties} proves the satisfaction of economic properties, namely efficiency, cost recovery and revenue adequacy \citep{Schweppe1988}, in the proposed market-clearing. This analytically supports that the move towards conic markets does not incur the loss of any economic properties compared to the prevalent LP-based markets.

From a practitioner's perspective, we illustrate the generality of our proposed market-clearing framework by defining a bid format for conic markets, enabling heterogeneous market participants to express their preferences. Our numerical studies highlight how the conic market encompasses an uncertainty-aware electricity market that efficiently remunerates the mitigation of uncertainty by the market participants, such that the real-time balance in the electricity system is ensured. \nt{A new class of flexibility products, called \textit{adjustment policies}, based on linear decision rules \citep{Holt1955} enable this. These policies are rules, agreed at the day-ahead market stage, that govern how flexibility providers respond to forecast error realizations during or close to the real-time operation.} We compare the proposed SOCP-based market-clearing proposal with two LP-based uncertainty-aware benchmarks, highlighting the advantages of moving towards a conic market framework.

\paragraph{Paper organization:}In \cref{sec:2_conic_market_clearing} we introduce the market setting, illustrate the relevance of SOC constraints via examples, introduce the bidding format, and present the general conic market-clearing  as an optimization problem. In \cref{sec:3_eqm_economic}, we analyze the spatial equilibrium underlying the optimization problem and discuss the economic properties constituting the market equilibrium. Next, \cref{sec:4_numerical_studies} presents numerical results on one of the market-clearing use cases by comparing an uncertainty-aware conic market with the available alternatives within the LP domain. Finally, \cref{sec:5_outlook} concludes by highlighting the key findings of this work and discusses future perspectives. \ref{app:sec_socp_duality} provides a concise background on SOCP duality, while we prove our theoretical results in \ref{app:sec_proofs}. We provide modeling examples and present the market-clearing problems employed in numerical studies in the Supplementary Material which serves as an electronic companion to the paper.

\paragraph{Notation:}The set of natural and real numbers is denoted by $\mathbb{N}$ and $\mathbb{R}$, respectively, whereas $\mathbb{R}_{+}$ and $\mathbb{R}_{-}$, respectively, denote the sets of non-negative and non-positive real numbers. Upper case alphabets with a script typeface, such as $\mathcal{A}$, represent sets, while vectors are denoted by lower case boldface and matrices by upper case boldface alphabets. For a vector $\bm{x}$, the operator $\bm{x}^{\top}$ denotes its transpose, $\norm{\bm{x}}$ represents its Euclidean norm and $\text{diag}(\bm{x})$ returns a diagonal matrix with vector $\bm{x}$ as the leading diagonal. We retrieve the $k$-th element of the vector $\bm{x}$ as the scalar $x_{k}$ and use the operator $[\cdot]_{k}$ to retrieve the $k$-th element of a general vector expression. $\mathbb{0}$ and $\mathbb{1}$ are vectors of zeros and ones; arithmetic operators $\leq$, $=$, and $\geq$ on vectors are understood element-wise. For a matrix $\mathbf{M} \in \mathbb{R}^{p \times q}$, $[\mathbf{M}]_{(:,k)} \in \mathbb{R}^{p}$ retrieves its $k$-th column while $[\mathbf{M}]_{(k,:)} \in \mathbb{R}^{1 \times q}$ retrieves its $k$-th row. The operator $\text{tr}(\mathbf{M})$ returns the trace of the matrix $\mathbf{M}$, while the expression $\mathbf{M} \succcurlyeq 0$ indicates its positive-semidefiniteness. Lastly, the operator $\otimes$ denotes the Kronecker product. %introduction
\section{A General Conic Market for Electricity} \label{sec:2_conic_market_clearing}
We discuss the setting of our conic market-clearing problem in \cref{subsec:2_1_market_setting}, followed by introducing SOC constraints in a market-clearing context in \cref{subsec:2_2_soc_constraints}. In \cref{subsec:2_3_nonSOC_convex_constraints}, we discuss the equality constraints and present the conic market bids in \cref{subsec:2_4_conic_market_bids}. Lastly, we formulate the general market-clearing problem in \cref{subsec:2_5_market_clearing} as an SOCP problem.

\subsection{Market Setting} \label{subsec:2_1_market_setting}
We consider a forward electricity market involving multiple discrete clearing periods within the finite time horizon of a single day. We focus on hourly electricity markets prevalent across the world and collect the hours of the day in a set $\mathcal{T}=\{1,2,\dots,T\}$, where $T=24$. 

\paragraph{Commodities:}
Without loss of generality, we assume that participants in the forward market compete at the \textit{day-ahead} stage. The hourly clearing periods at the day-ahead stage, therefore, correspond to the hours of the next day, which we collectively refer to as the \textit{real-time operation}. The two types of commodities to be traded are (i) energy and (ii) flexibility services. Both types of commodities are traded hourly in the day-ahead market, which is a purely financial market in the sense that the physical delivery of these commodities occurs during the real-time operation. The former commodity, i.e., energy, represents the quantity in MWh to be exchanged among the market participants during the real-time operation. The latter, i.e., flexibility services, refers to the exchanges that contribute to the supply-demand balance during the real-time operation. \nt{For instance, these exchanges could be a result of the activation of adjustment policies allocated to flexible market participants in response to an operational need that may arise during the real-time operation.} A potential occurrence of an imbalance between the total production and consumption of energy in the system during the real-time operation is such an operational need. As another example, flexibility services could also be traded for mitigation of a foreseeable congestion in parts of the electricity network. We define a set $\mathcal{P}= \{1,2,\dots,P\}$ to denote the $P$ commodities traded in the market.

\paragraph{Market participants:} 
Our market framework admits heterogeneous competing participants buying or selling one or more commodities in the market. We introduce the notation and discuss properties applicable to all participants here, while delegating the elaboration on various kinds of participants to \cref{subsec:2_2_soc_constraints}. Let the set $\mathcal{I} = \{1,2,\dots,I\}$ collect $I$ participants, such that $I \geq 2$. For each participant $i \in \mathcal{I}$, let $\bm{q}_{it} \in \mathbb{R}^{K_{i}}$ denote the $K_{i} \geq P$ number of decision variables at hour $t$. We retrieve the $k$-th element of the decision variable at hour $t$ as $q_{itk}\in\mathbb{R},\;\forall k\in\{1,2,\dots,K_{i}\}$. To facilitate a multi-period market-clearing, we stack $\bm{q}_{it}$ for the $T=24$ hours, extending each participant's decision vector to $\bm{q}_{i} \in \mathbb{R}^{K_{i}T}$. The first $P$ elements of the vector $\bm{q}_{it}$ represent contribution towards the $P$ commodities in hour $t$. For notational compactness, we introduce $\bm{q}_{ip} \in\mathbb{R}^{T},\;\forall p\in\mathcal{P}$ as a subvector of $\bm{q}_{i}$ extracting the hourly contributions by participant $i$ towards the trades of the $p$-th commodity over the $T$ hours. Apart from the contributions towards the commodity trades, each participant may have $K_{i}-P$ state variables at each hour $t$, which are involved in the participant's operational constraints. Let $c_{it}(\bm{q}_{it}): \mathbb{R}^{K_{i}} \mapsto \mathbb{R}$ denote the participant's cost function, such that each $c_{it}(\bm{q}_{it})$ is increasing, convex and twice-differentiable in $\bm{q}_{it}$ and satisfies $c_{it}(\mathbf{0})= 0 $. We adopt a sign convention that $c_{it}(\bm{q}_{it}) > 0$ applies for injection into the network and $c_{it}(\bm{q}_{it}) < 0$ for withdrawal, thereby representing the convex cost of injection and concave benefit of withdrawal. The temporally-separable structure of the cost function ensures its convexity while accommodating participants such as firms owning energy storage units, who toggle between being producers (discharging) and consumers (charging).

\paragraph{Electricity network:}
We represent the electricity network as a directed graph $(\mathcal{N}, \mathcal{L})$ formed by a set of nodes $\mathcal{N} = \{1,2,\dots,N\}$, each potentially hosting multiple market participants, and a set of lines $\mathcal{L}$ comprised of pairs of nodes $(n,n')$ that are connected. We define $\mathcal{I}_{n}\subseteq \mathcal{I},\;\forall n\in\mathcal{N}$ as the set of market participants connected to node $n$. The quantity of power flowing over each line and the direction of the flow is governed by \nt{nonlinear and non-convex power flow equations} during the real-time operation. Moreover, the flow across the network is limited by a maximum flow quantity in each line, known as the thermal limit or rated capacity of a power line.

\paragraph{System operator:}
A system operator ensures that (i) optimal market-clearing outcomes and efficient prices are achieved, and (ii) a continuous balance between consumption and generation is maintained during the real-time operation, while satisfying the transport limits of the underlying electricity network. Evidently, these roles correspond to the tasks of operating the market and operating the network, respectively. In this paper, we assume that the system operator is responsible for both the day-ahead market-clearing and the real-time operation, which is consistent with the prevalent organization in the United States. 

\paragraph{Competition and timeline:}
For the simplicity of exposition, this work assumes that no market participant acts strategically to exercise market power. Moreover, we overcome the non-convexity arising from integrality constraints by accepting offers (bids) from participants already committed to producing (consuming). In our market setting, the committed market participants submit their day-ahead supply offers and demand bids to the central system operator, before a predefined gate closure time, without any knowledge of the bids and offers from other participants. Thereafter, the market is cleared by the system operator matching the offers with the bids, while ensuring the feasibility of network constraints. Finally, cleared prices and quantities for all the commodities are publicly disclosed.

\paragraph{Payment mechanism:}
As in majority of electricity markets worldwide, we adopt a \textit{uniform pricing} scheme for pricing of energy, implying that all accepted supply offers and demand bids are cleared with a common price at a given location and time. Alternative payment mechanisms, e.g., pay-as-bid and Vickrey-Clarke-Groves pricing in a single commodity setting \citep{Vickrey1961} are other possibilities and our market-clearing problem is generalizable to admit them.
\subsection{SOC Constraints in a Market-Clearing Problem}\label{subsec:2_2_soc_constraints}
Our general market framework admits heterogeneous market participants of four kinds. First, we consider conventional power producers such as firms that own coal-fired, gas-fired, nuclear, and hydro power plants. These producers are \textit{dispatchable} at the day-ahead stage, i.e., they are able to plan their production during real-time operation with a high degree of certainty. Some are flexible, e.g., gas-fired and hydro, meaning that their planned production quantities can be modified during the real-time operation, if needed. On the contrary, some producers are relatively less flexible, e.g., coal-fired and nuclear power producers. Second, we consider firms owning weather-dependent renewable power production sources such as wind and solar power plants. These producers are \textit{non-dispatchable} and inflexible. Third, we consider energy consumers such as small-scale end-consumers or large industries. Some of these consumers may be flexible and can be counted as flexible resources. Last, we consider additional types of flexibility providers, which may not necessarily be conventional power producers or flexible consumers, for instance, firms owning energy storage units. 

SOC constraints applicable to these market participants are derived from a second-order cone, which is a convex set and is alternatively referred to as Lorentz cone or ice-cream cone. For the variable $\bm{q}_{i}$ of participant $i,\;\forall i \in \mathcal{I}$, an SOC constraint in its general form is given by
\begin{align}
  \norm{\mathbf{A}_{i}\bm{q}_{i} + \mathbf{b}_{i}} \leq \mathbf{d}^{\top}_{i}\bm{q}_{i} + e_{i} \quad \Leftrightarrow \quad \begin{bmatrix}\mathbf{A}_{i} \\ \mathbf{d}_{i}^{\top}\end{bmatrix}\bm{q}_{i} + \begin{bmatrix}\mathbf{b}_{i} \\ {e}_{i}\end{bmatrix}\;\in\; \mathcal{C}_{i}\;\subseteq\;\mathbb{R}^{m_{i}+1}\;, \label{eq:soc_constraint_example}
\end{align}
where $\mathcal{C}_{i}$ is a second-order cone of dimension $m_{i}+1$, where $m_{i}\in\mathbb{N}$. The dimensions $m_{i}+1$ of the cone reflect the relationship within the decision variables. Considering the heterogeneous mix of market participants involved, the dimensions of the cone in the SOC constraints are not necessarily identical among the various participants or even among the constraints of each participant. Parameters $\mathbf{A}_{i} \in \mathbb{R}^{m_{i} \times K_{i}T}$, $\mathbf{b}_{i} \in \mathbb{R}^{m_{i}}$, $\mathbf{d}_{i} \in \mathbb{R}^{K_{i}T}$ and $e_{i} \in \mathbb{R}$ embody the structural and geometrical information for each constraint. We use the two equivalent forms in \eqref{eq:soc_constraint_example} interchangeably, preferring the form with Euclidean norm while discussing the modeling of participant constraints and the conic form in the analytical proofs. 
\paragraph{Linear constraints:}
Any single-period or multi-period linear constraint arising from the operation of physical assets owned by a participant $i$ is represented by \eqref{eq:soc_constraint_example} with appropriate choice of parameters. Linear constraints are represented by the SOC constraint \eqref{eq:soc_constraint_example}, provided that $\mathbf{A}_{i}$ is a null matrix or the cone is dimensioned such that $m_{i}=0$. In the former case, \eqref{eq:soc_constraint_example} reduces to the linear constraint $ 0 \leq \mathbf{d}_{i}^{\top}\bm{q}_{i} + e_{i}$, which denotes a halfspace. Whereas in the latter case, \eqref{eq:soc_constraint_example} reformulates into a linear constraint $ 0 \leq \mathbf{d}_{i}^{\top}\bm{q}_{i} + e^{'}_{i}$, where $e^{'}_{i} = e_{i} - \norm{\mathbf{b}_{i}}$. For power producers, such constraints are, e.g., minimum or maximum production limits and ramping rate constraints that limit production change over subsequent hours. 
\paragraph{Quadratic constraints: }
While linear constraints are admissible in LP-based electricity markets, convex quadratic constraints are not. However, any convex quadratic operational constraint related to the assets of participant $i$ can be represented by \eqref{eq:soc_constraint_example} if $\mathbf{d}_{i} = \mathbf{0}$ and $e_{i} \geq 0$. Example EC.1 in the Supplementary Material illustrates the conic reformulation of such a quadratic constraint.

\paragraph{Other nonlinear constraints:}
In a more general sense, beyond linear and convex quadratic constraints, \eqref{eq:soc_constraint_example} captures the relationship among the $K_{i}T$ number of decision variables for each market participant, contributing towards asset-awareness of the market-clearing problem. For instance, such nonlinear constraints arise in the coordination between the electricity system and natural gas system, wherein the operational constraints of the natural gas system become relevant to the day-ahead electricity market-clearing problem, as discussed in the following.
\begin{example}[Electricity and gas system coordination]\label{example:natural_gas_network}
\nt{The interdependence between electricity and natural gas systems primarily arises due to the significant role played by gas-fired power plants in providing flexibility to the electricity system facing uncertainty. Congestion in the gas network during the real-time operation of the electricity system jeopardizes the availability of fuel for gas-fired power plants and therefore, adversely impacts the flexibility provision \citep{Byeon2020}. One of the crucial operational constraints of the gas system is the relation between gas flows and nodal pressures in the network. The steady-state flow of gas in the pipelines is represented by a non-convex quadratic equality constraint that relates the squared flow magnitude with the difference in squared pressures at the terminal nodes. A convex relaxation of this non-convex equality takes the form of SOC constraints, as illustrated in the following.}

\nt{Let $\varphi\in\mathbb{R}_{+}$ denote the flow of gas along a gas pipeline connecting two terminal nodes, a sending node $s$ and a receiving node $r$ with nodal pressures denoted by $\pi_{s}\in\mathbb{R}_{+}$ and $\pi_{r}\in\mathbb{R}_{+}$, respectively. Proposed by \cite{BorrazSanchez2016}, the convex relaxation for the non-convex equality between the flow along a pipeline and the pressures at the terminal nodes is expressed as
\begin{align}
    \varphi^{2}\;\leq\;\beta^{2}\;(\pi_{s}^{2} - \pi_{r}^{2}), 
    \label{eq:weymouth_equation_example}
\end{align}
where $\beta \in \mathbb{R}_{+}$ is a constant encoding the friction coefficient and geometry of pipelines. Ignoring all other pipelines and considering the gas network operator as an electricity market participant, we can denote its decision vector as $\bm{q}_{\text{GN}} = \begin{bmatrix} \varphi \quad \pi_{r} \quad \pi_{s}\end{bmatrix}^{\top}$ and can represent the constraint \eqref{eq:weymouth_equation_example} as an SOC constraint of the form \eqref{eq:soc_constraint_example} with parameters
\begin{align*}
    \mathbf{A} = \begin{bmatrix} \frac{1}{\beta} \quad 0  \quad 0 \\ 0 \quad 1 \quad0 \end{bmatrix}\;,\;\;\mathbf{b} = \mathbb{0}_{2}\;,\;\;\mathbf{d} = \begin{bmatrix}0 \quad 0 \quad 1\end{bmatrix}^{\top}\;\;\text{and}\;\;e=0,
\end{align*}
which is a three-dimensional SOC constraint, i.e., $m_{i}=2$. Similar SOC constraints are included for other pipelines in the gas network and over various hours, therefore enabling the consideration of gas network operational constraints within the conic electricity market-clearing problem.}

\nt{Coordination aside, the increased coupling leads to uncertainty propagation from the electricity system to the gas side. This implies that operational constraints involving state variables in the gas system are challenged by uncertain gas withdrawals from gas-fired power plants responding to uncertainty in the electricity system. Accurately modeling and controlling the variance of state variables entails the use of SOC constraints, as studied in \cite{VDvorkin2022}.}
\end{example}

Lastly, beyond asset-awareness, SOC constraints in their general form enable an uncertainty-aware market-clearing problem. Specifically, chance constraints enable endogenous modeling of uncertainty and risk faced by market participants and are analytically reformulated as SOC constraints under some mild conditions \citep{Nemirovski2007}, see Example \ref{example:chance_constraints} below.
\begin{example}[Chance Constraints]\label{example:chance_constraints}
Consider a market-clearing problem wherein, in addition to the nominal production quantities, a flexibility service is contracted from the flexibility providers in the market. A traded flexibility service is organized as an adjustment policy. As mentioned in \cref{subsec:2_1_market_setting}, these adjustment policies allow mitigation of the uncertainty realized during the real-time operation, while look-ahead decisions are made by the system operator at the day-ahead market-clearing stage. Such uncertainty could, for example, arise from imperfect forecasts for the production from weather-dependent renewable energy sources or from imperfect load forecasts for consumers. Assuming a single-period market-clearing problem for notational simplicity, let the set $\mathcal{W}=\{1,2,\dots,W\}$ collect the $W$ independent sources of uncertainty in the electricity system and vector $\bm{\xi}\in\mathbb{R}^{W}$ denote the random forecast errors representing this uncertainty. Assume that $\bm{\xi}$ follows a probability distribution $\mathbb{P}_{\xi}$, parameterized by the moments, mean $\bm{\mu}\in\mathbb{R}^{W}$ and covariance $\bm{\Sigma}\in\mathbb{R}^{W \times W}$, which are estimated by the system operator with access to a finite number of historical measurements. Under the chance-constrained optimization framework, the system operator allocates adjustment policies to flexible producers while allowing them to violate their operational constraints with a small probability $\hat{\varepsilon}\in [0,1]$. Assume again that participant $i$ has $K_{i}=2$ decision variables such that $\bm{q}_{i1}=\begin{bmatrix}\hat{q}_{i1} \;\; \alpha_{i1}\end{bmatrix}^{\top}\in\mathbb{R}^{2}$, where $\hat{q}_{i1}$ and $\alpha_{i1}$ are, respectively, the nominal production quantity and the adjustment policy. A chance constraint limiting the total production, i.e., the sum of nominal and adjustment, of the participant to its upper limit $\overline{Q}_{i}$ is written as
\begin{subequations}
\begin{align}
    \mathbb{P}_{\xi}\left(\begin{bmatrix}1 \quad \mathbb{1}^{\top} \bm{\xi}\end{bmatrix}\begin{bmatrix}{\hat{q}}_{i1} \\ \alpha_{i1} \end{bmatrix} \leq \overline{Q}_{i}\right) \geq (1-\hat{\varepsilon}), \label{eq:probabilistic_cc_example}
\end{align}
where the uncertainty is characterized by the total forecast error $\mathbb{1}^{\top}\bm{\xi} \in \mathbb{R}$. This probabilistic constraint reformulates to its analytic equivalent, based on \cite{Nemirovski2007}, as
\begin{align}
r_{\hat{\varepsilon}}\norm{\mathbf{X}\mathbb{1} \;\alpha_{i1}} \leq \overline{Q}_{i} - \hat{q}_{i1} - \mathbb{1}^{\top}\bm{\mu}\;\alpha_{i1},  \label{eq:analytical_reform_cc_example}
\end{align}
\end{subequations}
where all vectors of ones are $\mathbb{1}\in\mathbb{R}^{W}$, and  $\mathbf{X}\in\mathbb{R}^{W \times W}$ denotes a factorization of the covariance matrix such that $\bm{\Sigma} = \mathbf{X}\mathbf{X}^{\top}$. For instance, since $\mathbf{\Sigma} \succcurlyeq 0$ from the definition of covariance matrices, such a factorization can be obtained in a computationally efficient manner using Cholesky decomposition,  resulting in $\mathbf{X}$ having a lower-triangular structure. This decomposition is uniquely determined for covariance matrices that are full rank, i.e., all uncertainty sources are linearly independent, as assumed. The parameter $r_{\hat{\varepsilon}}\in\mathbb{R}_{+}$ is a safety parameter chosen by the system operator relying on the knowledge of the distribution $\mathbb{P}_{\xi}$, such that $r_{\hat{\varepsilon}}$ increases as ${\hat{\varepsilon}}$ reduces\footnote{When the random forecast errors $\bm{\xi}$ are assumed to be normally distributed, the safety parameter $r_{\hat{\varepsilon}}$ is given by the inverse cumulative distribution function of the standard Gaussian distribution evaluated at $(1-\hat{\varepsilon})$-quantile \citep{Nemirovski2007}. Dropping the assumption of normality, a more conservative choice of $r_{\hat{\varepsilon}}$ is obtained based on the so-called moment-based distributionally-robust chance constraints \citep{Wagner2008}, which considers the distribution $\mathbb{P}_{\xi}$ to lie inside an ambiguity set of all probability distributions characterized by the empirically-estimated moments $\bm{\mu}$ and $\bm{\Sigma}$. In that case, the safety parameter $r_{\hat{\varepsilon}}=\sqrt{\frac{1-\hat{\varepsilon}}{\hat{\varepsilon}}}$. A further generalization that considers even the moments of distribution $\mathbb{P}_{\xi} $ to lie in well-defined uncertainty sets such as ellipsoids \citep{Delage2010}, lead to a semidefinite constraint instead of the SOC constraint in \eqref{eq:analytical_reform_cc_example}, leading to a SDP-based market-clearing problem.}. Constraint \eqref{eq:analytical_reform_cc_example} is represented by the general SOC constraint \eqref{eq:soc_constraint_example} with parameters
\begin{align*}
    \mathbf{A} = \begin{bmatrix}\mathbb{0}\quad \mathbf{X}\mathbb{1}\end{bmatrix}\;,\;\;\mathbf{b} = \mathbb{0}\;,\;\;\mathbf{d} = -1/r_{\hat{\varepsilon}}\begin{bmatrix}1 \quad \mathbb{1}^{\top}\bm{\mu}\end{bmatrix}^{\top}\;\text{and}\;\;e=\overline{Q}_{i}/r_{\hat{\varepsilon}}\;,
\end{align*}
thereby resulting in an SOC constraint of dimension $W+1$. Further modeling details showcasing the analytical reformulation of more complicated constraints, such as inter-temporal chance constraints of market participants, are covered in the Supplementary Material. 
\end{example}

Finally, accounting for the potentially multiple SOC constraints faced by participant $i$, we introduce a set $\mathcal{J}_{i}=\{1,2,\dots,J_{i}\}$ collecting the $J_{i}$ constraints that extend \eqref{eq:soc_constraint_example} as
\begin{align}
   \norm{\mathbf{A}_{ij}\bm{q}_{i} + \mathbf{b}_{ij}} \leq \mathbf{d}^{\top}_{ij}\bm{q}_{i} + e_{ij},\;\forall j\in\mathcal{J}_{i}, \label{eq:multiple_SOC_constraints}
\end{align}
where each constraint with parameters $\mathbf{A}_{ij} \in \mathbb{R}^{m_{ij} \times K_{i}T}$, $\mathbf{b}_{ij} \in \mathbb{R}^{m_{ij}}$, $\mathbf{d}_{ij} \in \mathbb{R}^{K_{i}T}$ and $e_{ij} \in \mathbb{R}$ corresponds to a second-order cone $\mathcal{C} \subseteq \mathbb{R}^{m_{ij}+1}$. The feasibility region for \eqref{eq:multiple_SOC_constraints} is formed by the Cartesian product of $J_{i}$ second-order cones $\mathcal{C}_{i} = \prod_{j \in \mathcal{J}_{i}}\mathcal{C}_{ij} = \mathcal{C}_{i1} \times \cdots \times \mathcal{C}_{iJ_{i}}$, which is convex.

\subsection{Equality Constraints}\label{subsec:2_3_nonSOC_convex_constraints}
Each participant may be involved in trades corresponding to the $P$ commodities, subject to equality constraints that arise while considering temporal and spatial dynamics underlying their asset models. We model the equality constraints on the decision variable $\bm{q}_{i}$ of participant $i$ as $\mathbf{F}_{i}\bm{q}_{i} = \mathbf{h}_{i}$, where $\mathbf{F}_{i}\in\mathbb{R}^{R_{i} \times K_{i}T}$ and $\mathbf{h}_{i}\in\mathbb{R}^{R_{i}}$ are parameters encoding the $R_{i}$ equality constraints on $\bm{q}_{i}$, such that $R_{i}\leq K_{i}T$ and $\mathbf{F}_{i}$ has full row rank. 

The market-clearing conditions are also modeled as marginal equalities coupling the decisions of the participants such that supply-demand balance is ensured for each of the $P$ commodities traded in the market. Modeling these conditions, given the heterogeneous mix of market participants in our framework, requires the following definition.
\begin{definition}[Physical Fulfillment of Commodity Trades]\label{def:commodity_trades}
For participant $i \in \mathcal{I}$, the hourly injection or withdrawal towards commodity $p$ is given by $\mathbf{G}_{ip}\;\bm{q}_{ip} \in \mathbb{R}^{T}$, where $\mathbf{G}_{ip}\in\mathbb{R}^{T \times T}$ is a \textit{coupling matrix} formed by elements encoding the injection or withdrawal coefficients. 
\end{definition}

We now elaborate on this definition. The contribution by a dispatchable producer (either flexible or inflexible) towards the commodity representing energy is $\bm{q}_{ip} \in \mathbb{R}^{T}_{+}$, whereas for inflexible consumers, the contribution is $\bm{q}_{ip} \in \mathbb{R}^{T}_{-}$. For both kinds of market participants, the coupling matrices $\mathbf{G}_{ip}$ are identity matrices, i.e., $\mathbf{G}_{ip} = \text{diag}(\mathbb{1})$. Meanwhile, the contribution from energy storage units $\bm{q}_{ip}$ to the commodity corresponding to energy adopts different signs in the various hours depending on whether the storage unit is discharging (injection) or charging (withdrawal). Information on the conversion factors for participants from other sectors such as natural gas or district heating is also encoded within the entries of the matrix $\mathbf{G}_{ip}$. Finally, in addition to those traded system-wide, the market-clearing problem may entail commodities traded among a subset of all participants and thereby, reflect the agreements among that subset. $\mathbf{G}_{ip}$ may thus be null matrices for some commodities and for some participants.
\subsection{Bid Format in the Conic Electricity Market}\label{subsec:2_4_conic_market_bids}
For the sake of generality, we adopt a common bid format for the supply offers and demand bids, referring to them as supply and demand bids, respectively and define them in the following. 
\begin{definition}[Conic Market Bids]\label{def:conic_market_bids}
Let $\mathcal{B}_{i}$ denote a bid submitted by the market participant $i$ to the system operator. The bid $\mathcal{B}_{i}$ is a tuple defined as
\begin{align*}
    \mathcal{B}_{i}:= \Big( n_{i},\;\{\mathbf{A}_{ij}, \mathbf{b}_{ij}, \mathbf{d}_{ij}, e_{ij}\}_{j \in \mathcal{J}_i},\;\mathbf{F}_{i},\mathbf{h}_{i},\;\{\mathbf{G}_{ip}\}_{p \in \mathcal{P}},\; \{\mathbf{c}^{\text{Q}}_{it},\mathbf{c}^{\text{L}}_{it}\}_{t\in\mathcal{T}} \Big),
\end{align*}
where $n_{i} \in \mathcal{N}$ is the electricity network node at which the participant $i$ is located. Parameters $\mathbf{A}_{ij},\mathbf{b}_{ij},\mathbf{d}_{ij},e_{ij},\;\forall j\in\mathcal{J}_{i}$ are linked to the $J_{i}$ SOC constraints; $\mathbf{G}_{ip},\;\forall p\in\mathcal{P}$ correspond to the coupling matrices for the $P$ commodities; $\mathbf{F}_{i}$ and $\mathbf{h}_{i}$ correspond to the $R_{i}$ equality constraints; and lastly, $\mathbf{c}_{it}^{\text{Q}},\mathbf{c}_{it}^{\text{L}},\;\forall t\in\mathcal{T}$ denote the temporally-separated quadratic and linear bid prices.
\end{definition}
\begin{remark}[Conic Market Bids vs. Price-Quantity Bids]
The bids $\mathcal{B}_{i}$ are a generalization of the classical \textit{price-quantity} bids that form the backbone of various orders placed in currently-operational LP-based electricity markets, see \cite{NordPoolWeb2021} for example. As the name indicates, in addition to the participant's locational information, a price-quantity bid comprises a bid price representing the bidder's willingness-to-pay or willingness-to-receive for the associated quantity. In the proposed conic market bids, the quantities are formed by the parameters of the $j$-th SOC constraint $\{\mathbf{A}_{ij},\mathbf{b}_{ij},\mathbf{d}_{ij},e_{ij}\}$ admitting the simple linear characterization of quantities $\bm{q}_{i}$ when $\mathbf{A}_{ij}$ is a null matrix or when $m_{ij}=0$, while more intricate representations of the operational constraints are covered by a suitable choice of the SOC constraint parameters, as discussed in \cref{subsec:2_2_soc_constraints}. \nt{In contrast to price-quantity bids, the SOC constraint parameters enable market participants to explicitly reflect the SOC-representable nonlinearities in their costs and operational constraints. Furthermore, market participation in terms of supply (or demand) quantity $\bm{q}_{ip}$ towards a specific commodity $p\in\mathcal{P}$ is decoupled from the potential costs and constraints with associated with the state variables of the participants comprising $\bm{q}_{i}$. However, the proposed bid format requires a more complex exchange of information between the market participants and the market operator. In practice, a simple bid transformation software layer could convert the standard price-quantity bids by participants into the conic bid format, before the market is cleared.} 
\end{remark}
\subsection{Market-Clearing as an SOCP Problem} \label{subsec:2_5_market_clearing}
\paragraph{Network constraints:}
An accurate modeling of the physics of power flows in the electricity network introduces nonlinearities and non-convexities. Convexification of these power flow equations have utilized SOCP \citep{Kocuk2016} and SDP \citep{Lavaei2012} relaxations to ensure optimality while solving electricity system--related optimization problems. However, electricity market-clearing problems across the world adopt a linearized approximation of the power flow equations, relying on a number of assumptions \citep{Cain2012}. Consistent with current practice and for simplicity of exposition, our formulation adopts a variation of linearized power flow equations that leverages the so-called Power Transfer Distribution Factor (PTDF) matrix. \nt{Example EC.2 in Supplementary Material illustrates an extension to the market-clearing problem formulated in this work to include SOC-based convex relaxation of the non-convex AC power flow equations.}
\begin{definition}[Power flows using PTDF]\label{def:PTDF_powerflow}
With the set $\mathcal{I}_{n}\subseteq \mathcal{I}$ collecting market participants located at node $n$, the power flow $s^{a}_{\ell}$ along a line $\ell=(n,n')$ at hour $t$ is given by
\begin{align*}
s_{\ell}^{\text{a}} = \sum_{n \in \mathcal{N}} [\bm{\Psi}]_{(\ell,n)}\left(\sum_{i \in \mathcal{I}_{n}}\sum_{p\in\mathcal{P}}\;[\mathbf{G}_{ip}\bm{q}_{ip}]_{t}\right),
\end{align*}
where $\boldsymbol{\Psi} \in \mathbb{R}^{L \times N}$ denotes the PTDF matrix of the electricity network. Derived from the physical parameters of power lines comprising the network, entries of the PTDF matrix are the sensitivity of changes in the flow in any line to a unit injection at a given node. The expression $\sum_{i \in \mathcal{I}_{n}}\sum_{p\in\mathcal{P}}\;[\mathbf{G}_{ip}\bm{q}_{ip}]_{t} \in \mathbb{R}$ computes the algebraic sum of injections or withdrawals of all participants located at node $n$ towards the $P$ commodities at hour $t$. 
\end{definition}
\begin{remark}
All $P$ commodities share the common physical network for fulfillment of the trades, as reflected by the summation in the expression for power flows in Definition \ref{def:PTDF_powerflow}. \nt{Alternative formulations can be envisioned such that not all commodity trades rely on the physical network for fulfillment. For instance, flexibility services may include purely financial instruments to hedge risk in electricity markets \citep{Philpott2016, Mieth2021}.}
\end{remark}

\paragraph{Conic market-clearing problem, $\mathcal{M}^{\text{c}}$:}Without loss of generality, we assume that the participants have quadratic cost functions which satisfy the conditions in \cref{subsec:2_1_market_setting}. In the interest of computational and analytical simplicity, we reformulate the quadratic costs as SOC constraints following the approach in Example EC.1. \nt{Auxiliary decision variables $\bm{z}_{i} \in \mathbb{R}^{T},\forall i$ are introduced, representing the participants' hourly quadratic cost components.} The market-clearing problem, hereafter referred to as $\mathcal{M}^{\text{c}}$, is formulated as
\begin{subequations}\label{prob:Mcone}
\begin{align}
    \minimize{\bm{q}_{i},\bm{z}_{i}}\quad&\sum_{i \in \mathcal{I}}\;\sum_{t\in\mathcal{T}}\;\Big({z}_{it} + {\mathbf{c}_{it}^{\text{L}}}^{\top}\bm{q}_{it}\Big)\label{objfun_Mcone}\\
    \st\quad&\norm{\mathbf{C}^{\text{Q}}_{it}\bm{q}_{it}}^{2} \le  {z}_{it},\;\forall t,\;\forall i &:& (\bm{\mu}^{\text{Q}}_{it},\;\kappa^{\text{Q}}_{it},\;\nu^{\text{Q}}_{it}) \label{cons:Mcone_participants_obj_reform} \\
    &\norm{\mathbf{A}_{ij}\bm{q}_{i} + \mathbf{b}_{ij}} \leq \mathbf{d}^{\top}_{ij}\bm{q}_{i} + e_{ij},\;\forall j\in\mathcal{J}_{i},\;\forall i&\colon&(\bm{\mu}_{ij},\;{\nu}_{ij}) \label{cons:Mcone_participants_soc}\\
    &\mathbf{F}_{i}\bm{q}_{i} = \mathbf{h}_{i},\;\forall i &\colon&(\bm{\gamma}_{i}) \label{cons:Mcone_participants_eq}\\
    &\sum_{i \in \mathcal{I}} \mathbf{G}_{ip}\bm{q}_{ip} = \mathbf{0}_{T}, \; \forall p,&\colon&(\bm{\lambda}_{p})\label{cons:Mcone_coupling_constraints}\\
    &  \abs{\sum_{n \in \mathcal{N}} [\bm{\Psi}]_{(:,n)}\left(\sum_{i \in \mathcal{I}_{n}}\sum_{p\in\mathcal{P}}\;[\mathbf{G}_{ip}\bm{q}_{ip}]_{t}\right)} \leq \overline{\mathbf{s}},\;\forall t,&\colon&(\underline{\bm{\varrho}}_{t},\;\overline{\bm{\varrho}}_{t}) \label{cons:Mcone_network_flow}
\end{align}
\end{subequations}
where the objective \eqref{objfun_Mcone} minimizes social disutility (or maximizes social welfare) of the market-clearing problem over the time horizon, $t\in\mathcal{T}$. %The optimization variables are the decision variables of the market participants $\bm{q}_{i} \in \mathbb{R}^{K_{i}T}, \forall i$ and $\bm{z}_{i} \in \mathbb{R}^{T},\forall i$. 
Recall that $\bm{q}_{it} \in \mathbb{R}^{K_{i}}$ and $\bm{q}_{ip} \in \mathbb{R}^{T}$ are subsets of the $i$-th participant's decision vector $\bm{q}_{i} \in \mathbb{R}^{K_{i}T}$. The Lagrange multipliers associated with the constraints are shown in parentheses next to them. The constraint parameters, in their order of appearance, are $\mathbf{A}_{ij} \in \mathbb{R}^{m_{ij} \times K_{i}T}$, $\mathbf{b}_{ij} \in \mathbb{R}^{m_{ij}}$, $\mathbf{d}_{ij} \in \mathbb{R}^{K_{i}T}$, $e_{ij} \in \mathbb{R}$, $\mathbf{F}_{i} \in \mathbb{R}^{R_{i} \times K_{i}T}$, $\mathbf{h}_{i}\in\mathbb{R}^{R_{i}}$, $\mathbf{G}_{ip} \in \mathbb{R}^{T \times T}$, $[\bm{\Psi}]_{(:,n)} \in \mathbb{R}^{L}$ and $\overline{\mathbf{s}}\in\mathbb{R}^{L}$. The index $j\in\mathcal{J}_{i}$ refers to the $J_{i}$ SOC constraints of participant $i$ as discussed in \cref{subsec:2_2_soc_constraints}, while the index $p\in\mathcal{P}$ refers to the $P$ commodities traded among the participants. Constraints \eqref{cons:Mcone_participants_obj_reform}-\eqref{cons:Mcone_participants_eq} are participant-specific, while \eqref{cons:Mcone_coupling_constraints}-\eqref{cons:Mcone_network_flow} are related to the commodities exchanged in the market.

The temporally-separable quadratic and linear bid prices of participant $i$ are given by $\mathbf{c}_{it}^{\text{Q}}\;,\;\mathbf{c}_{it}^{\text{L}} \in \mathbb{R}^{K_{i}}$ and ${\mathbf{C}}_{it}^{\text{Q}} \in \mathbb{R}^{K_{i} \times K_{i}}$ is a factorization of the quadratic cost matrix such that $\text{diag}(\mathbf{c}^{\text{Q}}_{it})={{\mathbf{C}}^{\text{Q}}_{it}}^{\top}{{\mathbf{C}}_{it}^{\text{Q}}}$. Lagrange multipliers $\bm{\mu}^{\text{Q}}_{it} \in \mathbb{R}^{K_{i}},\;\kappa^{\text{Q}}_{it} \in\mathbb{R}_{+}\;\text{and}\;\nu^{\text{Q}}_{it}\in\mathbb{R}_{+},\; \forall t,\;\forall i$, are participant-specific dual variables associated with the rotated SOC constraints \nt{\eqref{cons:Mcone_participants_obj_reform} which are satisfied with equality at the optimal solution to problem $\mathcal{M}^{\text{c}}$.} Constraints \eqref{cons:Mcone_participants_soc} model the SOC constraints applicable to the market participants that enable asset- and uncertainty-awareness of the market-clearing problem. The tuple of dual variables associated with the SOC constraints \eqref{cons:Mcone_participants_soc} are $(\bm{\mu}_{ij},\nu_{ij})$, where $\bm{\mu}_{ij} \in \mathbb{R}^{m_{ij}}$ and $\nu_{ij} \in \mathbb{R}_{+}$. \nt{Unlike linear scalar inequalities which admit a single non-negative dual variable, SOC constraints \eqref{cons:Mcone_participants_obj_reform}-\eqref{cons:Mcone_participants_soc} admit tuples of dual variables, as shown while obtaining dual formulations for the SOC constraints in \ref{app:soc_duality_prelims}.} Lastly, the dual variable ${\bm{\gamma}}_{i} \in \mathbb{R}^{R_{i}}$ associates with the $R_{i}$ equality constraints for the participant $i$, see \cref{subsec:2_3_nonSOC_convex_constraints}. 

Constraints \eqref{cons:Mcone_coupling_constraints} are the system-wide balance constraints that couple the decisions of the market participants for each of the $P$ commodities traded in the market, such that the Lagrange multipliers associated with these constraints $\bm{\lambda}_{p}\in\mathbb{R}^{T}$ are interpretable as commodity prices. For instance, the Lagrange multiplier associated with the balance between electricity production and consumption is the nodal price of electricity. Lastly, constraints \eqref{cons:Mcone_network_flow} limit the magnitude of power flow in the lines to their rated capacity, $\overline{\mathbf{s}}$, as given by Definition \ref{def:PTDF_powerflow}. Given the element-wise absolute value operator in this constraint, we associate a pair of non-negative dual variables $\underline{\bm{\varrho}}_{t}\;,\;\overline{\bm{\varrho}}_{t} \in \mathbb{R}_{+}^{L}$ with it. Naturally, constraints \eqref{cons:Mcone_network_flow} can be altered to include asymmetric limits, i.e., the limits which depend on the direction of flow in the power lines, without significant change to the analytical results that follow. Further, note that instead of the chosen PTDF-based formulation, implementing an SOCP relaxation for the formulation of power flow equations affects the constraint \eqref{cons:Mcone_network_flow}. However, the overall market-clearing problem $\mathcal{M}^{\text{c}}$ remains within the SOCP framework. \nt{Finally, adopting the SOC relaxation for the non-convex AC power flow equations (Example EC.2 in Supplementary Material), constraints \eqref{cons:Mcone_network_flow} are replaced by SOC constraints while additional state variables are introduced to model the electricity network parameters. In that case, while \eqref{prob:Mcone} remains an SOCP problem, the pricing of commodities under such conic network constraints (presented in \cref{sec:3_eqm_economic}) merits further investigation.}
\begin{remark}[Strict Convexity]\label{remark:strict_convexity}
Observe that the optimization problem $\mathcal{M}^{\text{c}}$ has a strictly convex objective function \eqref{objfun_Mcone} provided every market participant incurs a non-zero quadratic cost at all hours for each of its $K_{i}$ variables. To see that, we recall that every positive semidefinite matrix is the Gram matrix for some set of vectors. For the quadratic cost matrix $\text{diag}(\mathbf{c}_{it}^{\text{Q}}) \succcurlyeq 0$,
\begin{align*}
    \bm{q}_{it}^{\top}\text{diag}(\mathbf{c}_{it}^{\text{Q}})\bm{q}_{it} =  \bm{q}_{it}^{\top}{\mathbf{C}_{it}^{\text{Q}}}^{\top}\mathbf{C}_{it}^{\text{Q}}\bm{q}_{it} = \norm{\mathbf{C}_{it}^{\text{Q}}\bm{q}_{it}}^{2},\;\forall t, \forall i,
\end{align*}
such that strict convexity is guaranteed only if the quadratic cost matrix is positive definite, i.e.,  $\text{diag}(\mathbf{c}_{it}^{\text{Q}})\succ 0$, which requires $c^{\text{Q}}_{itk} \neq 0,\;\forall k=1,2,\dots,K_{i},\;\forall t,\;\forall i$. 
\end{remark} %modeling
\section{Economic Interpretation and Equilibrium Analysis}\label{sec:3_eqm_economic}
We discuss the price formation process for the $P$ commodities traded in the market in \cref{subsec:3_1_conic_spatial_prices}, followed by the theoretical results that characterize the spatial price equilibrium underlying the centrally-solved market-clearing problem $\mathcal{M}^{c}$ in \cref{subsec:3_2_spatial_price_equilibrium}.
\subsection{Conic Spatial Prices for Commodities}\label{subsec:3_1_conic_spatial_prices}
Deriving optimal market-clearing prices for the problem $\mathcal{M}^{\text{c}}$ is not straightforward, since strong duality is not trivial to establish for an SOCP problem \citep{BenTal2001}. Unlike their LP counterparts where strong duality is guaranteed merely by the feasibility of the primal and dual problems as formulated in Farkas' lemma, strong duality in SOCP problems derives from the existence of \textit{strictly feasible} solutions to both the primal and dual problems \citep[Theorem 13]{Alizadeh2003}. Conventionally, ensuring the strict feasibility or even merely the feasibility of market-clearing problem is considered at a market design stage and is not of high relevance for the market operator. However, in the interest of generality and wider acceptance of a conic market-clearing framework, we address this crucial issue in our analytical results. 

Following Slater's constraint qualification for convex optimization problems, strict feasibility refers to the existence of points within the feasibility set of the primal problem where all inequalities - SOC and linear - are strictly satisfied. Additionally, a refinement to Slater's constraint qualification yields the so-called \textit{essentially strict feasibility} of an SOCP problem.
\begin{definition}[Essentially Strict Feasibility]\label{def:essentially_strict_feasibility}
The market-clearing problem $\mathcal{M}^{\text{c}}$ is essentially strictly feasible, if there exists a feasible solution tuple denoted by $(\bm{q}_{i}\;,\;\bm{z}_{i}),\;\forall i$ such that
\begin{align*}
   &\norm{\mathbf{C}^{\text{Q}}_{it}\bm{q}_{it}}^{2} < {z}_{it},\;\forall t,\;\forall i,\\
    &\norm{\mathbf{A}_{ij}\bm{q}_{i} + \mathbf{b}_{ij}} < \mathbf{d}^{\top}_{ij}\bm{q}_{i} + e_{ij},\;\forall j\in\mathcal{J}_{i},\;\forall i. 
\end{align*}
\end{definition}
Observe that essentially strict feasibility is a weaker requirement as compared to strict feasibility. However, it is necessary and sufficient for strong duality to hold for $\mathcal{M}^{\text{c}}$ since other inequalities \eqref{cons:Mcone_network_flow} are linear in decision variables \citep[\S 5.2.3]{Boyd2004}. We now provide the analytical results crucial to deriving optimal prices from the market-clearing problem $\mathcal{M}^{\text{c}}$.
\begin{theorem}[Strong Duality]\label{thm:strongduality} Let $\mathcal{D}^{\text{c}}$ denote the dual problem to the market-clearing problem $\mathcal{M}^{\text{c}}$. If the set of feasible solutions to the primal problem $\mathcal{M}^{\text{c}}$ is non-empty, then both the primal problem $\mathcal{M}^{\text{c}}$ and dual problem $\mathcal{D}^{\text{c}}$ are essentially strictly feasible, and consequently, strong duality holds for the primal-dual pair of problems $\mathcal{M}^{\text{c}}$ and $\mathcal{D}^{\text{c}}$.
\end{theorem}
Conditioned on feasibility of the primal market-clearing problem $\mathcal{M}^{\text{c}}$, Theorem \ref{thm:strongduality} enables economic interpretations of the market-clearing outcomes. Relying on classical Lagrangian duality theory, we provide an analytical expression for the optimal, spatially-differentiated nodal prices of the $P$ commodities in the following theorem.
\begin{theorem}[Conic Spatial Prices]\label{thm:theorem_priceformation} 
The solution to the conic market-clearing problem $\mathcal{M}^{\text{c}}$ results in optimal contributions $\bm{q}_{ip}^{\star},\;\forall i \in \mathcal{I}$ for the $p\in\mathcal{P}$ commodities and the market clears with optimal prices $\bm{\Pi}^{\star}_{p} \in \mathbb{R}^{N \times T}$ for the $p$-th commodity given by
\begin{align} 
    \bm{\Pi^{\star}}_{p} = {\bm{\Lambda}}^{\star}_{p} - \bm{\Psi}^{\top}(\overline{\bm{\rho}}^{\star} - \underline{\bm{\rho}}^{\star}), \enspace \forall p\in\mathcal{P}, \label{eq:nodal_price_formulation}
\end{align}
where $\underline{\bm{\rho}}^{\star}\;,\;\overline{\bm{\rho}}^{\star}\in\mathbb{R}^{L \times T}$ and ${\bm{\Lambda}}^{\star}_{p} \in \mathbb{R}^{N \times T}$ are variables with stacked columns of optimal dual variables $\underline{\bm{\varrho}}^{\star}_{t}$, $\overline{\bm{\varrho}}^{\star}_{t},\;\forall t$ and $\bm{\lambda}_{p}^{\star},\;\forall n$, respectively, over the $T$ market-clearing hours, i.e., 
\begin{align*}
    &\overline{\bm{\rho}}^{\star} = [\overline{\bm{\varrho}}^{\star}_{1}~\cdots~\overline{\bm{\varrho}}^{\star}_{T}],\; \underline{\bm{\rho}}^{\star} = [\underline{\bm{\varrho}}^{\star}_{1}~\cdots~\underline{\bm{\varrho}}^{\star}_{T}],\;\text{ and }\;\bm{\Lambda}_{p}^{\star} \coloneqq \mathbb{1}_{N}^{\top}\otimes\bm{\lambda}^{\star}_{p}.
\end{align*}
\end{theorem}
The commodity prices given by Theorem \ref{thm:theorem_priceformation} are analogous in structure to the LMPs resulting from the prevalent LP-based electricity market-clearing frameworks. Specifically, the optimal prices $\bm{\Pi}_{p}^{\star}$ comprise of a nodal price component for each commodity $\bm{\Lambda}_{p}^{\star}$ and a network price component which is non-zero only if congestion arises in the network as a consequence of the power flows along the lines to fulfill the trades.

\subsection{$\mathcal{M}^{\text{c}}$ as a Spatial Equilibrium Problem}\label{subsec:3_2_spatial_price_equilibrium}
First, we show the equivalence of the optimization problem $\mathcal{M}^{\text{c}}$ to a spatial equilibrium problem. Referring to the two roles played by the system operator discussed in  \cref{sec:1_introduction}, consider a virtual separation of the system operator into a \textit{market operator} and a \textit{network operator}. First, the market operator, acting as a \textit{price setter}, collects the conic market bids from market participants and is responsible for clearing the day-ahead market and for the real-time operation of the system. Second, the network operator, responsible for the physical fulfillment of the commodities, acts as a \textit{spatial arbitrager} to collect a non-zero revenue, \textit{congestion rent}, whenever the trades lead to congestion in the network.

Consider an equilibrium problem $\mathcal{E}^{\text{c}}$ comprised of a set of individual optimization problems of the $i\in\mathcal{I}$ market participants, the network operator and the market-clearing conditions. For market participant $i$ located at node $n_{i}\in\mathcal{N}$, let $\mathbf{W}_{ip}\in\mathbb{R}^{N \times T}$ denote the quantities of the commodity $p\in\mathcal{P}$ transacted (bought or sold) over the $N$ nodes at the $T$ hours. Under the perfect competition assumption, the commodity prices $\bm{\Pi}_{p},\;\forall p$ are exogenous to the participants' optimization problem. Therefore, participant $i$ maximizes her profit by deciding the optimal contributions $\bm{q}_{ip},\;\forall p\in\mathcal{P}$ commodities traded in the market, subject to her operational constraints, by solving the problem
\begin{subequations}
\begin{align}
    \maximize{\bm{q}_{i},\;\bm{z}_{i},\mathbf{W}_{ip}}\quad&\sum_{p\in\mathcal{P}} \text{tr}({{\bm{\Pi}_{p}}^{\top}}\mathbf{W}_{ip}) -  \sum_{t\in\mathcal{T}} \left({z}_{it} + {\mathbf{c}_{it}^{\text{L}}}^{\top}\bm{q}_{it}\right) \notag \\
    &\quad \quad -\sum_{t\in\mathcal{T}}{\bm{\omega}_{t}}^{\top} \Big(\sum_{p\in\mathcal{P}}\big([\mathbf{W}_{ip}]_{(:,t)} - \mathbb{I}_{n_{i}}[\mathbf{G}_{ip}\bm{q}_{ip}]_{t}\big)\Big) \label{objfun:i_prof_max}\\
    \st\quad&\norm{\mathbf{C}^{\text{Q}}_{it}\bm{q}_{it}}^{2} \le  {z}_{it},\;\forall t &:& (\bm{\mu}^{\text{Q}}_{it},\;\kappa^{\text{Q}}_{it},\;\nu^{\text{Q}}_{it}) \label{cons:i_prof_max_obj_ref}\\
    &\norm{\mathbf{A}_{ij}\bm{q}_{i} + \mathbf{b}_{ij}} \leq \mathbf{d}^{\top}_{ij}\bm{q}_{i} + e_{ij},\;\forall j\in\mathcal{J}_{i} &:&(\bm{\mu}_{ij},\;{\nu}_{ij})\label{cons:i_prof_max_soc}\\
    &\mathbf{F}_{i}\bm{q}_{i} = \mathbf{h}_{i} &:& (\bm{\gamma}_{i}) \label{cons:i_prof_max_eq}\\
    &\mathbf{W}_{ip}^{\top}\mathbb{1} = \mathbf{G}_{ip}\;\bm{q}_{ip}, \;\forall p \in \mathcal{P}, &:& (\bm{\widehat{\lambda}}_{ip}) \label{cons:transaction_product_equivalence} 
    \end{align}\label{prob:i_prof_max}
\end{subequations}
where the three terms comprising the objective function \eqref{objfun:i_prof_max} are the revenues generated by the transactions, the cost (utility) of production (consumption) and the third term is the transport cost for the physical fulfillment of the transactions at the nodes where they are delivered. Here, the indicator vector $\mathbb{I}_{n_{i}}\in\mathbb{R}^{N}$ contains the element corresponding to the location of participant $i$ as $1$, while all other elements are $0$. Constraints \eqref{cons:transaction_product_equivalence} ensure a balance of the commodity transactions with the contributions towards the trades as given by Definition \ref{def:commodity_trades}. Variables $\bm{\widehat{\lambda}}_{ip}\in\mathbb{R}^{T}$ are the Lagrange multipliers associated with these constraints. Note that the nodal commodity prices $\bm{\Pi}_{p}$ and the price of transmitting power $\bm{\omega}_{t}\in\mathbb{R}^{N},\;\forall t$ are set by the market operator and, as such, are considered as fixed by the participant.

Next, the network operator maximizes congestion rent, while fulfilling the trades of all the commodities in the market. Let $\mathbf{y}_{t} \in \mathbb{R}^{N},\;\forall t$ denote the net power injection comprising of all commodities and all market participants at the $N$ nodes at each hour. The network operator solves the following maximization problem subject to the network limits
\begin{align}
    \maximize{\mathbf{y}_{t}}\quad  \sum_{t\in\mathcal{T}}\;{\bm{\omega}_{t}}^{\top}\mathbf{y}_{t}\quad\st\quad  -\overline{\mathbf{s}} \leq \bm{\Psi}\mathbf{y}_{t}\leq \overline{\mathbf{s}},\quad\colon(\underline{\bm{\varrho}}_{t},\;\overline{\bm{\varrho}}_{t}),\label{prob:NO_prof_max}
\end{align}
where ${\bm{\omega}}_{t}$ are variables exogenous to the network operator and the constraints limit the power flows along the lines to their rated capacity in both directions.

Lastly, the market operator clears the market based on the following equalities
\begin{subequations}
\begin{align}
    &\sum_{p\in\mathcal{P}}[{\mathbf{Q}_{p}^{\text{inj}}}]_{(:,t)} = \mathbf{y}_{t},\;\forall t &:& (\bm{\omega}_{t}) \label{mo:transaction_cost_balance}\\
    &\sum_{i\in\mathcal{I}}\mathbf{G}_{ip}\bm{q}_{ip} = \mathbb{0},\;\forall p\in\mathcal{P}, &:& (\bm{\lambda}_{p})\label{mo:nodal_price_balance}
\end{align}\label{mo:equalities}
\end{subequations}
where the auxiliary variable $\mathbf{Q}_{p}^{\text{inj}}\in\mathbb{R}^{N \times T}$ denotes the commodity-specific net nodal injections
\begin{align}
    \mathbf{Q}_{p}^{\text{inj}} &\coloneqq \begin{bmatrix}\sum_{i \in \mathcal{I}_{1}}{(\mathbf{G}_{ip}\bm{q}_{ip})}^{\top} \quad \sum_{i \in \mathcal{I}_{2}}{(\mathbf{G}_{ip}\bm{q}_{ip})}^{\top} \quad \cdots \quad  \sum_{i \in \mathcal{I}_{N}}{(\mathbf{G}_{ip}\bm{q}_{ip})}^{\top} \end{bmatrix}^{\top}. \label{eq:def_Qp_inj}
\end{align}
The market-clearing equalities \eqref{mo:transaction_cost_balance} ensure the net injection at each of the nodes of the network is balanced by the transport service provided by the network operator. The shadow price of this constraint, $\bm{\omega}_{t}$, appears as a parameter in the network operator's maximization \eqref{prob:NO_prof_max} and is interpreted as the price of transmitting power from an arbitrary hub to the each of the nodes. As discussed previously, \eqref{mo:nodal_price_balance} ensure the system-wide balance of traded commodities.
\begin{theorem}[Competitive Spatial Equilibrium]\label{thm:comp_spatial_price_eqm}
The convex market-clearing problem $\mathcal{M}^{\text{c}}$ solved centrally by the system operator is equivalent to a competitive spatial equilibrium $\mathcal{E}^{\text{c}}$ comprised of market participants, $i\in\mathcal{I}$ each solving the profit maximization \eqref{prob:i_prof_max}, the network operator solving the congestion rent maximization \eqref{prob:NO_prof_max} and the market operator clearing the market by enforcing the equalities \eqref{mo:equalities}.
\end{theorem}
The proof for Theorem \ref{thm:comp_spatial_price_eqm} relies on the equivalence of the Karush-Kuhn-Tucker (KKT) optimality conditions of the two problems. 
\begin{corollary}[Existence and Uniqueness]\label{cor:existence_uniqueness}
The solution to the competitive spatial price equilibrium problem $\mathcal{E}^{\text{c}}$ exists and is unique in allocations $\bm{q}_{i}^{\star}$, provided all market participants bid with non-zero quadratic price components, i.e., $c_{itk}^{\text{Q}} \neq 0,\;\forall k=1,2,\dots,K_{i},\;\forall t,\;\forall i\in\mathcal{I}$.
\end{corollary}

While Corollary \ref{cor:existence_uniqueness} yields uniqueness of allocations  $\bm{q}_{i}^{\star}$, conditioned on the strict convexity of the objective function \eqref{objfun_Mcone}, no such guarantees on the uniqueness of the prices $\bm{\Pi}_{p}^{\star}$ can be given since the dual problem to \eqref{prob:Mcone} (formulated in \ref{app:dual_problem_formulation}) does not admit a strictly convex objective function. The conditions on the uniqueness of the allocations at equilibrium closely correspond to those in prevalent LP-based markets\footnote{Recently, \cite{Krebs2018} provided conditions on network connectivity, PTDF matrix parameters, and the specific structure of participants' cost functions under which unique allocations are obtained at market equilibrium with linear constraints. Further research in similar directions is needed to study such conditions for the proposed SOCP-based market.}. Next, we analyze the desired economic properties underlying the equilibrium $\mathcal{E}^{\text{c}}$. 
\begin{theorem}[Economic Properties]\label{thm:theorem_economic_properties}
The market-clearing problem $\mathcal{M}^{\text{c}}$ and its equivalent competitive spatial equilibrium $\mathcal{E}^{\text{c}}$ result in optimal allocations $\bm{q}_{i}^{\star},\;\forall i\in\mathcal{I}$ and spatial commodity prices $\bm{\Pi}_{p}^{\star},\;\forall p \in \mathcal{P}$ such that the following economic properties are attained at optimality:
\begin{enumerate}[label=(\roman*)]
    \item \textbf{Market efficiency:} Under the perfect competition assumption, social welfare is maximized, such that no participant desires to unilaterally deviate from the market-clearing outcomes.
    \item \textbf{Cost recovery:} Let the market bids $\mathcal{B}_{i}$ for each market participant $i\in\mathcal{I}$ be such that  $e_{ij}\geq\norm{\mathbf{b}_{ij}},\;\forall j\in\mathcal{J}_{i}$ and $\mathbf{h}_{i}=\mathbb{0}$. Then, the optimal allocations $\bm{q}_{i}^{\star},\;
    \forall i\in\mathcal{I}$ and optimal spatial commodity prices $\bm{\Pi}_{p}^{\star},\;\forall p\in\mathcal{P}$ ensure cost recovery for the market participants.
    \item \textbf{Revenue adequacy:} The market operator does not incur financial deficit at the end of the market-clearing horizon, i.e.,
      \begin{align*}
        \sum_{p\in\mathcal{P}}\sum_{i\in\mathcal{I}} \text{tr}({{\bm{\Pi}_{p}^{\star}}}^{\top}\mathbf{W}_{ip}^{\star}) - \sum_{t\in\mathcal{T}}{\bm{\omega}_{t}^{\star}}^{\top}\mathbf{y}_{t}^{\star} \geq 0.
    \end{align*}
\end{enumerate}
\end{theorem}
Theorem \ref{thm:theorem_economic_properties} characterizes the economic properties underlying the market-clearing outcomes from the conic market-clearing $\mathcal{M}^{\text{c}}$. While market efficiency is guaranteed under the perfect competition assumption, cost recovery is ensured for all participants under two conditions, which we elaborate in the following. First, as shown in the modeling examples in the Supplementary Material, the condition $e_{ij}\geq\norm{\mathbf{b}_{ij}}$ holds true for all participants in most practical settings, with the notable exception of market participants having a non-zero lower bound on their decision variables. The practical issue of non-guarantee of cost recovery for such market participants also prevails in currently-operational LP-based electricity markets. The second condition relates to the homogeneity of the linear equality constraints \eqref{cons:Mcone_participants_eq}, i.e., $\bm{q}_{i} = \mathbb{0},\;\forall i$ is a feasible solution to the market-clearing problem. Observe that, the feasibility of a zero-allocation solution, i.e., $\bm{q}_{i} = \mathbb{0},\;\forall i$ also satisfies the first condition, thereby indicating that these two conditions are equivalent. Lastly, the condition for revenue adequacy for the market operator requires that the net payments received from the market participants $i\in\mathcal{I}$ less the payments made by the market operator to the network operator towards transmission service is non-negative. 

\begin{remark}[Incentive Compatibility]
The incentive for actors to deviate from price-taking, perfectly-competitive behavior decreases to zero as the number of actors goes to infinity \cite{Roberts1976}. Thus, assuming a very large number of market participants, the conic market-clearing proposal tends towards incentive compatibility, i.e., at the limit participants bid according to their \textit{true} preferences. The conditions under which incentive compatiobility is satisfied by $\mathcal{M}^{\text{c}}$ is akin to the prevalent LP-based electricity markets, thereby preserving this desired economic property in the move towards an SOCP-based market-clearing framework.
\end{remark}

Beyond the desired economic properties discussed so-far, we analyze the robustness of the market-clearing outcomes resulting from $\mathcal{M}^{\text{c}}$ against small changes in parameters. Strong duality aside, in the context of a market-clearing problem, this property is crucial to be studied for SOCP problems. \textit{Robust solvability} refers to the property of an SOCP problem such that it remains solvable and obtains the optimal solution even when the problem parameters are changed by arbitrary small perturbations. Proposition 1.4.6 in \cite{BenTal2001} establishes that robust solvability is guaranteed by the strict feasibility of both primal and dual problems. The following corollary to Theorem \ref{thm:strongduality} formalizes robust solvability of the conic market-clearing problem.
\begin{corollary}[Robust Solvability of $\mathcal{M}^{\text{c}}$]\label{cor:robust_solvability} 
The market-clearing problem $\mathcal{M}^{\text{c}}$ is robust solvable, i.e., for the $p\in\mathcal{P}$ commodities traded in the market, both the optimal contributions $\bm{q}_{ip}^{\star},\;\forall i$ and the optimal prices $\bm{\Pi}_{p}^{\star}$ obtained, are robust against small perturbations in the parameters comprising the bids $\mathcal{B}_{i},\;\forall i \in \mathcal{I}.$
\end{corollary}
The proof to Corollary \ref{cor:robust_solvability} is a direct consequence of Theorem \ref{thm:strongduality}. This result ensures robust outcomes from a large-scale market-clearing problem admitting hundreds (possibly thousands) of market participants and their SOC constraints. Robust solvability status of the market-clearing problem implies that numerical pathologies arising from approximations, e.g., rounding-off errors, estimation of uncertain parameters, etc. do not cause solvability-related issues. %economic interpretations
\section{Numerical Studies}\label{sec:4_numerical_studies}
We perform numerical experiments on a 24-node electricity system wherein the various market participants, comprising of 6 wind power producers, 9 flexible and 3 inflexible power producers (PPs), 3 identical energy storage units (ESUs) owners, and 17 inflexible consumers, are connected at the nodes as shown in Figure \ref{fig:prices_and_network_Mcc}(a). The system data is adapted from \cite{Conejo2010} to include the wind farms and ESUs. We study the market-clearing outcomes under various renewable energy share (RES) paradigms\footnote{These paradigms are derived by suitably varying the installed capacity of the wind farms while dimensioning the ESUs such that the total available charging/discharging capacity remains fixed at 12.5\% of the total wind farm installed capacity. Such dynamic dimensioning of storage-related flexibility is crucial to ensure market-clearing feasibility for the high RES paradigms.}, ranging from 10\% to 60\% of the total energy demand from consumers met by the wind power producers. The wind power producers bid at zero prices to ensure acceptance of bids, and ESU owners bid at prices lower than the cheapest flexible PP. We assume that the inflexible loads exhibit perfect inelasticity of demand, rendering the social welfare maximization problem equivalent to finding cost-minimal dispatch, in expectation, for PPs and ESUs to provide energy and flexibility services needed to meet the net demand. Here, net demand refers to the energy demand from inelastic loads reduced by the production from wind farms during the real-time operation. There is no uncertainty in the inelastic demand. The expected net demand based on the day-ahead wind power production forecasts for the 50\% RES paradigm is shown in Figure \ref{fig:prices_and_network_Mcc}(b). Naturally, the market operator faces high uncertainty in the hours with smaller net demand due to the high share of weather-dependent renewable energy. To highlight the impact of network congestion on spatial prices and quantities, we consider two network configurations: (i) without any network bottlenecks and (ii) with network bottlenecks induced by reducing capacities of three transmission lines of the network, shown in blue in Figure \ref{fig:prices_and_network_Mcc}(a).
\begin{figure}
    \centering
     \tikzstyle{BOXY} = [rectangle, rounded corners = 5, minimum width=10, minimum height=10,text centered, draw=black, fill=white,line width=0.3mm,font=\footnotesize\linespread{0.5}\selectfont]
        \begin{tikzpicture}
        %Basic Market Operation
        \foreach \x in {0,0.15,...,3.45} 
          \draw [fill=gray!20,gray!20] (\x,0) rectangle (\x+0.1,0.6);
        \foreach \y in {3.8,3.95,...,7.25} 
        \draw [fill=gray!40,gray!40] (\y,0) rectangle (\y+0.1,0.6);
        %Energy Market Clearing
        \draw [fill=gray!60,gray!60] (1.65,0) rectangle (1.75,0.6);
        \draw [->,thick] (1.7,1.1) -- (1.7,0.7);
        %\draw [red,fill=red] (1.7,0.8) circle [radius=0.05];
        \node [align=left] (box1) [BOXY,above] at (0.25,1.1) {\textbf{\footnotesize{Day-ahead Market-Clearing}} \\ \footnotesize{Commodity 1: energy} \\
        \footnotesize{Commodity 2: adjustment policy (flexibility)}};
  
        %IntraDay Market
        \draw [fill,gray] (3.5,0.8) circle [radius=0.05];
        \foreach \z in {3.85,4,...,7.25}
        	\draw [fill,gray] (\z,0.8) circle [radius=0.05];
        	
        %Activation
    %    \draw [fill=green!20,green!20] (5.6,0) rectangle (5.7,0.6);
        \draw [->,thick] (5.65,1.3) -- (5.65,0.9);
        % \draw [red,fill=red] (5.65,0.8) circle [radius=0.05];
        \node [align=left] (box1) [BOXY,above] at (6,1.1) {\textbf{\footnotesize{Real-time Operation}} \\ $\bullet$ \footnotesize{energy physically delivered} \\ $\bullet$ \footnotesize{hourly-activated policies}};
        
        %Labels
        \draw[thick] (3.7,-0.2) -- (3.7,0.9);
        \node[align=center] at (2,.3){\footnotesize Day: ($D-1$)};
        \node[align=center] at (5.2,.3){\footnotesize Day: $D$};
        \node[below] at (3.85,0){\tiny 1};
        \node[below] at (5.65,0){\tiny t};
        \node[below] at (5.6,-0.2){\footnotesize Hours};
        \node[below] at (7.3,0){\tiny 24};
        \end{tikzpicture}
        \vspace{-0.5cm}
    \caption{Illustration of commodities traded in an uncertainty-aware electricity market}
    \label{fig:commodities_market_timeline}
    \vspace{-0.25cm}
\end{figure}
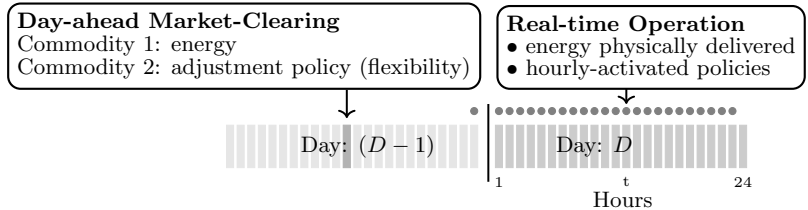
Based on this setup, \cref{subsec:4_1_market_analysis} demonstrates the uncertainty-awareness of our market-clearing proposal, followed by an analysis of the market-clearing outcomes and conic spatial equilibrium prices. In \cref{subsec:4_2_compare_LP_benchmarks}, we compare the proposed SOCP market-clearing problem with two uncertainty-aware alternatives within the LP domain. While we discuss our numerical results solely in the context of an uncertainty-aware market framework in this section, the examples provided in the Supplementary Material can be used to extend the market-clearing problem to include asset- and/or network-awareness.
\subsection{SOCP-based Uncertainty-aware Energy and Flexibility Market}\label{subsec:4_1_market_analysis}
Recalling Example \ref{example:chance_constraints}, we consider a chance-constrained electricity market-clearing problem wherein two commodities, energy and adjustment policies, are cleared by the system operator. \nt{Figure \ref{fig:commodities_market_timeline} illustrates such a two-commodity market, highlighting the day-ahead clearing and the activities during the real-time operation.} The commodities are traded such that an uncertainty-aware spatial price equilibrium is achieved. 

\paragraph{\nt{Adjustment policies as a flexibility service:}}Optimal adjustment policies allocated to flexibility providers enable the mitigation of uncertainty realized during the real-time operation, while look-ahead decisions are made by the system operator at the day-ahead market-clearing stage. Activated during the real-time operation, these policies are in \textit{per unit} and imply the contribution of each flexibility provider towards mitigating the potential real-time imbalance in the system. For example, a flexibility provider may, in the day-ahead market, be allocated a policy corresponding to 10\% adjustment, implying that it contributes to mitigating 10\% of any type of imbalance (either over- or under-supply) during the real-time operation. \nt{Contrary to the prevalent capacity-based reserves \citep{Gonzalez2014}, further discussed in \cref{subsec:4_2_compare_LP_benchmarks}, such adjustment policies when considered in a two-commodity market clearing setup, tightly couple the flexibility provider's actual operational constraints with the delivery of the flexibility service. Consequently, the flexibility procured is dimensioned optimally and can be priced dynamically, consistent with the actual flexibility needs in the electricity system \citep{Ratha2019}. First proposed by \cite{Warrington2013} within a robust optimization framework, such policies are preferable over capacity-based reserves since they are temporally coupled and therefore, potentially exploit the spatio-temporal correlation of forecast errors. This is highly relevant in a multi-period market setting with flexibility providers facing temporally-coupled constraints. Finally, from a practical standpoint, adopting these policies implies that in addition to payments for energy, flexibility providers are paid for their flexibility service upfront, i.e., at the day-ahead stage.}

\paragraph{\nt{Optimal policies in a chance-constrained market:}} \nt{Employing chance-constrained stochastic programs to optimize the adjustment policies in contrast to scenario-based or robust stochastic programs is beneficial from an economic interpretation point of view. First, chance-constrained stochastic programs can be safely approximated by deterministic programs under mild conditions, e.g., the feasibility region formed by the inequalities is polyhedral and the adjustment policies are affine in random variables \citep{Nemirovski2012}. Computational tractability aside, such a problem reformulation is particularly suitable for market settings since the clearing outcomes (commodity quantities and prices) are deterministic, analytically expressable, and therefore, more acceptable to market participants. Furthermore, the convex approximations become exact when the uncertainty follows a Gaussian distribution. Second, the economic properties of the market equilibrium, discussed in \cref{sec:3_eqm_economic}, hold not only for the expected allocations obtained but also for every realization drawn from the probability distribution underlying the system operator's uncertainty model. Referring to Example \ref{example:chance_constraints}, this implies that, for a given choice of $r_{\hat{\varepsilon}}$, Theorem~\ref{thm:theorem_economic_properties} holds for the expected value and every realization $\widehat{\bm{\xi}}$ drawn from the distribution $\mathbb{P}_{\xi}$ characterized by mean $\bm{\mu}$ and $\bm{\Sigma}$ \citep{Dvorkin2020}. These advantages make chance-constrained optimization a suitable framework to incorporate uncertainty in electricity markets\footnote{\nt{In addition to the assumption that distributional knowledge is available to the system operator (or can be estimated from historical data), the modeling of chance constraints adopted in this paper is agnostic to the degree of constraint violations. Future modeling extensions can be made to provide bounds on the cost of expected constraint violations, e.g., adopting the approach by \cite{Chen2010} which remains within the SOCP framework. Finally, beyond the fully stochastic optimization frameworks, quasi-stochastic market-clearing approaches such as \cite{Mays2021}, have recently been proposed to alleviate the adverse impacts of decisions made under uncertainty in day-ahead electricity markets. Although formulated as LP (or MILP) problems, such approaches attempt to approximate the fully stochastic market equilibrium relying on parameters exogenous to the market, e.g., administrative choices made by system operators, outcomes from stochastic reliability problems solved ahead of the day-ahead market, etc.}}.} Considering the novelty of admitting adjustment policies as opposed to the conventional flexible capacity to meet the uncertain net demand, we provide insights into the endogenous consideration of uncertainty and quantify the flexibility payments in the following. The modeling of market participants, chance-constrained market-clearing problem and its SOCP reformulation (which we refer to hereafter as $\mathcal{M}^{\text{cc}}$) is provided in the Supplementary Material.

\begin{figure}[tb!]\scriptsize
\begin{minipage}[]{\columnwidth}
\centering
\input{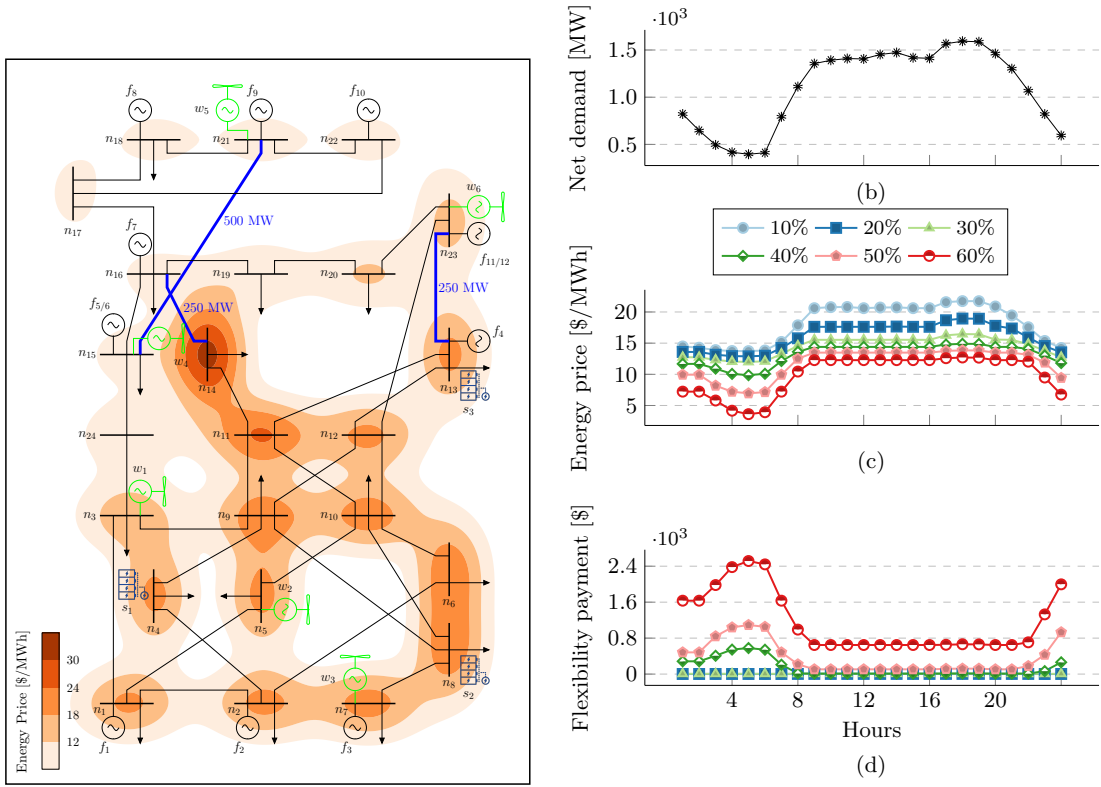}
~
\begin{tikzpicture}
  \pgfplotsset{set layers}
    \pgfplotsset{
       ylabel right/.style={
            after end axis/.append code={
                \node [rotate=90, anchor=north] at (rel axis cs:1.15,0.5) {#1};
            }   
        }
    }
 \pgfplotsset{
        % initialize Dark2
        cycle list/Paired,
        % combine it with 'mark list*':
        cycle multiindex* list={
            mark list*\nextlist
            Paired\nextlist
        },
    }
    \begin{groupplot}[
        group style={
        group size=1 by 3,
        % y descriptions at=edge left,
        horizontal sep = 100pt,
        vertical sep   = 50pt,
    },
  ]
 
% Net Demand Plot
\nextgroupplot[
    width=8cm, height=3.5cm,
    ylabel={\footnotesize Net demand [MW]},
    scaled y ticks=base 10:-3,
    xtick={4,8,12,16,20,24},
    ytick={500, 1000, 1500},
    yticklabels={\footnotesize{0.5}, \footnotesize{1.0},\footnotesize{1.5}},
    xticklabels={,,},
    ymajorgrids=true,
    ylabel shift = 1 pt,
    legend style={
            at={(0.6,1.25)},
            anchor=north east},
    grid style=dashed,
    axis x line*=bottom,
    axis y line*=left
]
\addplot [black, mark=10-pointed star] table [x=hours, y=netdemand, col sep=comma] {netdemand_output.csv};

\nextgroupplot[
    width=8cm, height=3.5cm,
    ylabel={\footnotesize Energy price [\$/MWh]},
    xtick={4,8,12,16,20,24},
    xticklabels={,,},
    ytick={5, 10, 15, 20},
    yticklabels={\footnotesize{5}, \footnotesize{10},\footnotesize{15},\footnotesize{20}},
    ylabel shift = 3 pt,
    legend style={
                at={(0.8,1.6)},
                anchor=north east},
    legend columns = 3,
    ymajorgrids=true,
    grid style=dashed,
    axis x line*=bottom,
    axis y line*=left,
    every axis plot/.append style={thick}]
    \addplot table [x=hours, y=res10, col sep=comma] {el_prices_with_res.csv};
    \addlegendentry{\scriptsize$10\%$};
    \addplot table [x=hours, y=res20, col sep=comma] {el_prices_with_res.csv};
    \addlegendentry{\scriptsize$20\%$};
    \addplot table [x=hours, y=res30, col sep=comma] {el_prices_with_res.csv};
    \addlegendentry{\scriptsize$30\%$};
    \addplot table [x=hours, y=res40, col sep=comma] {el_prices_with_res.csv};
    \addlegendentry{\scriptsize$40\%$};
    \addplot  table [x=hours, y=res50, col sep=comma] {el_prices_with_res.csv};
    \addlegendentry{\scriptsize$50\%$};
    \addplot table [x=hours, y=res60, col sep=comma] {el_prices_with_res.csv};
    \addlegendentry{\scriptsize$60\%$};

\nextgroupplot[
    width=8cm, height=3.5cm,
    xlabel={\footnotesize Hours},
    ylabel={\footnotesize Flexibility payment [\$]},
    xtick={4,8,12,16,20,24},
    xticklabels ={\footnotesize{4},\footnotesize{8},\footnotesize{12},\footnotesize{16},\footnotesize{20}},
    ytick={0, 800,1600,2400},
    yticklabels={\footnotesize{0}, \footnotesize{0.8},\footnotesize{1.6},\footnotesize{2.4}},
    scaled y ticks=base 10:-3,
    ymajorgrids=true,
    grid style=dashed,
    axis x line*=bottom,
    axis y line*=left,
    every axis plot/.append style={thick}]
\addplot table [x=hours, y=res10, col sep=comma] {re_prices_with_res.csv};
\addplot table [x=hours, y=res20, col sep=comma] {re_prices_with_res.csv};
\addplot table [x=hours, y=res30, col sep=comma] {re_prices_with_res.csv};
\addplot table [x=hours, y=res40, col sep=comma] {re_prices_with_res.csv};
\addplot  table [x=hours, y=res50, col sep=comma] {re_prices_with_res.csv};
\addplot table [x=hours, y=res60, col sep=comma] {re_prices_with_res.csv};

\end{groupplot}
\node at (3.2cm, -0.4cm) {\footnotesize (b)};
\node at (3.2cm, -4.2cm) {\footnotesize (c)};
\node at (3.2cm, -8.5cm) {\footnotesize (d)};
\end{tikzpicture}
\end{minipage}
\caption{(a) 24-node electricity network showing a visualization of spatial prices of energy for the network configuration with bottlenecks, (b) expected net demand for the 50\% RES paradigm, (c) system-wide prices for energy and (d) the total hourly flexibility payments for various RES paradigms} 
\label{fig:prices_and_network_Mcc}
\vspace{-0.4cm}
\end{figure}
\paragraph{Impact of congestion and uncertainty on prices:}
The density plot in Figure \ref{fig:prices_and_network_Mcc}(a) visualizes the impact of network bottlenecks on the day-ahead energy prices for hour 23 under the 50\% renewable energy share paradigm. Figures \ref{fig:prices_and_network_Mcc}(c) and \ref{fig:prices_and_network_Mcc}(d) show the commodity prices for the network configuration without bottlenecks for the various RES paradigms. Observe that with higher shares of renewable energy, the payment made by the market operator towards flexibility increases, complementary to the gradual reduction in the energy price due to wind farms bidding with zero prices. Overall, increasing uncertainty faced at the day-ahead market-clearing stage leads to lower energy prices while the payments towards flexibility services increase, thereby resulting in the right market signals for investments in flexibility over the long run. Note that, since the adjustment polices are quantified in per unit, the hourly flexibility payments shown in Figure \ref{fig:prices_and_network_Mcc}(d) correspond to total payments made by the market operator towards flexibility, adopting an allocation determined by the adjustment policies of individual flexibility providers and as such, following a \textit{differentiated pricing} scheme. We now discuss the allocation of adjustment policies and provide further insights into the pricing of flexibility.
\input{plots/FigureCompareSOCPDispatchNetworkconfig}
\paragraph{Flexibility allocation and payments:}
For the 50\% RES paradigm, Figures \ref{fig:socp_dispatch_networkconfig}(a)-\ref{fig:socp_dispatch_networkconfig}(f) show the optimal allocation of dispatch and adjustment policies to the PPs ($f1,\;f2,\dots,\;f12$) and to the ESUs ($s1,\;s2,\;s3$) for selected hours of the day for both network configurations. First, observe that non-zero adjustment policies are only allocated to flexibility providers that are also dispatched for the commodity energy, which is consistent with the requirement that both over- and under-supply imbalances during the real-time operation are mitigated by the flexibility delivered. Second, the network configuration with bottlenecks mandates the allocation of adjustment policies to more number of flexible power producers, as network congestion is expected to impact the flexibility delivery during real-time operation. However, note that ESUs are not allocated adjustment policies in this configuration. This is explained by (i) the availability of flexible PPs in favorable locations of the network with respect to congested power lines, and (ii) the inter-temporal constraints and end-of-day energy balance requirement for ESUs (see Supplementary Material). Indeed, for the paradigm with 60\% RES (not shown in the Figure), ESUs in the case with network bottlenecks are allocated non-zero adjustment policies to contribute towards flexibility provision, with the market-clearing problem choosing a more expensive flexibility allocation, offset by the reduced cost of energy provision. Finally, observe in Figure \ref{fig:socp_dispatch_networkconfig}(b) that the adjustment policies may take negative values, implying that the action (increasing/decreasing) production by power producers may be in opposition to the system requirement (under-/over-supply), provided it leads to cost-optimal flexibility provision under congested network conditions.

Figures \ref{fig:socp_dispatch_networkconfig}(g)-\ref{fig:socp_dispatch_networkconfig}(h) show the flexibility payments made to flexibility providers under the two network configurations. Flexibility payments are, in general, higher for the case with network bottlenecks as compared to the case without any bottlenecks. Moreover, as previously-discussed in reference to Figure \ref{fig:prices_and_network_Mcc}(d), flexibility payments are higher for hours with high production from wind farms. In the short run, this incentivizes market participants to bid their flexibility in these hours. To further analyze how flexibility is valued and paid for by the market operator, we introduce an ex-post parameter called \textit{flexibility payment rate} (FPR). Defined for each flexibility provider $i$, FPR is the rate in $\$/\text{MWh}$ at which \nt{it} is paid for the flexibility service:
\begin{align}
    \text{FPR}_{it} = \frac{[\bm{\Pi}_{p}^{\star}]_{(n_{i},t)}\times\alpha^{\star}_{it}}{\abs{\hat{q}^{\star}_{it} - \overline{q}_{it}}},
\end{align}
where $[\bm{\Pi}_{p}^{\star}]_{(n_{i},t)}$ retrieves the price of the commodity flexibility service at the node $n_{i}$ where the participant $i$ is located, $\hat{q}_{it}^{\star}$ and $\alpha^{\star}_{it}$ denote the nominal dispatch and adjustment policy allocated to the participant $i$, respectively. The quantity $\overline{q}_{it}$ is the dispatch under the perfect forecast case, i.e., when the day-ahead market-clearing is deterministic. We obtain $\overline{q}_{it}$ by solving the market-clearing problem $\mathcal{M}^{\text{cc}}$ assuming day-ahead forecasts are realized perfectly during the real-time operation, such that all adjustment policies are set to zero, i.e., $\alpha_{it}=0,\;\forall t,\;\forall i\in\mathcal{I}$. Figures \ref{fig:socp_dispatch_networkconfig}(i) - \ref{fig:socp_dispatch_networkconfig}(j) show the FPRs for the various participants for the two network configurations. Observe that, equal segments within each bar indicate equal FPRs for all flexibility providers for a given hour. First, we note that, in general, more flexibility providers being allocated non-zero adjustment policies in the network configuration with bottlenecks leads to lower FPRs for the flexibility providers contrary to the one without bottlenecks. A notable exception is hour 7 in the case without bottlenecks, wherein ESUs are paid for the flexibility at very high rate. This is a consequence of two contributing factors: (i) ESUs are dispatched with small nominal quantities in the hours 7 and 8 (see Figure \ref{fig:socp_dispatch_networkconfig}(e)) which implies that their ability to provide both charging and discharging flexibility is valued highly, and (ii) the net demand follows a steep rise in the hours 6-8 (see Figure \ref{fig:prices_and_network_Mcc}(b)) and as a result the market-clearing problem faces a scarcity of not only flexible capacity, but also the ramping ability needed to meet this change in uncertainty. In contrast, in hour 8, while the ESUs are still nominally dispatched close to zero, the flexibility is no longer scarce in the system as the net demand rises sufficiently enough to lead to the nominal dispatch of power producer $f1$, thereby homogenizing the FPRs again. Overall, the FPR for a flexibility provider depends on a number of factors, including the level of uncertainty perceived by the system (quantified by the forecast error covariance matrix as well as the day-ahead forecasts), the location of the flexibility provider, congestion in the network and whether other flexibility providers are available. 
\subsection{Comparison with LP-based Uncertainty-aware Benchmarks}
\label{subsec:4_2_compare_LP_benchmarks}
Next, using numerical simulations corresponding to realizations of the uncertain wind power production, we compare the market-clearing outcomes and performance of the proposed SOCP-based uncertainty-aware market-clearing problem $\mathcal{M}^{\text{cc}}$ to alternatives within the LP-domain. 
\begin{figure}
    \centering
\tikzstyle{BOXY} = [rectangle, rounded corners = 5, minimum width=10, minimum height=10,text centered, draw=black, fill=white,line width=0.3mm, font=\footnotesize]
    \begin{tikzpicture}[scale=0.75, every node/.style={scale=0.75}]
    \node [align=center] (box1) [BOXY] {Forecast error \\ samples};
    \node [align=center] (box2) [BOXY, right of=box1, xshift=2.5cm, yshift=2cm, fill=maincolor!10] {Mean,\\covariance};
    \node [align=center] (box3) [BOXY, right of=box1, xshift=2.5cm, yshift=0cm, fill=secondcolor!10] {Minimum reserve \\requirement};
    \node [align=center] (box4) [BOXY, right of=box1, xshift=2.5cm, yshift=-2cm, fill=thirdcolor!40] {Scenarios};
    
    \node [align=center] (box5) [BOXY, right of=box2, xshift=2cm, yshift=0cm,fill=maincolor!10] {SOCP \\ chance-constrained\\market clearing};
    \node [align=center] (box6) [BOXY, right of=box3, xshift=2cm, yshift=0cm, fill=secondcolor!10] {LP \\ deterministic\\market clearing};
    \node [align=center] (box7) [BOXY, right of=box4, xshift=2cm, yshift=0cm, fill=thirdcolor!40] {LP \\ scenario-based \\market clearing};
    
    \node [align=center] (box8) [BOXY, right of=box5, xshift=3cm, yshift=0cm] {LP \\ deterministic\\flexibility activation};
    \node [align=center] (box9) [BOXY, right of=box6, xshift=3cm, yshift=0cm] {LP \\ deterministic \\real-time markets};
    \node [align=center] (box10) [BOXY, right of=box7, xshift=3cm, yshift=0cm] {LP \\ deterministic \\real-time markets};
    
     %paths
    \draw [black,->,=>stealth,line width=0.3mm] (box1) -- (box2.west);
    \draw [black,->,=>stealth,line width=0.3mm] (box1) -- (box3);
    \draw [black,->,=>stealth,line width=0.3mm] (box1) -- (box4.west);

    \draw [black,->,=>stealth,line width=0.3mm] (box2) -- (box5);
    \draw [black,->,=>stealth,line width=0.3mm] (box3) -- (box6);
    \draw [black,->,=>stealth,line width=0.3mm] (box4) -- (box7);
    
    \draw [black,->,=>stealth,line width=0.3mm] (box5) -- (box8);
    \draw [black,->,=>stealth,line width=0.3mm] (box6) -- (box9);
    \draw [black,->,=>stealth,line width=0.3mm] (box7) -- (box10);
    
    %Texts
    \node [align=left, right of = box8, xshift=1.5cm] {\Large $\mathcal{M}^{\text{cc}}$};
    \node [align=left, right of = box9, xshift=1.5cm] {\Large $\mathcal{R}1$};
    \node [align=left, right of = box10, xshift=1.5cm] {\Large $\mathcal{R}2$};
    
    \node [align=left, above of = box2, yshift=0.25cm, gray] {PRE-MARKET};
    \node [align=left, above of = box5, yshift=0.25cm, gray] {DAY-AHEAD};
    \node [align=left, above of = box8, yshift=0.25cm, gray] {REAL-TIME};
    
    \end{tikzpicture}
    \caption{Illustration of the SOCP market-clearing and the LP-based reference problems}
    \label{fig:conicMC_and_benchmarks}
    \vspace{-0.5cm}
\end{figure}
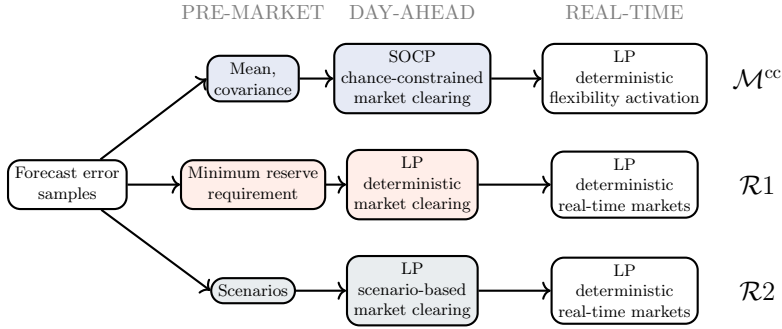

We consider two LP-based market-clearing problems as benchmarks: $\mathcal{R}1$ and $\mathcal{R}2$. While it is most closely related to the currently-operational electricity markets, the deterministic market-clearing problem $\mathcal{R}1$ considers uncertainty during the day-ahead clearing stage by commissioning flexibility in the form of reserve capacity. The procured capacity is subsequently activated in real-time electricity markets, cleared closer to physical delivery. To ensure a cost-optimal allocation of the reserve capacity, the market operator enforces a \textit{exogenously-determined} minimum reserve requirement to procure flexible capacity from flexibility providers. Market-clearing problem $\mathcal{R}2$ considers uncertainty based on day-ahead scenarios for realizations of uncertain renewable energy production. The market operator seeks to maximize the expected social welfare under uncertainty, resulting in a day-ahead schedule which is then adjusted during real-time operation for each of the foreseen scenarios. Similarly to $\mathcal{M}^{\text{cc}}$, this approach considers uncertainty endogenously as parameterized by the scenarios considered while solving the market-clearing problem. However, leaving aside the fundamental question of trusting the scenario-generating agent, this approach mandates a large number of scenarios to appropriately represent the uncertainty, thereby, limiting its practical adoption from a computational perspective. Figure \ref{fig:conicMC_and_benchmarks} illustrates the market-clearing activities, starting with pre-market processing of forecast error samples to generate statistical moments (mean and covariance), to dimension the reserve requirement and to generate scenarios, for $\mathcal{M}^{\text{cc}}$, $\mathcal{R}1$ and $\mathcal{R}2$, respectively. In our numerical studies, we impose a minimum reserve requirement such that the probability of demand curtailment, due to unavailability of flexible capacity to be dispatched during the real-time operation, is less than 5\% for the forecast error samples considered. Moreover, observe that the real-time operation in  $\mathcal{M}^{\text{cc}}$ does not involve pricing and can be done without re-optimization provided uncertainty bounds defined by statistical moments are reliable, whereas both $\mathcal{R}1$ and $\mathcal{R}2$ involve real-time markets, whose outcomes are used to price flexibility services. As with $\mathcal{M}^{\text{cc}}$, the market-clearing formulations and further details on the two reference problems $\mathcal{R}1$ and $\mathcal{R}2$ are provided in the Supplementary Material. 
\begin{figure}
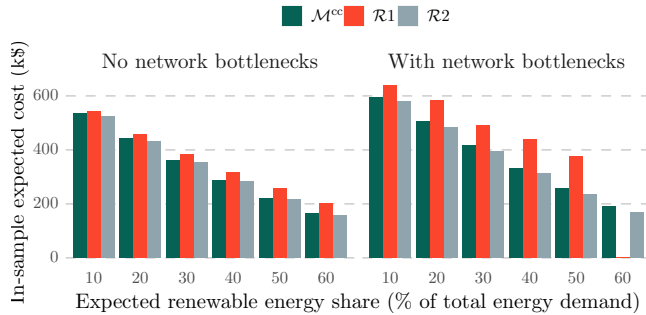

    \centering
    \include{plots/CS1ExpCostWithRES}
    \caption{In-sample market-clearing cost in the various RES paradigms considered}
    \label{fig:cs1_insample_vs_RES}
    \vspace{-0.4cm}
\end{figure}
\paragraph{In-sample market-clearing results:} First, we study the \textit{in-sample} performance of the market-clearing problems. In-sample refers to the forecast samples used in the pre-market stage to obtain the parameters and scenarios for the market-clearing problems $\mathcal{M}^{\text{cc}}$ and the benchmarks $\mathcal{R}1$ and $\mathcal{R}2$, see Figure \ref{fig:conicMC_and_benchmarks}. For the 6 wind farms in the network, we consider 100 forecast error samples drawn from a multivariate Gaussian distribution having zero mean and a standard deviation of 10\% of the nominal day-ahead forecast values. Consequently, the safety parameter $r_{\varepsilon}$ is the inverse cumulative distribution function of the standard Gaussian distribution evaluated at $(1-\varepsilon)$--quantile. For $\mathcal{M}^{\text{cc}}$, we fix the system operator's constraint violation probability at $\varepsilon = 0.05$. Figure \ref{fig:cs1_insample_vs_RES} shows a comparison of the expected day-ahead market-clearing cost for the three market-clearing problems under the two network configurations in the various RES paradigms. Observe that, with increasing share of renewable energy, due to its exogenous consideration of uncertainty, market-clearing problem $\mathcal{R}1$ performs increasingly worse compared to the other problems for both the network configurations, leading to infeasibility of market-clearing problem at 60\% renewable energy share for the case with network bottlenecks. Furthermore, market-clearing problems $\mathcal{M}^{\text{cc}}$ and $\mathcal{R}2$ result in comparable in-sample expected cost. However, it is worth noting that, contrary to $\mathcal{M}^{\text{cc}}$, $\mathcal{R}2$ does not provide any guarantees on the feasibility of market-clearing problem when faced with scenarios beyond those considered as in-sample\footnote{Depending on the constraint violation probability $\varepsilon$ and the number of decision variables, \cite[Theorem 4]{Alamo2015} provides an analytical expression for the number of scenarios that must be considered while solving $\mathcal{R}2$ to obtain the same probabilistic feasibility guarantee as $\mathcal{M}^{\text{cc}}$. For the network considered in this numerical study, this evaluates to $>80,000$ scenarios, thereby rendering problem $\mathcal{R}2$ unable to provide such a guarantee while being cleared within the desired day-ahead market-clearing solve times (typically, about 1-2 hours).}. We discuss the out-of-sample performance further in the following. The first two rows in Table \ref{tab:cost_and_time} provide a comparison for the cost and computation times of the various market-clearing problems in the 50\% RES paradigm.
\paragraph{Out-of-sample performance:}
To examine the out-of-sample performance, we perform a deterministic real-time market clearing  such that the day-ahead decisions made for each of the market-clearing problems are fixed while adjustments are made in real-time to meet the uncertainty realization. For the market-clearing problem $\mathcal{M}^{cc}$, this implies that the flexible market participants adjust their production in real-time strictly in accordance to the adjustment policies assigned at the day-ahead market-clearing stage. On the other hand, for the LP-based reference market-clearing problems $\mathcal{R}1$ and $\mathcal{R}2$, the adjustments in real-time are allowed up to the flexible capacity limits obtained at the day-ahead stage. Consistent with the prevalent practice on handling real-time adjustments in electricity markets, we introduce a 10\% premium over the bid prices in the day-ahead market-clearing stage to incentivize the market participants to adjust their schedule during real-time operation. Since some wind forecast scenarios may lead to infeasibility of the market-clearing problem due to unavailability of flexibility during the real-time operation, we introduce contingency actions in the market-clearing that correspond to load shedding and wind curtailment respectively. Variables corresponding to these contingency actions are penalized in the market-clearing objective such that they are the last resort in ensuring supply-demand balance. 
\begin{table}
\renewcommand{\arraystretch}{0.75} % this reduces the vertical spacing between rows
\linespread{1.0}\selectfont\centering
\begin{center}
\caption{In-sample and out-of-sample performance of market-clearing problems for the 50\% RES paradigm}
\vspace{0.1cm}
\begin{tabular}{lccccccc}
    \toprule
    \multirow{2}{*}{\footnotesize{Parameter}} & \multirow{2}{*}{\footnotesize{Unit}} & \multicolumn{3}{c}{\footnotesize{No network bottlenecks}} & \multicolumn{3}{c}{\footnotesize{With network bottlenecks}}\\
    \cmidrule(r){3-5} \cmidrule(r){6-8}
    & & \footnotesize{$\mathcal{M}^{\text{cc}}$} & \footnotesize{$\mathcal{R}1$} & \footnotesize{$\mathcal{R}2$} & \footnotesize{$\mathcal{M}^{\text{cc}}$} & \footnotesize{$\mathcal{R}1$} &
    \footnotesize{$\mathcal{R}2$}\\
    \toprule
    \footnotesize{In-sample expected cost} & \footnotesize{\$1000} & \footnotesize{221.85} & \footnotesize{257.85} & \footnotesize{216.49} & \footnotesize{256.62} & \footnotesize{377.56} & \footnotesize{235.16} \\
    \footnotesize{Market-clearing computation time} & \footnotesize{s} & \footnotesize{2.08} & \footnotesize{0.22} & \footnotesize{178.23} & \footnotesize{2.38} & \footnotesize{0.27} & \footnotesize{215.5}\\
    \midrule
    \footnotesize{Out-of-sample expected cost} & \footnotesize{\$1000}& \footnotesize{221.76} &\footnotesize{259.57} &\footnotesize{224.29} &\footnotesize{256.51} & \footnotesize{396.86} &\footnotesize{244.28} \\
    \footnotesize{Out-of-sample infeasibility} & \footnotesize{\%} & \footnotesize{0.2} &\footnotesize{3.2} &\footnotesize{0} &\footnotesize{0.2} &\footnotesize{0.6} &\footnotesize{8.8}\\
    \footnotesize{Load shedding probability} & \footnotesize{\%} &\footnotesize{0} &\footnotesize{0} &\footnotesize{0} &\footnotesize{0} &\footnotesize{1.0} &\footnotesize{0.4} \\
    \footnotesize{Wind curtailment probability} & \footnotesize{\%} &\footnotesize{0} &\footnotesize{0} &\footnotesize{46.0} &\footnotesize{0} &\footnotesize{0} &\footnotesize{50.8}\\
    \bottomrule
\end{tabular}
\label{tab:cost_and_time}
\end{center}
\vspace{-0.65cm}
\end{table}

\begin{figure}
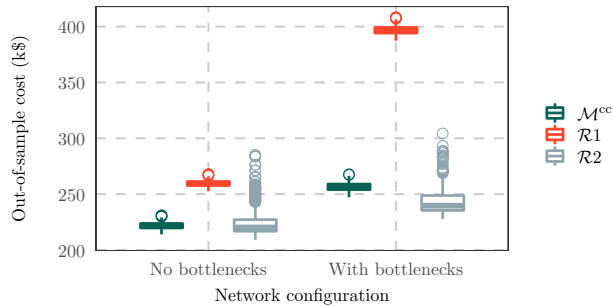

    \centering
    \include{plots/CS1OOSCostComparison}
    \caption{Out-of-sample market-clearing cost comparison in the 50\% RES paradigm}
    \label{fig:cs1_out_of_samplecost}
    \vspace{-0.4cm}
\end{figure}

We prepare a test dataset comprised of 500 wind forecast scenarios, distinct from those used for in-sample simulations, drawn from the same multivariate Gaussian distribution. The box and whiskers plots in Figure \ref{fig:cs1_out_of_samplecost} show the distribution of the resulting out-of-sample cost for the market-clearing problems under the 50\% RES paradigm. For each box, the central line indicates the median, whereas the ends denote the 25th and 75th percentiles. The whiskers extend upto 1.5 times the inter-quartile range, while remaining values are shown as outliers in the form of rings. The second part of Table \ref{tab:cost_and_time} provides the results from the out-of-sample simulations. First, referring to Figure \ref{fig:cs1_out_of_samplecost}, observe that the ordering in terms of expected cost is preserved from the in-sample simulations, i.e., $\mathcal{R}1$ leads to higher costs. We note that market-clearing problem $\mathcal{R}2$ results in highly variable out-of-sample costs compared to the SOCP market-clearing problem $\mathcal{M}^{\text{cc}}$. Second, the out-of-sample expected cost for $\mathcal{M}^{\text{cc}}$ is close to the in-sample case, mentioned in Table \ref{tab:cost_and_time}. On the other hand, market-clearing problem $\mathcal{R}1$ exhibits an increase in cost of 0.75\% and 5.34\% over their in-sample values for the network configurations with and without bottlenecks, respectively. Similarly, problem $\mathcal{R}2$ results in out-of-sample expected cost which exceeds that in-sample expected cost by 3.54\% and 4.1\%, respectively for the network configurations with and without bottlenecks. Finally, both the LP-based market-clearing problems endure load shedding or wind curtailment when faced with unknown forecast samples. These results highlight how the proposed SOCP-based uncertainty-aware market-clearing problem outperforms the prevalent LP-based alternatives. Next, we provide comparative insights on the day-ahead market-clearing outcomes for the three problems.
\paragraph{Day ahead-market-clearing outcomes:}
\begin{figure}[tb!]\scriptsize
\definecolor{Mcc}{RGB}{7,98,86}
\definecolor{R1}{RGB}{252,69,45}
\definecolor{R2}{RGB}{185,199,204}
\definecolor{defcol}{RGB}{90,90,255}
\begin{minipage}[]{\columnwidth}
\centering
\begin{tikzpicture}
\pgfplotsset{set layers}

\pgfplotsset{
   ylabel right/.style={
        after end axis/.append code={
            \node [rotate=90, anchor=north] at (rel axis cs:1.15,0.5) {#1};
        }   
    }
}

% EL Prices Plot
\begin{axis}[
    width=8cm, height=3.5cm,
    xlabel={\footnotesize Hours},
    ylabel={\footnotesize Energy price [\$/MWh]},
    xtick={4,12,20},
    xticklabels ={\footnotesize{4},\footnotesize{12},\footnotesize{20}},
    ytick={8, 11, 14},
    yticklabels={\footnotesize{8}, \footnotesize{11},\footnotesize{14}},
    legend style={
                at={(0.9,1.25)},
                anchor=north east},
    legend columns = 3,
    ymajorgrids=true,
    grid style=dashed,
    axis x line*=bottom,
    axis y line*=left]
\addplot [Mcc,line width=2pt] table [x=hours, y=Mcc, col sep=comma] {elprices_output.csv};
\addlegendentry{\scriptsize$\mathcal{M}^{\text{cc}}$};
\addplot [R1,line width=2pt]  table [x=hours, y=R1, col sep=comma]  {elprices_output.csv};
\addlegendentry{\scriptsize$\mathcal{R}1$};
\addplot [R2,line width=2pt]  table [x=hours, y=R2, col sep=comma]  {elprices_output.csv};
\addlegendentry{\scriptsize$\mathcal{R}2$};
\end{axis}
\node at (3.25,-1.1) {(a)};
\end{tikzpicture} ~
\begin{tikzpicture}
\definecolor{clr1}{RGB}{49,130,189}
\definecolor{clr2}{RGB}{215,25,28}
\pgfplotsset{set layers}

\pgfplotsset{
   ylabel right/.style={
        after end axis/.append code={
            \node [rotate=90, anchor=north] at (rel axis cs:1.15,0.5) {#1};
        }   
    }
}
% Net Demand Plot
\begin{axis}[
    width=8cm, height=3.5cm,
    xlabel={\footnotesize Hours},
    ylabel={\footnotesize Flexibility payment [\$]},
    xtick={4,12,20},
    xticklabels ={\footnotesize{4},\footnotesize{12},\footnotesize{20}},
    ytick={800, 1600, 2400},
    yticklabels={\footnotesize{0.8}, \footnotesize{1.6},\footnotesize{2.4}},
    ymajorgrids=true,
    scaled y ticks=base 10:-3,
    legend style={
            at={(0.6,1.25)},
            anchor=north east},
    legend columns = 2,
    grid style=dashed,
    axis x line*=bottom,
    axis y line*=left
]
\addplot [Mcc,line width=2pt] table [x=hours, y=Mcc, col sep=comma] {compare_Mcc_R1.csv};
\addlegendentry{\scriptsize$\mathcal{M}^{\text{cc}}$};
\addplot [R1,line width=2pt] table [x=hours, y=R1, col sep=comma] {compare_Mcc_R1.csv};
\addlegendentry{\scriptsize$\mathcal{R}1$};
\end{axis}

\node at (3.25,-1.1) {(b)};
\end{tikzpicture}\\
\begin{tikzpicture}
\definecolor{clr1}{RGB}{49,130,189}
\definecolor{clr2}{RGB}{215,25,28}
\pgfplotsset{set layers}

\pgfplotsset{
   ylabel right/.style={
        after end axis/.append code={
            \node [rotate=90, anchor=north] at (rel axis cs:1.15,0.5) {#1};
        }   
    }
}
% Generators dispatch plot
\begin{axis}[
    ybar=0.75pt, 
    bar width=3.75pt, 
    enlargelimits=0.15,
    width=8cm, height=3.5cm,
    xlabel={\footnotesize Hours},
    ylabel={\footnotesize PP dispatch [MW]},
    ymin=400,ymax=1600,
    ytick={500, 1000, 1500},
    yticklabels={\footnotesize{500}, \footnotesize{1000},\footnotesize{1500}},
    legend pos=north west,
    ymajorgrids=true,
    grid style=dashed,
    legend style={
        %nodes={scale=0.5, transform shape},
        at={(0.75,1.2)},
        anchor=north east},
    legend columns = 3,
    axis x line*=bottom,
    axis y line*=left
]
 \addplot [Mcc, fill=Mcc!40, fill opacity=1] table [x=hours, y=Mcc, col sep=comma] {gens_output.csv};
 \addplot [R1, fill=R1!40, fill opacity=1]   table [x=hours, y=R1, col sep=comma] {gens_output.csv};
 \addplot [R2, fill=R2!40, fill opacity=1]   table [x=hours, y=R2, col sep=comma] {gens_output.csv};
 \legend{\scriptsize$\mathcal{M}^{\text{cc}}$,\scriptsize$\mathcal{R}1$,\scriptsize$\mathcal{R}2$}
\end{axis}
\node at (3.25,-1.1) {(c)};
\end{tikzpicture}~
\begin{tikzpicture}
\definecolor{clr1}{RGB}{49,130,189}
\definecolor{clr2}{RGB}{215,25,28}
\pgfplotsset{set layers}

\pgfplotsset{
   ylabel right/.style={
        after end axis/.append code={
            \node [rotate=90, anchor=north] at (rel axis cs:1.15,0.5) {#1};
        }   
    }
}

% ESRs dispatch plot
\begin{axis}[
    ybar=0.75pt, 
    bar width=3.75pt, 
    enlargelimits=0.15,
    width=8cm, height=3.5cm,
    xlabel={\footnotesize Hours},
    ylabel={\footnotesize ESU dispatch [MW]},
    ymin=-220,ymax=220,
    ytick={-200, 0, 200},
    yticklabels={\footnotesize{-200}, \footnotesize{0},\footnotesize{200}},
    legend pos=north west,
    ymajorgrids=true,
    grid style=dashed,
    legend style={
        at={(0.75,1.2)},
        anchor=north east},
    legend columns = 3,
    axis x line*=bottom,
    axis y line*=left
]
 \addplot [Mcc, fill=Mcc!40, fill opacity=1] table [x=hours, y=Mcc, col sep=comma] {esrs_output.csv};
 \addplot [R1, fill=R1!40, fill opacity=1]   table [x=hours, y=R1, col sep=comma] {esrs_output.csv};
 \addplot [R2, fill=R2!40, fill opacity=1]   table [x=hours, y=R2, col sep=comma] {esrs_output.csv};
 \legend{\scriptsize$\mathcal{M}^{\text{cc}}$,\scriptsize$\mathcal{R}1$,\scriptsize$\mathcal{R}2$}
\end{axis}
\node at (3.25,-1.1) {(d)};
\end{tikzpicture}
\end{minipage}
\caption{Day-ahead market-clearing outcome for various problems: (a) hourly system-level energy price, (b) hourly flexibility payment, (c) dispatch of power producers (PPs) and (d) dispatch of energy storage units (ESUs)} 
\label{fig:prices_and_dispatch}
\vspace{-0.25cm}
\end{figure}
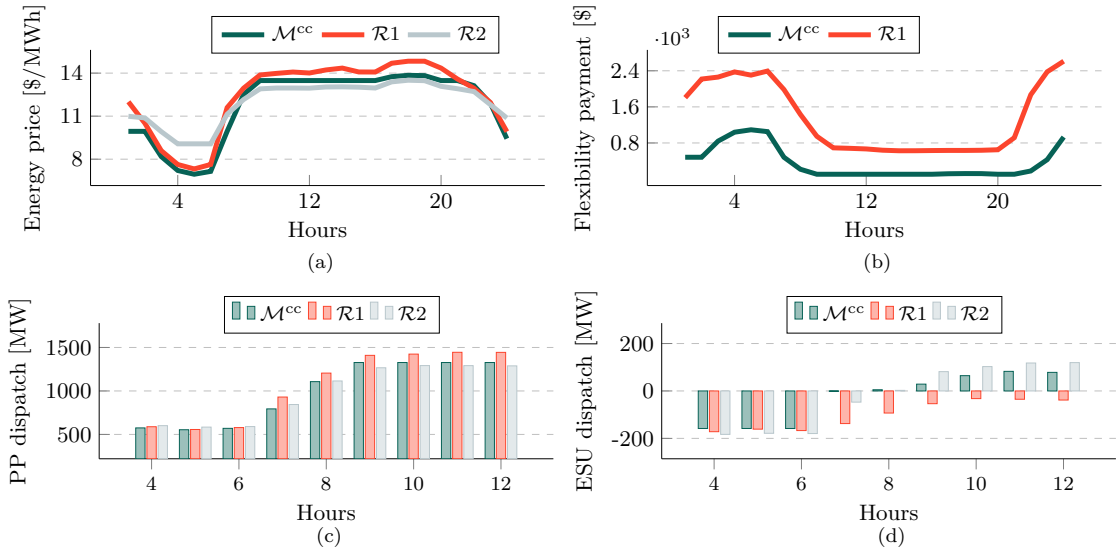
For the 50\% RES paradigm, Figure \ref{fig:prices_and_dispatch} compares the day-ahead market-clearing outcomes among the three problems for the network configuration without any bottlenecks. Observe that, as expected, the energy price in Figure \ref{fig:prices_and_dispatch}(a) closely follows the swings in net demand, shown in Figure \ref{fig:prices_and_network_Mcc}(b). Figure \ref{fig:prices_and_dispatch}(b) compares the flexibility payment at the day-ahead market-clearing stage for the problems $\mathcal{M}^{\text{cc}}$ and $\mathcal{R}1$. In contrast to $\mathcal{M}^{\text{cc}}$ wherein flexibility payments are a result of the dual solution to the market-clearing problem, flexibility payments in $\mathcal{R}1$ correspond to the primal solution, i.e., the payments made towards the reserve capacity to meet the minimum reserve requirement. The difference in hourly flexibility payments illustrate the challenge in dimensioning the capacity-based reserves properly. Note that, in absence of any specific capacity reservation or a meaningful flexibility balance at the day-ahead clearing stage, $\mathcal{R}2$ mandates the real-time markets for pricing of flexibility services and is therefore omitted from the comparison\footnote{A real-time balance is indeed enforced in $\mathcal{R}2$, such that real-time adjustments in each scenario is aligned with the supply-demand balance. However, the price convergence between day-ahead energy prices and real-time prices is a desired property to maximize welfare, as discussed in \cite{Zavala2017}. This price convergence is in fact a characteristic of our implementation $\mathcal{R}2$, but it limits the ability to explicitly extract values corresponding to flexibility payment either from the primal or dual solutions to the market-clearing problem.}. Figures \ref{fig:prices_and_dispatch}(c) and \ref{fig:prices_and_dispatch}(d) show the scheduled dispatch of PPs and ESUs, respectively, for a selection of hours of the day. On account of the minimum reserve requirement associated with market-clearing problem $\mathcal{R}1$, in contrast with the other two problems which consider uncertainty endogenously, the dispatch involves higher generation in hours 7 through 12. The excess generation, which partly explains the higher market-clearing cost associated with $\mathcal{R}1$, is used to charge the ESUs in the network such that sufficient flexible capacity is available to meet the minimum reserve requirement. This effect further underscores the importance of endogenous consideration of uncertainty within the market-clearing framework. 
 %numerical results
\section{Concluding Remarks}\label{sec:5_outlook}
We proposed and analyzed a new conic formulation for forward markets, which we applied to the specific case of electricity markets. Our contribution is threefold. From a market design perspective, we revisited the spatial price equilibrium problem beyond the LP framework by formulating the market-clearing problem within an SOCP framework. From a theoretical perspective, we relied on Lagrangian duality theory for SOCP problems to characterize the solution to the spatial price equilibrium problem involving rational and self-interested market participants in terms of optimality and robustness to parameter perturbation. We derived analytical expressions for conic spatial prices of the traded commodities and provided analytical proofs to demonstrate that moving towards conic markets for electricity does not incur the loss of any desired economic properties, i.e., the proposed conic market retains market efficiency, cost recovery and revenue adequacy, under common assumptions also applicable to LP-based markets. Finally, from a practitioner's perspective, we illustrated the generality of our proposed market-clearing framework by defining a bid format for conic markets.

Our conic market proposal admits SOC-representable nonlinearities underlying the costs and constraints of market participants and the system operator to enable an uncertainty-, asset- and network-aware market-clearing problem. This step in the evolution of electricity markets is crucial for the successful transition of electricity systems worldwide from dispatchable, fossil fuel--based system towards a weather-dependent, renewable energy--based system, supported by a heterogeneous mix of energy and flexibility providers. Our numerical studies leverage in-sample and out-of-sample simulations to showcase the benefits of the proposed SOCP-based market-clearing problem over the LP-based benchmarks in terms of social welfare and its invariability, endogenous representation of uncertainty, appropriate dimensioning of flexibility need, and guarantees on the feasibility of the market-clearing problem. We provide insights into payments associated with the flexibility procurement based on adjustment policies, which are central to our uncertainty-aware SOCP-based market-clearing proposal.

This paper opens up a variety of important directions for future work. First, considering the generality of the proposal, it is potentially interesting to model market-clearing use cases that involve one or more of the attributes related to uncertainty-, asset- and network-awareness. For instance, the modeling examples and the theoretical results developed in this work can be used to study the coordination between the transmission and distribution system operators in an uncertainty- and network-aware setting. This enables the flexibility available at the distribution level to be appropriately harnessed and priced within the market framework while considering a physically-accurate network constraint representation. Along similar lines is the use case studying the market-based coupling of the electricity and natural gas sectors. Here, the focus of the analysis potentially broadens from merely an optimal network- and asset-aware flexibility procurement to one that designs well-aligned incentives and risk measures to ensure a reliable and resilient system operation in both the interdependent sectors. Second, the conic market bids introduced in \cref{subsec:2_4_conic_market_bids} merit further evaluation in terms of acceptance by market participants, alignment with market regulatory agencies as well as adaptations needed in the prevalent market-clearing algorithms to adopt them. Intended to accelerate their practical adoption, such evaluations could take shape as market simulations and participant-specific sensitivity studies to systematically quantify the costs and benefits of moving towards conic electricity markets. Third, the theoretical results discussed in the paper can be extended to include market participants that act strategically. One can then leverage conic complementarity programs to investigate how exercising market power influences the market-clearing outcomes, specifically in the context of the flexibility payments involved in uncertainty-aware use case analyzed in \cref{sec:4_numerical_studies}, and more generally regarding the satisfaction of desired economic properties. \nt{Further, beyond the operational economic properties analyzed in this work, it is of interest to evaluate the satisfaction of long-run equilibrium properties, such as long-run cost recovery of participants.} Another important direction to pursue is the extension of the analytical results we derived on the Lorentz cone to the cone of semidefinite matrices, which results in an SDP-based market-clearing framework. Further generalization over the LP framework aside, such an extension could potentially give rise to markets that consider uncertainty under the paradigm of distributionally-robust chance-constrained optimization and that admit an SDP-based convexification approach for non-convex power flows in the electricity network. Finally, beyond the electricity system, the theoretical results in this work are of potential interest in competitive settings involving physical or non-physical systems wherein cost- and constraint-based nonlinearities currently mandate the use of approximation techniques via linearization.

%% Unit commitment constraints %conclusion

\appendix
\section{Lagrangian Duality for SOCP Problems}\label{app:sec_socp_duality}
\setcounter{equation}{0}
In \ref{app:soc_duality_prelims}, we provide a concise theoretical background of SOCP duality, while directing interested readers to \cite{Lobo1998}, \cite{BenTal2001} and \cite{Alizadeh2003} for details. The dual problem to market-clearing problem $\mathcal{M}^{\text{c}}$ is provided in \ref{app:dual_problem_formulation}.
\subsection{Preliminaries}\label{app:soc_duality_prelims}
Consider an arbitrary SOCP problem in variable $\mathbf{x}\in\mathbb{R}^{N}$ of the form
\begin{align}
    \minimize{\mathbf{x}}\quad  \mathbf{c}^{\top}\mathbf{x} \quad\st\quad \norm{\mathbf{A}_{j}x + \mathbf{b}_{j}}\leq \mathbf{d}_{j}^{\top}\mathbf{x} + e_{j}, \quad \forall j=1,2,\dots,J, \label{example:SOC_duality}
\end{align}
with parameters $\mathbf{c}\in\mathbb{R}^{N}$, $\mathbf{A}_{j}\in\mathbb{R}^{k_{j}\times N}$, $\mathbf{b}_{j}\in\mathbb{R}^{k_{j}}$, $\mathbf{d}_{j}\in\mathbb{R}^{N}$ and $e_{j}\in\mathbb{R}$. Based on \eqref{eq:soc_constraint_example}, the SOC constraint in problem \eqref{example:SOC_duality} may be expressed as
\begin{align*}
    \begin{bmatrix}\mathbf{A}_{j} \\ \mathbf{d}_{j}^{\top} \end{bmatrix}\mathbf{x} + \begin{bmatrix}\mathbf{b}_{j}\\e_{j} \end{bmatrix} \in \mathcal{C}_{j} \Leftrightarrow -\begin{bmatrix}\mathbf{A}_{j}\mathbf{x}+\mathbf{b}_{j} \\ \mathbf{d}_{j}^{\top}\mathbf{x}+e_{j}\end{bmatrix} \leqslant_{\mathcal{C}_{j}} 0,
\end{align*}
where $\mathcal{C}_{j}\subseteq\mathbb{R}^{k_{j}+1}$ is the standard second-order cone and generalized inequality $\leqslant_{\mathcal{C}_{j}}$ denotes a partial ordering over the cone $\mathcal{C}_{j}$. The Lagrangian function for \eqref{example:SOC_duality} writes as
\begin{align*}
    \Theta(\mathbf{x},\bm{\lambda}_{1},\dots,\bm{\lambda}_{J}) = \mathbf{c}^{\top}\mathbf{x} - \sum_{j=1}^{J}\bm{\lambda}_{j}^{\top} \left( \begin{bmatrix}\mathbf{A}_{j} \\ \mathbf{d}_{j}^{\top} \end{bmatrix}\mathbf{x} + \begin{bmatrix}\mathbf{b}_{j}\\e_{j} \end{bmatrix} \right),
\end{align*}
where $\bm{\lambda}_{j}\in\mathbb{R}^{k_{j}+1}$ for $j=1,2,\dots,J$ are Lagrange multipliers. The dual function is given by
\begin{align*}
    g(\bm{\lambda}_{1},\dots,\bm{\lambda}_{J}) = \begin{cases} -\sum_{j=1}^{J}\bm{\lambda}_{j}^{\top}\begin{bmatrix}\mathbf{b}_{j} \\ e_{j} \end{bmatrix}, \quad &\mathbf{c} = \sum_{j=1}^{J}\begin{bmatrix}\mathbf{A}_{j} \\ \mathbf{d}_{j}^{\top} \end{bmatrix}^{\top}\bm{\lambda}_{j}, \\
    -\infty, \quad &\text{otherwise}.
    \end{cases}
\end{align*}
Lastly, the dual problem to \eqref{example:SOC_duality} after substituting $\bm{\lambda}_{j} = \begin{bmatrix}\bm{\mu}_{j}^{\top} \quad \nu_{j} \end{bmatrix}^{\top}$, such that $\bm{\mu}_{j}\in\mathbb{R}^{k_{j}}$ and $\nu_{j}\in\mathbb{R}_{+}$ are auxiliary dual variables, can be written as
\begin{subequations}
\begin{align}
\maximize{\bm{\mu}_{j},\nu_{j}}\quad&-\sum_{j=1}^{J}(\mathbf{b}_{j}^{\top}\bm{\mu}_{j} + e_{j}\nu_{j})\\
\st\quad&\sum_{j=1}^{J}(\mathbf{A}_{j}^{\top}\bm{\mu}_{j} + \mathbf{d}_{j}\nu_{j}) = \mathbf{c}\\
&\norm{\bm{\mu}_{j}} \leq \nu_{j}, \enspace \forall j=1,2,\dots,J.
\end{align}
\end{subequations}
Similarly, we write an arbitrary SOCP problem with rotated SOC constraints as
\begin{align}
    \minimize{\mathbf{x}}\quad  \mathbf{c}^{\top}\mathbf{x} \quad\st\quad \norm{\mathbf{A}_{j}\mathbf{x} + \mathbf{b}_{j}}^{2}\leq \mathbf{d}_{j}^{\top}\mathbf{x} + e_{j}, \quad \forall j=1,2,\dots,J. \label{ex:rotated_SOC}
\end{align}
From the definition of a rotated second-order cone \citep{MosekApS2021}, we have the equivalence
\begin{align*}
    \norm{\mathbf{A}_{j}\mathbf{x} + \mathbf{b}_{j}}^{2}\leq \mathbf{d}_{j}^{\top}\mathbf{x} + e_{j} \Leftrightarrow 
    - \begin{bmatrix} \mathbf{A}_{j}\mathbf{x} + \mathbf{b}_{j} \\ \mathbf{d}_{j}^{\top}\mathbf{x} + e_{j} \\ \frac{1}{2}
    \end{bmatrix} \leqslant_{\mathcal{C}^{\text{R}}_{j}} 0,
\end{align*}
where $\mathcal{C}^{\text{R}}_{j} \subseteq \mathbb{R}^{k_{j}+2}$ denotes the rotated second-order cone. As before, defining Lagrange multipliers $\bm{\lambda}_{j} = [\bm{\mu}_{j}^{\top} \quad \nu_{j} \quad \kappa_{j}]^{\top} \in \mathbb{R}^{k_{j}+2}$ for $j=1,2,\dots,J$, such that $\bm{\mu}_{j}\in\mathbb{R}^{k_{j}}$, $\nu_{j}\in\mathbb{R}_{+}$ and $\kappa_{j}\in\mathbb{R}_{+}$, the Lagrangian function for the problem \eqref{ex:rotated_SOC} can be written as
\begin{align*}
    \Theta(\mathbf{x},\bm{\mu}_{1},\dots,\bm{\mu}_{J},\nu_{1},\dots,\nu_{j},\kappa_{1},\dots,\kappa_{J}) = \mathbf{c}^{\top}\mathbf{x} - \sum_{j=1}^{J}(\bm{\mu}_{j}^{\top}\mathbf{A}_{j}\mathbf{x} + \bm{\mu}_{j}^{\top}\mathbf{b}_{j} + \frac{1}{2}\kappa_{j} + \nu_{j}\mathbf{d}^{\top}_{j}\mathbf{x} + \nu_{j}e_{j}),
\end{align*}
from which the dual problem is derived as
\begin{subequations}
\begin{align}
\maximize{\bm{\mu}_{j},\nu_{j},\kappa_{j}}\quad&-\sum_{j=1}^{J}(\mathbf{b}_{j}^{\top}\bm{\mu}_{j} + e_{j}\nu_{j} + \frac{1}{2}\kappa_{j})\\
\st\quad&\sum_{j=1}^{J}(\mathbf{A}_{j}^{\top}\bm{\mu}_{j} + \mathbf{d}_{j}\nu_{j}) = \mathbf{c}\\
&\norm{\bm{\mu}_{j}}^{2} \leq \kappa_{j}\nu_{j}, \enspace \forall j=1,2,\dots,J.
\end{align}
\end{subequations}
\subsection{Dual Problem to Problem $\mathcal{M}^{\text{c}}$}\label{app:dual_problem_formulation}
Let $\mathcal{D}^{\text{c}}$ denote the dual problem to the market-clearing problem $\mathcal{M}^{\text{c}}$. We now derive $\mathcal{D}^{\text{c}}$ relying on the dualization approach in \ref{app:soc_duality_prelims}. Let $\Theta$($\bm{q}_{i}$,$\;\bm{z}_{i}$,$\;\bm{\mu}_{it}^{\text{Q}},\;\kappa_{it}^{\text{Q}},\;\nu_{it}^{\text{Q}}$,$\;\bm{\mu}_{ij}$,$\;\nu_{ij}$, $\;\bm{\gamma}_{i}$,$\;\bm{\lambda}_{p}$,$\;\underline{\bm{\varrho}}_{t}$,$\;\overline{\bm{\varrho}}_{t}$) denote the Lagrangian function of $\mathcal{M}^{\text{c}}$ given by
\begin{align}
&\resizebox{0.95\textwidth}{!}{$
\begin{aligned}
    \Theta &=
    \sum_{i\in\mathcal{I}}\sum_{t\in\mathcal{T}} \left(z_{it} + {\mathbf{c}_{it}^{\text{L}}}^{\top}\bm{q}_{it} \right) - \sum_{i \in \mathcal{I}}\sum_{t\in\mathcal{T}}\left({\bm{\mu}^{\text{Q}}_{it}}^{\top}\mathbf{C}_{it}^{\text{Q}}\bm{q}_{it} + \frac{1}{2}\kappa_{it}^{\text{Q}} + \nu_{it}^{\text{Q}}z_{it} \right) \notag \\
    & \quad -  \sum_{i \in \mathcal{I}}\sum_{j\in\mathcal{J}_{i}} \left(\bm{\mu}_{ij}^{\top}(\mathbf{A}_{ij}\bm{q}_{i} + \mathbf{b}_{ij}) + \nu_{ij}(\mathbf{d}_{ij}^{\top}\bm{q}_{i} + e_{ij}) \right) + \bm{\gamma}^{\top}(\mathbf{F}_{i}\bm{q}_{i} - \mathbf{h}_{i}) - \sum_{p\in\mathcal{P}}\bm{\lambda}_{p}^{\top}\sum_{i \in \mathcal{I}}\mathbf{G}_{ip}\bm{q}_{ip} \notag \\
    & \quad + \sum_{t\in\mathcal{T}} \left( \overline{\bm{\varrho}}_{t}^{\top} \left(\sum_{n \in \mathcal{N}} [\bm{\Psi}]_{(:,n)}\left(\sum_{i \in \mathcal{I}_{n}}\sum_{p\in\mathcal{P}}\;[\mathbf{G}_{ip}\bm{q}_{ip}]_{t}\right) - \overline{\mathbf{s}}\right) - \underline{\bm{\varrho}}_{t}^{\top}\left(\sum_{n \in \mathcal{N}} [\bm{\Psi}]_{(:,n)}\left(\sum_{i \in \mathcal{I}_{n}}\sum_{p\in\mathcal{P}}\;[\mathbf{G}_{ip}\bm{q}_{ip}]_{t}\right) + \overline{\mathbf{s}}\right) \right). \notag
    \end{aligned}
    $}
\end{align}
For notational convenience, let cost vectors for various hours be stacked to obtain $\mathbf{c}^{\text{L}}_{i} \in \mathbb{R}^{K_{i}T}$ and a block diagonal matrix $\mathbf{C}_{i}^{\text{Q}} = \mathbf{C}_{i1}^{\text{Q}} \oplus \cdots \oplus \mathbf{C}_{iT}^{\text{Q}} \in \mathbb{R}^{K_{i}T \times K_{i}T}$. Let $\bm{\mu}^{\text{Q}}_{i} \in \mathbb{R}^{K_{i}T}$ and $\bm{\nu}^{\text{Q}}_{i}\in\mathbb{R}^{T}$ denote the stacked dual variables $\bm{\mu}_{it}^{\text{Q}}$ and $\nu_{it}^{\text{Q}}$, respectively. Let $\widehat{\bm{\lambda}}_{i}\in\mathbb{R}^{K_{i}T}$ denote the stacked version of an auxiliary dual variable $\widehat{\bm{\lambda}}_{it} \in \mathbb{R}^{K_{i}}$ for each market participant $i$ with its $k$-th element given by
\begin{align}
 \widehat{\lambda}_{itk} = \begin{cases} [\mathbf{G}_{ip}\bm{\lambda}_{p}]_{t},~&\text{if}\;k = p,\\ 0,~&\text{otherwise}. \end{cases} \label{eq:lambda_it_def}
\end{align}
This auxiliary dual variable represents the contribution of each market participant towards the trades of the $P$ commodities in the market. Finally, we define an auxiliary variable representing the $i$-th participant's contribution towards network congestion, denoted as $\widehat{\bm{\varrho}}_{it}\in\mathbb{R}^{K_{i}}$, given by
\begin{align}\label{eq:varrho_it_def}
  \widehat{\varrho}_{itk} = \begin{cases}  \mathbb{1}^{\top}[\mathbf{G}_{ip}]_{(:,t)} \mathbb{I}_{n_{i}} \bm{\Psi}^{\top} (\overline{\bm{\varrho}}_{t} - \underline{\bm{\varrho}}_{t}),~&\text{if}\;k = p,\\ 0,~&\text{otherwise}, \end{cases}
\end{align}
where $\mathbb{1}\in\mathbb{R}^{T}$ and $\mathbb{I}_{n_{i}}\in \mathbb{R}^{N}$ is an indicator vector having the element corresponding to the node $n_{i}\in\mathbb{N}$ where participant $i$ is located as 1, while all other elements are zero. The variables $\widehat{\bm{\varrho}}_{it}$ are then stacked to obtain $\widehat{\bm{\varrho}}_{i}\in\mathbb{R}^{K_{i}T}$. With these substitutions, we write the dual problem $\mathcal{D}^{\text{c}}$ as
\begin{subequations}
\begin{align}
\maximize{\Xi}\quad&-\sum_{i \in \mathcal{I}} \sum_{j\in\mathcal{J}_{i}}(\mathbf{b}_{ij}^{\top}\bm{\mu}_{ij} + e_{ij}\nu_{ij}) - \sum_{i\in\mathcal{I}}\sum_{t\in\mathcal{T}}\frac{1}{2}\kappa_{it}^{\text{Q}}- \sum_{t\in\mathcal{T}}{\overline{\mathbf{s}}}^{\top}(\underline{\bm{\varrho}}_{t} + \overline{\bm{\varrho}}_{t})\label{objfun_generic_conic_dual}\\
\st\quad& \mathbb{1} - \bm{\nu}_{i}^{\text{Q}} = \mathbb{0},\;\forall i\\
    &\mathbf{c}^{\text{L}}_{i} - \mathbf{C}_{i}^{Q}\bm{\mu}_{i}^{\text{Q}} - \sum_{j\in\mathcal{J}_{i}}(\mathbf{A}_{ij}^{\top}\bm{\mu}_{ij} + \mathbf{d}_{ij}\nu_{ij}) +\mathbf{F}_{i}^{\top}\bm{\gamma}_{i} - \widehat{\bm{\lambda}}_{i} + \widehat{\bm{\varrho}}_{i} = \mathbb{0}, \enspace \forall i \label{cons:participants_decision_dual}\\
    &\norm{\bm{\mu}^{\text{Q}}_{it}}^{2} \leq \kappa_{it}^{\text{Q}}{\nu}_{it}^{\text{Q}},\;\forall t,\;\forall i \label{cons:participant_dual_feasibility_cost_reform}\\
    &\norm{\bm{\mu}_{ij}} \leq \nu_{ij},\;\forall j\in\mathcal{J}_{i},~\forall i. \label{cons:participant_dual_feasibility_soc}\\
    & \eqref{eq:lambda_it_def} - \eqref{eq:varrho_it_def},\;\forall k=1,2,\dots,K_{i},\;\forall t,\;\forall i,
\end{align}\label{prob:Dc}
\end{subequations}
which is an SOCP problem in variables $\Xi = \{\bm{\mu}^{\text{Q}}_{i},\;\kappa_{it}^{\text{Q}},\;\bm{\nu}_{i}^{\text{Q}},\;\bm{\mu}_{ij},\;\nu_{ij},\;\bm{\gamma}_{i},\;\bm{\lambda}_{p}, \;\widehat{\bm{\lambda}}_{i},\;\underline{\bm{\varrho}}_{t}, \;\overline{\bm{\varrho}}_{t},\;\widehat{\bm{\varrho}}_{i} \}.$
\section{Proofs}\label{app:sec_proofs}
\subsection*{Proof of Theorem 1} %\ref{thm:strongduality}
\label{app:strongduality_proof}
First, we prove the existence of essentially strictly feasible solutions to the market-clearing problem $\mathcal{M}^{\text{c}}$ and its dual $\mathcal{D}^{\text{c}}$, inspired by the theory in \cite{Lobo1998}. From Definition \ref{def:essentially_strict_feasibility}, finding strictly feasible solutions to the primal market-clearing problem $\mathcal{M}^{\text{c}}$ reduces to finding tuples $(\mathbf{q_{i}},\;\bm{z}_{i}),\;\forall i$, which strictly satisfy the SOC constraints \eqref{cons:Mcone_participants_obj_reform}-\eqref{cons:Mcone_participants_soc}. Likewise, proving strict dual feasibility reduces to finding tuples $(\bm{\mu}_{it}^{\text{Q}},\;\kappa_{it}^{\text{Q}},\;\nu_{it}^{\text{Q}}),\;\forall t,\;\forall i$ and $(\bm{\mu}_{ij},\;\nu_{ij}),\;\forall j\in\mathcal{J}_{i},\;\forall i$ that strictly satisfy the SOC constraints \eqref{cons:participant_dual_feasibility_cost_reform}-\eqref{cons:participant_dual_feasibility_soc}, respectively. Relying on a variant of the big-M method widely applied to LP problems \citep{Fortuny-Amat1981}, we require the following auxiliary result to find strictly feasible solutions to $\mathcal{D}^{\text{c}}$.
\begin{lemma}Given the feasibility of the market-clearing problem $\mathcal{M}^{\text{c}}$, for each market participant $i\in\mathcal{I}$, there exists a large enough finite scalar bound $\overline{D}^{Q}_{i}\in\mathbb{R}$ on the Euclidean norm of $\bm{q}_{i}$ given by  $\norm{\bm{q}_{i}} \leq \overline{D}^{Q}_{i}$, such that the optimal solution to $\mathcal{M}^{\text{c}}$, denoted by $(\bm{q}_{i}^{\star},\;\bm{z}_{i}^{\star})$, is unchanged with addition of the norm bounds.\label{app_prop_bounds}
\end{lemma}
\proof For each participant $i\in\mathcal{I}$, the $K_{i}$ number of decision variables at each hour $t$ constituting $\bm{q}_{it}$ are either the commodity contributions $[\bm{q}_{ip}]_{t}$ for the $p\in\mathcal{P}$ physical commodities discussed in \cref{subsec:2_1_market_setting} or any physical state variables pertaining to the participants. Hence, $\bm{q}_{it},\;\forall t$ are finite and both bounded above and below. Observe that, a bounded $\bm{q}_{it}$ results in a bounded $z_{it}$ due to \eqref{cons:Mcone_participants_obj_reform}. Given feasible points $\bm{q}_{i},\;\forall i$, one could arbitrarily choose scalars $\overline{D}^{Q}_{i}$ such that $\overline{D}^{Q}_{i} \geq \norm{\bm{q}_{i}},\;\forall i$ and iteratively solve $\mathcal{M}^{\text{c}}$ for values of $\overline{D}^{Q}_{i}$ large enough such that the optimal solution $(\bm{q}_{i}^{\star},\;\bm{z}_{i}^{\star}),\;\forall i$ no longer changes over the subsequent iterations. Such an iterative scheme would result in the norm bounds $\overline{D}^{Q}_{i},\forall i$, thereby completing the proof. \endproof

Due to Definition \ref{def:essentially_strict_feasibility} and Lemma \ref{app_prop_bounds}, we write a reduced form of market-clearing problem $\mathcal{M}^{\text{c}}$ while retaining only the SOC constraints with additional norm bound constraints
\begin{subequations}\label{prob:Mcone_reduced}
\begin{align}
    \minimize{\bm{q}_{i},\bm{z}_{i}}\quad&\sum_{i \in \mathcal{I}}\;\sum_{t\in\mathcal{T}}\;({z}_{it} + {\mathbf{c}_{it}^{\text{L}}}^{\top}\bm{q}_{it})\\
    \st\quad&\norm{\mathbf{C}^{\text{Q}}_{it}\bm{q}_{it}}^{2} \le  {z}_{it},\;\forall t,\;\forall i &:& (\bm{\mu}^{\text{Q}}_{it},\;\kappa^{\text{Q}}_{it},\;\nu^{\text{Q}}_{it}) \label{cons:str_feas_rot_soc} \\
    &\norm{\mathbf{A}_{ij}\bm{q}_{i} + \mathbf{b}_{ij}} \leq \mathbf{d}^{\top}_{ij}\bm{q}_{i} + e_{ij},\;\forall j\in\mathcal{J}_{i},\;\forall i&\colon&(\bm{\mu}_{ij},\;{\nu}_{ij}) \label{cons:str_feas_soc}\\
    &\norm{\bm{q}_{i}} \leq \overline{D}^{Q}_{i},\;\forall i,&\colon&(\overline{\bm{\mu}}_{i},\;\overline{{\nu}}_{i}) 
\end{align}
\end{subequations}
where Lagrangian multipliers $\overline{\bm{\mu}}_{i}\in\mathbb{R}^{K_{i}T}$ and $\overline{\nu}_{i}\in\mathbb{R}_{+}$ are associated with the norm bound constraints. Following the approach in \ref{app:soc_duality_prelims}, the dual problem to \eqref{prob:Mcone_reduced} in variables $\Xi^{\text{R}} = \{\bm{\mu}^{\text{Q}}_{i},\;\kappa_{it}^{\text{Q}},\;\bm{\nu}_{i}^{\text{Q}},\;\bm{\mu}_{ij},\;\nu_{ij},\;\overline{\bm{\mu}}_{i},\;\overline{\nu}_{i}\}$ writes as
\begin{subequations}\label{prob:Mcone_reduced_dual}
\begin{align}
\maximize{\Xi^{\text{R}}}\quad&-\sum_{i \in \mathcal{I}} \sum_{j\in\mathcal{J}_{i}}(\mathbf{b}_{ij}^{\top}\bm{\mu}_{ij} + e_{ij}\nu_{ij}) - \sum_{i\in\mathcal{I}}\sum_{t\in\mathcal{T}}\frac{1}{2}\kappa_{it}^{\text{Q}}- \sum_{i\in\mathcal{I}}\overline{D}^{Q}_{i}\overline{\nu}_{i}\\
\st\quad& \mathbb{1} - \bm{\nu}_{i}^{\text{Q}} = \mathbb{0},\;\forall i \label{eq:dualfeas_reduced_z}\\
    &\mathbf{c}^{\text{L}}_{i} - \mathbf{C}_{i}^{Q}\bm{\mu}_{i}^{\text{Q}} - \sum_{j=1}^{J_{i}}(\mathbf{A}_{ij}^{\top}\bm{\mu}_{ij} + \mathbf{d}_{ij}\nu_{ij}) + \overline{\bm{\mu}}_{i} = \mathbb{0}, \enspace \forall i \label{eq:dualfeas_reduced_q}\\
    &\norm{\bm{\mu}^{\text{Q}}_{it}}^{2} \leq \kappa_{it}^{\text{Q}}\nu_{it}^{\text{Q}},\;\forall t,\;\forall i \\
    &\norm{\bm{\mu}_{ij}} \leq \nu_{ij},\;\forall j\in\mathcal{J}_{i},~\forall i\\
    &\norm{\bm{\overline{\mu}}_{i}} \leq \overline{\nu}_{i},\;\forall i.
\end{align}
\end{subequations}
Proving the existence of strictly feasible points for \eqref{prob:Mcone_reduced_dual} is straightforward. Choosing any arbitrary vectors $\bm{\mu}_{ij},\;\forall j \in \mathcal{J}_{i},\;\forall i$ and $\bm{\mu}_{it}^{\text{Q}},\;\forall t,\;\forall i$, we can compute the values of $\nu_{ij} > \norm{\bm{\mu}_{ij}}$ and $\kappa_{it}^{\text{Q}} > \norm{\bm{\mu}_{it}^{\text{Q}}}$, while $\nu_{it}^{\text{Q}} = 1$ from \eqref{eq:dualfeas_reduced_z}. Now, the variable $\overline{\bm{\mu}}_{i}$ follows from the equality \eqref{eq:dualfeas_reduced_q} and consequently, $\overline{\nu}_{i}$ can be any number larger than $\norm{\bm{\overline{\mu}}_{i}}$. Hence, we have found at least one strictly feasible solution to \eqref{prob:Mcone_reduced_dual}, which proves essentially strict feasibility of the dual problem $\mathcal{D}^{\text{c}}$, given that Lemma \ref{app_prop_bounds} holds.

Next, we prove essentially strict feasibility of the primal market-clearing problem $\mathcal{M}^\text{c}$ employing the so-called Phase-I method, discussed in \citet[\S 11.4]{Boyd2004}. Consider an SOCP problem in variables $(\bm{q}_{i},\;\bm{z}_{i},\;x_{i},\;x_{i}^{\text{Q}})$ where $x_{i}\in\mathbb{R}$ and $x_{i}^{\text{Q}}\in\mathbb{R}$ are arbitrary slack variables
\begin{subequations}\label{prob:Mcone_augmented}
\begin{align}
    \minimize{{\bm{q}_{i},\bm{z}_{i},x_{i},x_{i}^{\text{Q}}}}\quad&\sum_{i \in \mathcal{I}}\;\sum_{t\in\mathcal{T}}\;({z}_{it} + {\mathbf{c}_{it}^{\text{L}}}^{\top}\bm{q}_{it}) + \sum_{i \in \mathcal{I}}\;(x_{i} + x_{i}^{\text{Q}})\\
    \st\quad&\norm{\mathbf{C}^{\text{Q}}_{it}\bm{q}_{it}}^{2} \le  {z}_{it} + x_{i}^{\text{Q}},\;\forall t,\;\forall i \label{eq:str_feas_soc_rot_primal} \\
    &\norm{\mathbf{A}_{ij}\bm{q}_{i} + \mathbf{b}_{ij}} \leq \mathbf{d}^{\top}_{ij}\bm{q}_{i} + e_{ij} + x_{i},\;\forall j\in\mathcal{J}_{i},\;\forall i. \label{eq:str_feas_soc_primal}
\end{align}
\end{subequations}
Observe that, obtaining strictly solutions to the primal problem \eqref{prob:Mcone_augmented} and its dual problem is straightforward. For instance, choosing
\begin{subequations}
\begin{align}
&\bm{q}_{i} = \mathbb{0},\; x_{i} > \maximize{j\in\mathcal{J}_{i}}\;\norm{b_{ij}} - e_{ij},\;\forall i,\; \label{eq:strictly_feasible_primal_1}\\
&\text{any}\; \bm{z}_{i}\in\mathbb{R}^{T}_{+}\;\text{and}\;x_{i}^{\text{Q}} > \maximize{t\in\mathcal{T}}\;-{z}_{it},\;\forall i,  \label{eq:strictly_feasible_primal_2}
\end{align}\label{eq:strictly_feasible_primal}
\end{subequations}
gives a strictly feasible primal solution to \eqref{prob:Mcone_augmented}. Likewise, a strictly feasible solution for the dual problem to \eqref{prob:Mcone_augmented} can be found by an approach (omitted from presentation, for the sake of brevity) discussed above, i.e., by adding non-binding upper bound constraints on the primal variables $(x_{i},x_{i}^{\text{Q}}),\;\forall i$. Consequently, \citet[Theorem 13]{Alizadeh2003} establishes strong duality for \eqref{prob:Mcone_augmented} due to strict primal and dual feasibility. We now provide an auxiliary result relating the augmented problem \eqref{prob:Mcone_augmented} to the primal market-clearing problem $\mathcal{M}^{\text{c}}$. 
\begin{lemma}\label{app_prop_augmented}
For the tuple $({\bm{q}}_{i}^{\star},\;{\bm{z}}_{i}^{\star},\;{x}_{i}^{\star},\;{{x}_{i}^{\text{Q}}}^{\star}),\;\forall i$ denoting the optimal solution to problem \eqref{prob:Mcone_augmented}, only one of the following conditions holds:
\begin{enumerate}[label=(\roman*)]
    \item if ${x}^{\star}_{i} < 0,\;\forall i$ and ${{x}_{i}^{\text{Q}}}^{\star} < 0,\;\forall i$, then $({\bm{q}}_{i}^{\star},\;{\bm{z}}_{i}^{\star}),\;\forall i$ is strictly feasible for the problem $\mathcal{M}^{\text{c}}$, or
    \item if for any participant $i$, ${x}^{\star}_{i} = 0$ or ${{x}_{i}^{\text{Q}}}^{\star} = 0$, then for that participant $({\bm{q}}_{i}^{\star},\;{\bm{z}}_{i}^{\star})$ is strictly feasible for problem $\mathcal{M}^{\text{c}}$, provided $e_{ij} > \norm{\mathbf{b}_{ij}},\;\forall j\in\mathcal{J}_{i}$ and the quadratic cost components $\mathbf{c}^{\text{Q}}_{it}\neq \mathbb{0},\;\forall t$.
\end{enumerate}
\end{lemma}
\proof First, observe that if the tuple of optimal solution solution $({\bm{q}}_{i}^{\star},\;{\bm{z}}_{i}^{\star},\;{x}_{i}^{\star},\;{{x}_{i}^{\text{Q}}}^{\star}),\;\forall i$ to problem \eqref{prob:Mcone_augmented} is such that for any $i$, $x_{i}^{\star} > 0$ or ${{x}_{i}^{\text{Q}}}^{\star} > 0$, then it implies that problem $\mathcal{M}^{\text{c}}$ is infeasible, which contradicts our assumption in Theorem \ref{thm:strongduality}. Hence, for feasibility of the problem \eqref{prob:Mcone_augmented}, we have $x_{i} \leq 0,\;\forall i$ and $x_{i}^{\text{Q}}\leq 0,\;\forall i$. Now, from \eqref{eq:str_feas_soc_rot_primal} and \eqref{eq:str_feas_soc_primal}, it is clear that if both $x_{i}^{\star} < 0,\;\forall i$ and ${x_{i}^{\text{Q}}}^{\star} < 0,\;\forall i$ hold, we have found $(\bm{q}_{i}^{\star},\;\bm{z}_{i}^{\star}),\;\forall i$ which strictly satisfy the SOC inequalities in $\mathcal{M}^{\text{c}}$, i.e.,  \eqref{cons:Mcone_participants_obj_reform} and \eqref{cons:Mcone_participants_soc}. Conversely, if the optimal solution is obtained such that for any $i$, $x_{i}^{\star}=0$ or ${x_{i}^{\text{Q}}}^{\star} = 0$, then $(\bm{q}_{i}^{\star},\;\bm{z}_{i}^{\star})$ strictly satisfies the SOC inequalities corresponding to the $i$-th participant in $\mathcal{M}^{\text{c}}$ under some mild conditions derived from \eqref{eq:strictly_feasible_primal}. Note that ${x_{i}^{\text{Q}}}^{\star} = 0 > \maximize{t}\;-z_{it}$ holds true only if $z_{it} > 0,\;\forall t$. This is ensured by non-zero quadratic cost components of the market participant $i$ at all hours, i.e., $\mathbf{c}_{it}^{\text{Q}} \neq \mathbb{0},\;\forall t$. Moreover if any $\mathbf{c}_{it}^{\text{Q}} = \mathbb{0}$, then the constraint \eqref{cons:Mcone_participants_obj_reform} corresponding to that hour is trivially satisfied and therefore, can be possibly eliminated. Lastly, from $x_{i}^{\star} = 0 > \maximize{j\in\mathcal{J}_{i}}\;\norm{b_{ij}} - e_{ij}$, we obtain $\norm{b_{ij}} - e_{ij} < 0,\forall j\in\mathcal{J}_{i}$, which is then a requirement for $\bm{q}_{i}^{\star} = \mathbb{0}$ from \eqref{eq:strictly_feasible_primal_1} to be strictly feasible for \eqref{cons:Mcone_participants_soc}. This completes the proof. \endproof

From Lemma \ref{app_prop_augmented}, solving problem \eqref{prob:Mcone_augmented} leads to $(\bm{q}_{i},\;\bm{z}_{i}),\;\forall i$ that are essentially strictly feasible for $\mathcal{M}^{\text{c}}$ under some mild conditions. Evident from the Examples and formulations in the Supplementary Material, these conditions are met in the practical implementations of $\mathcal{M}^{\text{c}}$. We have now proved that an essentially strictly feasible solution to primal problem $\mathcal{M}^{\text{c}}$ exists. 

Essentially strict feasibility of the primal and dual problems is necessary and sufficient for strong duality to hold for the primal-dual pair of problems $\mathcal{M}^{\text{c}}$ and $\mathcal{D}^{\text{c}}$ \citep[Theorem 1.4.4]{BenTal2001}, thereby completing the proof.

\subsection*{Proof of Theorem 2} %\ref{thm:theorem_priceformation}
\label{app:theorem_priceformation}
The proof follows from the partial Lagrangian function of $\mathcal{M}^{\text{c}}$, obtained by keeping the constraints \eqref{cons:Mcone_participants_obj_reform}-\eqref{cons:Mcone_participants_eq} and relaxing the others, which we write as
\begin{align*}
    \widehat{\Theta} &= - \sum_{p\in\mathcal{P}}\bm{\lambda}_{p}^{\top}\sum_{i \in \mathcal{I}}\mathbf{G}_{ip}\bm{q}_{ip} + \sum_{t\in\mathcal{T}} \overline{\bm{\varrho}}_{t}^{\top} \left(\sum_{n \in \mathcal{N}} [\bm{\Psi}]_{(:,n)}\left(\sum_{i \in \mathcal{I}_{n}}\sum_{p\in\mathcal{P}}\;[\mathbf{G}_{ip}\bm{q}_{ip}]_{t}\right) - \overline{\mathbf{s}}\right) \notag \\ 
    & \qquad\qquad - \sum_{t\in\mathcal{T}} \underline{\bm{\varrho}}_{t}^{\top}\left(\sum_{n \in \mathcal{N}} [\bm{\Psi}]_{(:,n)}\left(\sum_{i \in \mathcal{I}_{n}}\sum_{p\in\mathcal{P}}\;[\mathbf{G}_{ip}\bm{q}_{ip}]_{t}\right) + \overline{\mathbf{s}}\right).
\end{align*}
With the substitution of the auxiliary variable $\mathbf{Q}_{p}^{\text{inj}},\;\forall p\in\mathcal{P}$ defined in \eqref{eq:def_Qp_inj} followed by a rearrangement of terms, the partial Lagrangian equivalently writes as follows
\begin{align}
&\resizebox{0.92\textwidth}{!}{$
\begin{aligned}
    \widehat{\Theta} &=  \underbrace{-\sum_{p\in\mathcal{P}} \bm{\lambda}_{p}^{\top}{\mathbf{Q}_{p}^{\text{inj}}}^{\top} \mathbb{1}}_{\text{System-wide balance}} + \underbrace{\sum_{t\in\mathcal{T}} \overline{\bm{\varrho}}_{t}^{\top} \left(\sum_{p\in\mathcal{P}} \bm{\Psi} \; [{\mathbf{Q}_{p}^{\text{inj}}}]_{(:,t)}\right) -  \sum_{t\in\mathcal{T}} \overline{\bm{\varrho}}_{t}^{\top} \overline{\mathbf{s}} - \sum_{t\in\mathcal{T}} \underline{\bm{\varrho}}_{t}^{\top}\left(\sum_{p\in\mathcal{P}} \bm{\Psi} \; [{\mathbf{Q}_{p}^{\text{inj}}}]_{(:,t)}\right) - \sum_{t\in\mathcal{T}} \underline{\bm{\varrho}}_{t}^{\top} \overline{\mathbf{s}}}_{\text{Line flow limits}}, 
    \end{aligned}
    $}\label{proofs:partial_lagrangian}
\end{align}
where the vector of ones $\mathbb{1}\in\mathbb{R}^{N}$. Here, the partial Lagrangian comprises of terms corresponding to the system-wide balance equalities and the line flow limit inequalities, as indicated. In the sense of \cite{Bohn1984}, the spot price of electricity for consumers at a given hour is analytically expressed as the sum of shadow price of the system balance constraint and the sensitivity of changes in the demand to the flow in the capacity-constrained transmission lines. We denote these optimal prices $\bm{\Pi}_{p}\in\mathbb{R}^{N \times T}$ for the $p$-th commodity and provide an analytical expression in the following. We derive an expression for the spot price of commodities as the negative sensitivity of the partial Lagrangian \eqref{proofs:partial_lagrangian} to the nodal injections as
\begin{align}
\frac{\partial\widehat{\Theta}}{\partial[{\mathbf{Q}_{p}^{\text{inj}}}]_{(:,t)}} = -\Big(-\mathbb{1}[\bm{\lambda}_{p}]_{t} + \bm{\Psi}^{\top}(\overline{\bm{\varrho}}_{t} - \underline{\bm{\varrho}}_{t})\Big),\;\forall p,\;~\forall t, \label{eq:partial_lagrangian_sensitivity}
\end{align}
where the negative sign originates from the sign convention adopted in this work, such that the consumption (withdrawals) are given by $\bm{q}_{ip}\in\mathbb{R}^{T}_{-}$. For compactness of expression, we extend \eqref{eq:partial_lagrangian_sensitivity} to all hours by defining auxiliary variables $\underline{\bm{\rho}},~\overline{\bm{\rho}}\in\mathbb{R}^{L \times T}$ and $\bm{\Lambda}_{p}\in\mathbb{R}^{N\times T}$, such that $\overline{\bm{\rho}} = [\overline{\bm{\varrho}}_{1}~\cdots~\overline{\bm{\varrho}}_{T}]$, $\underline{\bm{\rho}} = [\underline{\bm{\varrho}}_{1}~\cdots~\underline{\bm{\varrho}}_{T}]$ and $\bm{\Lambda}_{p} = \mathbb{1}^{\top} \otimes \bm{\lambda}_{p}$. Lastly, we express the spatial prices as
\begin{align} 
    \mathbf{\Pi}_{p} = {\bm{\Lambda}}_{p} - \bm{\Psi}^{\top}(\overline{\bm{\rho}} - \underline{\bm{\rho}}), \;\forall p, \label{eq:final_sensitivity_relation}
\end{align}
which are analogous in structure to conventional LMPs prevalent in LP-based market-clearing problems. Theorem \ref{thm:strongduality} proves strong duality for the market-clearing problem $\mathcal{M}^{\text{c}}$, thereby ensuring the optimality of the spatial prices when the dual variables in \eqref{eq:final_sensitivity_relation} are replaced by their values at the optimal solution.

\subsection*{Proof of Theorem 3.} %\ref{thm:comp_spatial_price_eqm} 
To prove the equivalence of the optimization problem $\mathcal{M}^{\text{c}}$ with the equilibrium $\mathcal{E}^{\text{c}}$, we show that the KKT conditions of the equilibrium problem are identical to those of the optimization. First, for the network operator's congestion rent maximization problem \eqref{prob:NO_prof_max}, let $\Theta_{\text{NO}}$ denote the Lagrangian function of the problem \eqref{prob:NO_prof_max}. We write the KKT conditions defining the optimal solution as
\begin{itemize}
    \item \textit{Stationarity condition:}
    \begin{subequations}
    \begin{align}
        \frac{\partial \Theta_{\text{NO}}}{\partial \mathbf{y}_{t}} = \bm{\omega}_{t} + \bm{\Psi}^{\top}(\overline{\bm{\varrho}}_{t} - \underline{\bm{\varrho}}_{t}) = \mathbb{0},\; \forall t \label{KKTcond:yt_1}
\end{align} 
    \item \textit{Primal feasibility, dual feasibility and complementarity conditions:}
    \begin{align}
        -&\bm{\Psi}\mathbf{y}_{t} \leq \overline{\mathbf{s}};\quad\underline{\bm{\varrho}}_{t} \geq \mathbb{0};\quad \underline{\bm{\varrho}}_{t} \odot (\bm{\Psi}\mathbf{y}_{t} + \overline{\mathbf{s}}) = \mathbb{0} \label{KKTcond:yt_2}\\
        &\bm{\Psi}\mathbf{y}_{t} \leq \overline{\mathbf{s}};\quad\overline{\bm{\varrho}}_{t} \geq \mathbb{0};\quad \overline{\bm{\varrho}}_{t} \odot (\bm{\Psi}\mathbf{y}_{t} - \overline{\mathbf{s}}) = \mathbb{0},\label{KKTcond:yt_3}
    \end{align}\label{KKTcond:yt_all}
\end{subequations}
where $\odot$ denotes the Hadamard (element-wise) product operator.
\end{itemize}
For each participant $i\in\mathcal{I}$, let $\Theta_{i}(\bm{q}_{i},\;\bm{z}_{i},\;\bm{\mu}_{it}^{\text{Q}},\;\kappa_{it}^{\text{Q}},\;\nu_{it}^{\text{Q}},\;\bm{\mu}_{ij},\;\nu_{ij},\;\bm{\gamma}_{i},\;\bm{\widehat{\lambda}}_{ip}$) denote the Lagrangian function for problem \eqref{prob:i_prof_max} given by
\begin{align}
    \Theta_{i} &= \sum_{p\in\mathcal{P}} \text{tr}(\bm{\Pi}_{p}^{\top}\mathbf{W}_{ip}) - \sum_{t\in\mathcal{T}}\Big(z_{it} + {\mathbf{c}_{it}^{\text{L}}}^{\top}\bm{q}_{it}\big) - \sum_{t\in\mathcal{T}}\bm{\omega}_{t}^{\top}\big(\sum_{p\in\mathcal{P}}[\mathbf{W}_{ip}]_{(:,t)}\big) + \sum_{t\in\mathcal{T}}\bm{\omega}_{t}^{\top}\big(\sum_{p\in\mathcal{P}}\mathbb{I}_{n_{i}}[\mathbf{G}_{ip}\bm{q}_{ip}]_{t}\big) \notag \\
    & \quad + \sum_{t\in\mathcal{T}}\Big(\bm{\mu}_{it}^{\text{Q}}\mathbf{C}_{it}^{\text{Q}}\bm{q}_{it} + \frac{1}{2}\kappa_{it}^{\text{Q}} + \nu_{it}^{\text{Q}}z_{it}\Big) + \sum_{j\in\mathcal{J}_{i}}\Big(\bm{\mu}_{ij}^{\top}\big(\mathbf{A}_{ij}\bm{q}_{i} + \mathbf{b}_{ij}\big) + \nu_{ij}(\mathbf{d}_{ij}^{\top}\bm{q}_{i} + e_{ij}\big)\Big) \notag \\
    & \quad \quad - \bm{\gamma}_{i}^{\top}(\mathbf{F}_{i}\bm{q}_{i} - \mathbf{h}_{i}) + \widehat{\bm{\lambda}}_{ip}^{\top}\Big(\mathbf{G}_{ip}\bm{q}_{ip} - \mathbf{W}_{ip}^{\top}\mathbb{1}\Big) \label{proofs:lagrangian_prob_i_prof_max}
\end{align}
Using the auxiliary dual variables $\bm{\mu}_{i}^{\text{Q}},\;\bm{\nu}_{i}$ and parameters $\mathbf{c}_{i}^{\text{L}},\;\mathbf{C}_{i}^{\text{Q}}$ defined previously in \ref{app:dual_problem_formulation}, we write the KKT optimality conditions for a participant $i\in\mathcal{I}$ as
\begin{subequations}
\begin{itemize}
    \item \textit{Stationarity conditions:}
    \begin{align}
    \frac{\partial \Theta_{i}}{\partial \bm{q}_{i}} &= -\mathbf{c}_{i}^{\text{L}} + \mathbf{C}_{i}^{\text{Q}}\bm{\mu}_{i}^{\text{Q}} +  \sum_{j\in\mathcal{J}_{i}}\big(\mathbf{A}_{ij}^{\top}\bm{\mu}_{ij} + \mathbf{d}_{ij}\nu_{ij}\big) -\mathbf{F}_{i}^{\top}\bm{\gamma}_{i} + \frac{\partial}{\partial \bm{q}_{i}}\Big(\bm{\widehat{\lambda}}_{ip}^{\top}\big(\mathbf{G}_{ip}\bm{q}_{ip}\big)\Big) \notag \\
    &\qquad\qquad\qquad\qquad\qquad\qquad\qquad\qquad + \frac{\partial}{\partial \bm{q}_{i}}\Big(\sum_{t\in\mathcal{T}}\bm{\omega}_{t}^{\top}\big(\mathbb{I}_{n_{i}}\sum_{p\in\mathcal{P}}[\mathbf{G}_{ip}\bm{q}_{ip}]_{t}\big)\Big) = \mathbb{0} \label{KKT:qi_1_preliminary} \\
    \frac{\partial \Theta_{i}}{\partial \mathbf{W}_{ip}} &= \mathbf{\Pi}_{p} - \bm{\widehat{\lambda}}_{ip}\mathbb{1}^{\top} - \bm{\Omega} = \mathbf{0}, \;\forall p\in\mathcal{P} \label{KKT:qi_2_preliminary},\\
    \frac{\partial \Theta_{i}}{\partial \bm{z}_{i}} &= \mathbb{1} - \bm{\nu}_{i} = \mathbb{0} \label{KKT:qi_3}
    \end{align}
where $\bm{\Omega}\in\mathbb{R}^{N\times T} \coloneqq \begin{bmatrix}\bm{\omega}_{1} &\bm{\omega}_{2} &\cdots &\bm{\omega}_{T}\end{bmatrix}$, such that the term $\sum_{t\in\mathcal{T}}\bm{\omega}_{t}^{\top}\big(\sum_{p\in\mathcal{P}}[\mathbf{W}_{ip}]_{(:,t)}\big)$ in the Lagrangian \eqref{proofs:lagrangian_prob_i_prof_max} reduces to $\text{tr}(\bm{\Omega}^{\top}\mathbf{W}_{ip})$. Furthermore, from \eqref{KKTcond:yt_1}, we have $\bm{\omega}_{t} = -\bm{\Psi}^{\top}(\overline{\bm{\varrho}}_{t} - \underline{\bm{\varrho}}_{t})$ at the optimal solution. With this and from the definitions of $\widehat{\bm{\lambda}}_{i}$ and $\widehat{\bm{\varrho}}_{i}$ in \eqref{eq:lambda_it_def}  and \eqref{eq:varrho_it_def}, respectively, \eqref{KKT:qi_1_preliminary} reduces to
    \begin{align}
    \frac{\partial \Theta_{i}}{\partial \bm{q}_{i}} &= -\mathbf{c}_{i}^{\text{L}} + \mathbf{C}_{i}^{\text{Q}}\bm{\mu}_{i}^{\text{Q}} +  \sum_{j\in\mathcal{J}_{i}}\big(\mathbf{A}_{ij}^{\top}\bm{\mu}_{ij} + \mathbf{d}_{ij}\big) - \mathbf{F}_{i}^{\top}\bm{\gamma}_{i} + \widehat{\bm{\lambda}}_{i} - \widehat{\bm{\varrho}}_{i} = \mathbb{0} \label{KKT:qi_1}
    \end{align}
Similarly, the stationarity condition \eqref{KKT:qi_2_preliminary} reduces to
\begin{align}
    \frac{\partial \Theta_{i}}{\partial \mathbf{W}_{ip}} = \mathbf{\Pi}_{p} - \bm{\widehat{\lambda}}_{ip}\mathbb{1}^{\top} + \bm{\Psi}^{\top}(\overline{\bm{\rho}} - \underline{\bm{\rho}}) = \mathbf{0}, \;\forall p\in\mathcal{P} \label{KKT:qi_2},
\end{align}
following the definitions of auxiliary dual variables $\overline{\bm{\rho}}$ and $\underline{\bm{\rho}}$ in Theorem \ref{thm:theorem_priceformation}.
\item \textit{Primal feasibility, dual feasibility and complementarity conditions:}
\begin{align}
 &\resizebox{0.8\textwidth}{!}{$
    \begin{aligned}
    &\norm{\mathbf{C}^{\text{Q}}_{it}\bm{q}_{it}}^{2} \le  {z}_{it}\;;\;\;\norm{\bm{\mu}_{it}^{\text{Q}}}^{2} \leq \kappa_{it}^{\text{Q}}\nu_{it}^{\text{Q}},\;\kappa_{it}^{\text{Q}}\geq 0,\nu_{it}^{\text{Q}}\geq 0\;;\;\;\begin{bmatrix}
   {\bm{\mu}_{it}^{\text{Q}}}^{\top} \quad\nu_{it}^{\text{Q}} \quad\kappa_{it}^{\text{Q}}
    \end{bmatrix}
    \begin{bmatrix}
        \mathbf{C}_{it}^{\text{Q}}\bm{q}_{it} \\ z_{it} \\ \frac{1}{2}
    \end{bmatrix}  = 0, \;\; \forall t \end{aligned}
    $}\label{KKT:qi_4}\\
    &\resizebox{0.96\textwidth}{!}{$
    \begin{aligned}
    &\norm{\mathbf{A}_{ij}\bm{q}_{i} + \mathbf{b}_{ij}} \leq \mathbf{d}_{ij}^{\top}\bm{q}_{i} + e_{ij}\;;\;\; \norm{\bm{\mu}_{ij}} \leq \nu_{ij},\;\nu_{ij}\geq 0\;;\;\; \begin{bmatrix} {\bm{\mu}}^{\top}_{ij}  \quad \nu_{ij} \end{bmatrix}  \Big(\begin{bmatrix} \mathbf{A}_{ij} \\ \mathbf{d}_{ij}^{\top}\end{bmatrix} \bm{q}_{i} + \begin{bmatrix} \mathbf{b}_{ij} \\ e_{ij} \end{bmatrix}\Big) = 0, \;\; \forall j \in \mathcal{J}_{i} \end{aligned} 
    $}\label{KKT:qi_5}\\
    &\mathbf{F}_{i}\bm{q}_{i} = \mathbf{h}_{i};\quad \bm{\gamma}_{i} \text{ free} \label{KKT:qi_6}\\
    &\eqref{eq:lambda_it_def} - \eqref{eq:varrho_it_def},\;\forall k=1,2,\dots,K_{i},\;\forall t,\;\forall i. \label{KKT:qi_7}
\end{align}
\end{itemize}
\end{subequations}
Equations \eqref{KKTcond:yt_all}, \eqref{KKT:qi_3}-\eqref{KKT:qi_7}, $\forall i \in \mathcal{I}$ and \eqref{mo:equalities} form the KKT optimality conditions for the equilibrium problem $\mathcal{E}^{\text{c}}$. Observe that, in addition to the primal and dual feasibility involving SOC constraints, \eqref{KKT:qi_4}-\eqref{KKT:qi_5} ensure the conic complementarity condition is met at the optimal solution \citep[Theorem 16]{Alizadeh2003}. 

To see that the KKT optimality conditions for the optimization problem $\mathcal{M}^{\text{c}}$ are identical to those of the equilibrium, observe that at the optimal solution:
\begin{enumerate}[label=(\roman*)]
    \item ${\widehat{\bm{\lambda}}}^{\star}_{ip},\;\forall i\in\mathcal{I}$ appearing in \eqref{KKT:qi_2} are attained such that ${\widehat{\bm{\lambda}}}^{\star}_{ip} = \bm{\lambda}_{p}^{\star},\;\forall p\in\mathcal{P},\;\forall i\in\mathcal{I}$, where $\bm{\lambda}_{p}^{\star},\;\forall p\in\mathcal{P}$ corresponds to the system-wide price of the commodities. Therefore, \eqref{KKT:qi_2} provides the analytical expression for conic spatial prices as formulated in Theorem \ref{thm:theorem_priceformation}, and
    \item $\mathbf{y}^{\star}_{t},\;\forall t$ appearing in \eqref{KKTcond:yt_2}-\eqref{KKTcond:yt_3} are obtained such that $\mathbf{y}_{t}^{\star} = \sum_{p\in\mathcal{P}}[{\mathbf{Q}_{p}^{\text{inj}}}^{\star}]_{(:,t)},\;\forall t$. This follows from the market-clearing condition \eqref{mo:transaction_cost_balance}. Note that the auxiliary variable ${\mathbf{Q}_{p}^{\text{inj}}}$ appears in the partial Lagrangian of problem $\mathcal{M}^{\text{c}}$, as shown \eqref{proofs:partial_lagrangian}.
\end{enumerate}
This establishes the equivalence of the KKT optimality conditions for the centrally-solved optimization problem $\mathcal{M}^{\text{c}}$ and the equilibrium problem $\mathcal{E}^{\text{c}}$, thereby completing the proof.
\endproof

\subsection*{Proof of Corollary 1}%\ref{cor:existence_uniqueness}
% \comms{\textbf{Existence:} 
% \begin{itemize}
%     \item \textbf{Strategy sets:} Strategy sets for all participants, the network operator and the market operator are convex. However, compactness of strategy sets is not assumed for market participants $i\in\mathcal{I}$, although Lemma \ref{app_prop_bounds} solves this to some extent and can be extended. For the network operator, compactness is given by the feasibility. For the market operator, as prices can be negative due to congestion in the network, the strategy set is not given to be compact.
%     \item Continuity of game map is given by problem design
% \end{itemize}} 

To prove existence of solutions to the competitive spatial price equilibrium problem $\mathcal{E}^{\text{c}}$, we recall \citet[Theorem 1]{Rosen1965} which provides an existence result for solutions to equilibrium problems conditioned on the convexity and compactness of each participant's strategy sets and the continuity of their payoff functions. For the market participants $i\in\mathcal{I}$, while convexity and closure of the strategy sets are given by the feasibility region defined by equalities and non-strict inequalities \eqref{cons:i_prof_max_obj_ref}-\eqref{cons:transaction_product_equivalence},  Lemma \ref{app_prop_bounds} proves that $(\bm{q}_{i},\;\bm{z}_{i}),\;\forall i$ are bounded. Note that the quantities bought or sold by the $i$-th participant $\mathbf{W}_{ip},\;\forall p$ are bounded, conditioned on the boundedness of $\bm{q}_{i}$ due to \eqref{cons:transaction_product_equivalence}. Consequently, the strategy sets of the participants are convex as well as closed and bounded, thereby satisfying the convexity and compactness conditions. Further, each participant's cost function \eqref{objfun:i_prof_max} is continuous in the decision variables. For the network operator, convexity and compactness of the strategy set is given due to the network limits and continuity of the payoff function is satisfied by the linear objective function in \eqref{prob:NO_prof_max}. Therefore, from \citet[Theorem 1]{Rosen1965}, this proves that at least one solution exists for the spatial price equilibrium problem $\mathcal{E}^{\text{c}}$. To derive conditions under which at most one solution exists to problem $\mathcal{E}^{\text{c}}$, we refer to the equivalence of equilibrium problem $\mathcal{E}^{\text{c}}$ to the convex market-clearing optimization problem $\mathcal{M}^{\text{c}}$ established by Theorem \ref{thm:comp_spatial_price_eqm}. From this equivalence, the uniqueness of allocations $\bm{q}_{i}^{\star},\;\forall i $ at the equilibrium relies on the strict convexity of the problem $\mathcal{M}^{\text{c}}$, which we characterize in Remark \ref{remark:strict_convexity}. However, observe that, the uniqueness of solutions to the dual market-clearing problem $\mathcal{D}^{\text{c}}$ is not given due to lack of the strict convexity property of the dual objective \eqref{objfun_generic_conic_dual}. Therefore, uniqueness of spatially-differentiated conic prices $\bm{\Pi}_{p}^{\star},\;\forall p$ given by \eqref{eq:nodal_price_formulation} is not guaranteed. This completes the proof. 

\subsection*{Proof of Theorem 4}%\ref{thm:theorem_economic_properties}
\begin{enumerate}[label=(\roman*)]
    \item \textbf{Market efficiency:} An efficient market maximizes social welfare, such that no participant unilaterally deviates from the market-clearing outcomes since each participant maximizes her profit at the optimal allocations $\bm{q}_{i}^{\star}$ and prices $\bm{\Pi}_{p}^{\star},\;\forall p$. Under the assumption of perfectly competitive market participants, this is given if the KKT optimality conditions of the centrally-solved optimization problem $\mathcal{M}^{\text{c}}$ and equilibrium problem $\mathcal{E}^{\text{c}}$ involving rational and self-interested actors are identical, which we have established in Proof of Theorem \ref{thm:comp_spatial_price_eqm}. 
    \item \textbf{Cost recovery:} Mathematically, the non-negativity of payoff for the market participants holds true, if at the optimal solution
    \begin{align}
     &\resizebox{0.94\textwidth}{!}{$
    \begin{aligned}
        \sum_{p\in\mathcal{P}} \text{tr}({{\bm{\Pi}_{p}^{\star}}^{\top}}{\mathbf{W}^{\star}_{ip}}) -  \sum_{t\in\mathcal{T}} \left({z}^{\star}_{it} + {\mathbf{c}_{it}^{\text{L}}}^{\top}\bm{q}^{\star}_{it}\right) -\sum_{t\in\mathcal{T}}{\bm{\omega}^{\star}_{t}}^{\top} \Big(\sum_{p\in\mathcal{P}}\big([\mathbf{W}^{\star}_{ip}]_{(:,t)} - \mathbb{I}_{n_{i}}[\mathbf{G}_{ip}\bm{q}^{\star}_{ip}]_{t}\big)\Big) \geq 0,\;\forall i\in\mathcal{I}. \end{aligned}
        $}\label{econ_pro:cost_recovery_condition}
    \end{align}
    To prove that \eqref{econ_pro:cost_recovery_condition} holds, we derive dual problems to each participant's profit maximization problem \eqref{prob:i_prof_max}. The dual problem in variables $\Xi_{i} = \{\bm{\mu}^{\text{Q}}_{i},\;\kappa_{it}^{\text{Q}},\;\bm{\nu}_{i}^{\text{Q}},\;\bm{\mu}_{ij},\;\nu_{ij},\;\bm{\gamma}_{i}\}$ writes as
    \begin{align}
    \forall i \in \mathcal{I}
    \begin{cases}
    \minimize{\Xi_{\text{i}}}\quad&\sum_{j\in\mathcal{J}_{i}}(\mathbf{b}_{ij}^{\top}\bm{\mu}_{ij} + e_{ij}\nu_{ij}) + \sum_{t\in\mathcal{T}}\frac{1}{2}\kappa_{it}^{\text{Q}} + \bm{\gamma}_{i}^{\top}\mathbf{h}_{i} \\
    \st\quad& \eqref{KKT:qi_3} - \eqref{KKT:qi_2} \\
            & \norm{\bm{\mu}_{it}^{\text{Q}}}^{2} \leq \kappa_{it}^{\text{Q}}\nu_{it}^{\text{Q}},\;\forall t\\
            & \norm{\bm{\mu}_{ij}} \leq \nu_{ij},\;\forall j\in\mathcal{J}_{i}\\
            &\kappa_{it}^{\text{Q}}\geq 0,\;\nu_{it}^{\text{Q}}\geq 0,
    \end{cases}\label{prob:Participants_prof_max_dual}
    \end{align}
    Using Lemmas \ref{app_prop_bounds} and \ref{app_prop_augmented} provided in the Proof of Theorem \ref{thm:strongduality}, existence of strictly feasible primal-dual solutions to the participant's profit maximization problem \eqref{prob:i_prof_max} and its dual \eqref{prob:Participants_prof_max_dual} is established (omitted from presentation, for the sake of brevity). Consequently, from Theorem \ref{thm:strongduality}, strong duality holds for this primal-dual pair, thereby enforcing the primal problem \eqref{prob:i_prof_max} and its dual \eqref{prob:Participants_prof_max_dual} to attain identical objective function values at the optimal solution. Condition \eqref{econ_pro:cost_recovery_condition} is equivalent to the non-negativity of the dual objective in \eqref{prob:Participants_prof_max_dual} at the optimal solution, which we now analyze in the following.
    
    Observe that, the second term of the objective function of the dual problem \eqref{prob:Participants_prof_max_dual} is non-negative from the primal feasibility condition $\kappa_{it}\geq 0,\forall i$. The first term, enclosed in parenthesis, is non-negative if $e_{ij}\geq\norm{\mathbf{b}_{ij}},\;\forall j\in\mathcal{J}_{i}$. This stems from the Cauchy-Schwarz Inequality, i.e., we have
    \begin{align*}
        |\mathbf{b}_{ij}^{\top}\bm{\mu}_{ij}| \leq \norm{\mathbf{b}_{ij}}\norm{\bm{\mu}_{ij}} \leq \norm{\mathbf{b}_{ij}}\nu_{ij},\;\forall j \in\mathcal{J}_{i},
    \end{align*}
    where the last inequality is due to the primal feasibility condition $\norm{\bm{\mu}_{ij}}\leq \nu_{ij},\;\forall j\in\mathcal{J}_{i}$. With the condition $e_{ij}\geq\norm{\mathbf{b}_{ij}},\;\forall j\in\mathcal{J}_{i}$, we have $|\mathbf{b}_{ij}^{\top}\bm{\mu}_{ij}| \leq e_{ij}\nu_{ij},\;\forall j\in\mathcal{J}_{i}$. Under the condition $e_{ij}\geq\norm{\mathbf{b}_{ij}}$, $e_{ij}\nu_{ij}\geq 0\;\forall j\in\mathcal{J}_{i}$; therefore, the first term in the objective is bounded below by 0. The variable $\bm{\gamma}_{i}$ is free, therefore the third term in the objective is guaranteed to be non-negative if $\mathbf{h}_{i}=\mathbb{0}$. This completes the proof of cost recovery for the market participants $i\in\mathcal{I}$ under the conditions: (i) $e_{ij}\geq\norm{\mathbf{b}_{ij}},\;\forall j\in\mathcal{J}_{i},\;\forall i$ and (ii) $\mathbf{h}_{i} = \mathbb{0},\;\forall i$. 
    \item \textbf{Revenue adequacy:} Mathematically, the market operator is revenue adequate if
      \begin{align}
        \underbrace{\sum_{p\in\mathcal{P}}\sum_{i\in\mathcal{I}} \text{tr}({{\bm{\Pi}_{p}^{\star}}}^{\top}\mathbf{W}_{ip}^{\star})}_{\text{Term A}} \;\; \underbrace{-\sum_{t\in\mathcal{T}}{\bm{\omega}_{t}^{\star}}^{\top}\mathbf{y}_{t}^{\star}}_{\text{Term B}} \geq 0, \label{proof:rev_adeq_condition}
    \end{align}
    where Term A refers to the net payments received from the $i\in\mathcal{I}$ market participants for the $p\in\mathcal{P}$ commodities and Term B refers to the payments made to the network operator towards transmission service. To reduce notational complexity, we drop the superscript $\star$ denoting optimal values in the proof that follows, yet they are always implied. Using \eqref{eq:partial_lagrangian_sensitivity}, we expand Term A and rearrange the summation operators to reflect the dependence of variables 
    \begin{align}
        \resizebox{0.94\textwidth}{!}{$
        \begin{aligned}
        \sum_{p\in\mathcal{P}}\sum_{i\in\mathcal{I}} \text{tr}({{\bm{\Pi}_{p}}}^{\top}\mathbf{W}_{ip}) &= \sum_{p\in\mathcal{P}}\sum_{i\in\mathcal{I}}\sum_{t\in\mathcal{T}}\Big(\Big(\mathbb{1}[\bm{\lambda}_{p}]_{t} - \bm{\Psi}^{\top}(\overline{\bm{\varrho}}_{t} - \underline{\bm{\varrho}}_{t})\Big)^{\top}[\mathbf{W}_{ip}]_{(:,t)}\Big) \\
        &= \sum_{p\in\mathcal{P}}\sum_{i\in\mathcal{I}}\sum_{t\in\mathcal{T}}\Big([\bm{\lambda}_{p}]_{t}\mathbb{1}^{\top}[\mathbf{W}_{ip}]_{(:,t)}\Big) - \sum_{t\in\mathcal{T}}\Big(\Big(\bm{\Psi}^{\top}(\overline{\bm{\varrho}}_{t} - \underline{\bm{\varrho}}_{t})\Big)^{\top}\sum_{p\in\mathcal{P}}\sum_{i\in\mathcal{I}}[\mathbf{W}_{ip}]_{(:,t)}\Big).
        \end{aligned} $}\label{proof:rev_adeq_termA}
    \end{align}
    Gathering the equalities \eqref{cons:transaction_product_equivalence} representing the transaction quantities with the injections (or withdrawals) for the $I$ participants, we have $\mathbf{W}_{ip}^{\top}\mathbb{1} = \mathbf{G}_{ip}\bm{q}_{ip},\;\forall i$. Since this equality holds individually for each participant, we can add them for all participants to get $\sum_{i\in\mathcal{I}}\mathbb{1}^{\top}\mathbf{W}_{ip} = \sum_{i\in\mathcal{I}}(\mathbf{G}_{ip}\bm{q}_{ip})^{\top}$, such that
    \begin{align}
        \sum_{i\in\mathcal{I}}\mathbb{1}^{\top}[\mathbf{W}_{ip}]_{(:,t)} = \sum_{i\in\mathcal{I}}[\mathbf{G}_{ip}\bm{q}_{ip}]_{t} = 0, \quad \forall t, \label{proof:rev_adeq_transaction_bal}
    \end{align}
    where the second equality results from the market-clearing condition \eqref{mo:nodal_price_balance} that holds at optimality. Hence, the first term in \eqref{proof:rev_adeq_termA} vanishes. The second term in \eqref{proof:rev_adeq_termA} is non-zero only if there is a congestion in the grid, i.e., any $\overline{\bm{\varrho}}_{t}\neq0$ or $\underline{\bm{\varrho}}_{t}\neq0$.
    Next, we expand Term B in \eqref{proof:rev_adeq_condition}, using the stationarity condition \eqref{KKTcond:yt_1} and the market-clearing condition \eqref{mo:transaction_cost_balance}, as 
    \begin{align}
    -\sum_{t\in\mathcal{T}}\bm{\omega}_{t}^{\top}\mathbf{y}_{t} = \sum_{t\in\mathcal{T}}\Big(\bm{\Psi}^{\top}(\overline{\bm{\varrho}}_{t} - \underline{\bm{\varrho}}_{t})\Big)^{\top}\sum_{p\in\mathcal{P}}[\mathbf{Q}_{p}^{\text{inj}}]_{(:,t)}
    \end{align}
    which is non-zero only if there is congestion in the network, i.e., the network operator earns congestion rent. We remove the summation over the hours by using the expression for the conic spatial prices given by Theorem \ref{thm:theorem_priceformation} and the auxiliary variables defined therein, to rewrite Term B of \eqref{proof:rev_adeq_condition} as
    \begin{align}
    -\sum_{t\in\mathcal{T}}\bm{\omega}_{t}^{\top}\mathbf{y}_{t} &= \sum_{t\in\mathcal{T}}\Big(\bm{\Psi}^{\top}(\overline{\bm{\varrho}}_{t} - \underline{\bm{\varrho}}_{t})\Big)^{\top}\sum_{p\in\mathcal{P}}[\mathbf{Q}_{p}^{\text{inj}}]_{(:,t)} \notag \\
    &= \sum_{p\in\mathcal{P}}\text{tr}\Big((\bm{\Psi}^{\top}(\overline{\bm{\rho}} - \overline{\bm{\rho}}))^{\top} \mathbf{Q}_{p}^{\text{inj}}\Big) = \sum_{p\in\mathcal{P}}\text{tr}\Big((\bm{\Lambda}_{p} - \bm{\Pi}_{p})^{\top} \mathbf{Q}_{p}^{\text{inj}}\Big) . \label{proof:rev_adeq_termB}
    \end{align}
    Similarly, using \eqref{proof:rev_adeq_transaction_bal}, we rewrite Term A of \eqref{proof:rev_adeq_condition} after rearrangement as
    \begin{align}
    \sum_{p\in\mathcal{P}}\sum_{i\in\mathcal{I}} \text{tr}({{\bm{\Pi}_{p}}}^{\top}\mathbf{W}_{ip}) &= - \sum_{t\in\mathcal{T}}\Big(\Big(\bm{\Psi}^{\top}(\overline{\bm{\varrho}}_{t} - \underline{\bm{\varrho}}_{t})\Big)^{\top}\sum_{p\in\mathcal{P}}\sum_{i\in\mathcal{I}}[\mathbf{W}_{ip}]_{(:,t)}\Big) \notag \\
    &= -\sum_{p\in\mathcal{P}}\text{tr}\Big((\bm{\Lambda}_{p} - \bm{\Pi}_{p})^{\top}(\sum_{i\in\mathcal{I}}\mathbf{W}_{ip})\Big). \label{proof:rev_adeq_termA_final}
    \end{align}
    Observe that $\sum_{i\in\mathcal{I}}\mathbf{W}_{ip}$ sums the transaction quantities over the nodes and periods for all participants for a given commodity $p$, which is indeed equal to the commodity-specific net nodal injections given by \eqref{eq:def_Qp_inj}, i.e., $\mathbf{Q}_{p}^{\text{inj}} = \sum_{i\in\mathcal{I}}\mathbf{W}_{ip},\;\forall p$. Therefore, the terms in \eqref{proof:rev_adeq_termB} and \eqref{proof:rev_adeq_termA_final} cancel each other out, thereby satisfying the revenue adequacy condition for the market operator at the optimal solution. As a result, we have shown that there exists a budget balance for the market operator under the proposed conic market framework, i.e., the market-operator does not accrue any surplus revenue.
\end{enumerate} %appendix proofs

\bibliographystyle{elsarticle-harv} 
\bibliography{references}
\end{document}

%%%FRONT MATTER
\RUNAUTHOR{Ratha et al.}

\RUNTITLE{Moving from Linear to Conic Markets for Electricity}

\TITLE{Moving from Linear to Conic Markets for Electricity}

\ARTICLEAUTHORS{%
\AUTHOR{Anubhav Ratha, Pierre Pinson, H\'el\`ene Le Cadre, Ana Virag, Jalal Kazempour}
} % end of the block

\ABSTRACT{%
This document serves as an electronic companion (EC) for the paper ``Moving from Linear to Conic Markets for Electricity". In \cref{app:sec_modeling_examples}, we provide modeling examples that illustrate how second order cone (SOC) constraints enable future electricity markets to be uncertainty-, asset- and network-aware. In \cref{app:sec_uncertainty_aware}, we present the market-clearing problems of the proposed SOCP-based uncertainty-aware electricity market as well as the linear programming (LP) based market-clearing benchmarks. Numbered sections and equations throughout the electronic companion correspond to those in the paper, while a prefix `EC.' denotes these elements within this document.
}%
\maketitle
%% Here starts the e-companion (EC)
%%%%%%%%%%%%%%%%%%%%%%%%%%%%%%%%%%%%%%%%%%%%%%%%%%%%%%%%%%
\ECSwitch
%\ECDisclaimer
%%%%%%%%%%%%%%%%%%%%%%%%%%%%%%%%%%%%%%%%%%%%%%%%%%%%%%%%%%
\vspace{-1.25cm}
\section{Modeling Examples}\label{app:sec_modeling_examples}
In addition to the examples discussed in the paper, this section demonstrates the versatility of SOC constraints in modeling the various asset- and network-related nonlinearities faced by electricity markets. While Example \ref{example:quad_cost_reformulation} covers the general case of including participants with quadratic costs in electricity markets, Example \ref{example:electricity_network} demonstrates the network-awareness of the market framework proposed in the paper by showing how it can be extended to include the SOCP relaxation of nonlinear and non-convex power flow equations.
\begin{example}[Quadratic cost]\label{example:quad_cost_reformulation}
Consider participant $i$ having a quadratic cost (utility) of production (consumption), such that the cost function at each hour $t$ is given by $c_{it}(\mathbf{q}_{it}) = \mathbf{q}_{it}^{\top}\text{diag}(\mathbf{c}^{\text{Q}}_{it})\mathbf{q}_{it} + {\mathbf{c}^{\text{L}}_{it}}^{\top}\mathbf{q}_{it}$, where $\mathbf{c}^{\text{Q}}_{it} \in \mathbb{R}^{K_{i}}$ and $\textbf{c}_{it}^{\text{L}}\in \mathbb{R}^{K_{i}}$ denote the quadratic and linear cost coefficients, respectively. Let ${\mathbf{C}}_{it}^{\text{Q}} \in \mathbb{R}^{K_{i} \times K_{i}}$ be defined as a factorization of the quadratic cost matrix such that $\text{diag}(\mathbf{c}^{\text{Q}}_{it})={{\mathbf{C}}^{\text{Q}}_{it}}^{\top}{{\mathbf{C}}_{it}^{\text{Q}}}$. The existence of such a factorization is given since $\text{diag}(\mathbf{c}^{\text{Q}}_{it})\succcurlyeq 0,\;\forall t,\;\forall i\in\mathcal{I}$. The quadratic objective function of the participant $i$ is equivalently written as 
\begin{subequations}
\begin{align}
     \minimize{\mathbf{q}_{it},z_{it}}\quad&\sum_{t\in\mathcal{T}} \Big( {z}_{it} + {\mathbf{c}_{it}^{\text{L}}}^{\top}\mathbf{q}_{it} \Big) \label{eq:obj_reformulation_example}\\
    \st\quad&\norm{{\mathbf{C}}_{it}^{\text{Q}}\mathbf{q}_{it}}^{2} \le  {z}_{it},\;\forall t, \label{eq:ex_quad_cost_reformulation}
\end{align}
\end{subequations}
where $z_{it} \in \mathbb{R}_{+}$ is an auxiliary variable, resulting in the linear objective function \eqref{eq:obj_reformulation_example}. The rotated SOC constraint \eqref{eq:ex_quad_cost_reformulation} is a special form of the general SOC constraint (1), see Mosek ApS (2021). %\eqref{eq:soc_constraint_example}, see \cite{MosekApS2021}. 
We illustrate this for single-period market-clearing, i.e., $T=1$, assuming that participant $i$ has $K_{i}=2$ decisions variables, such that $\mathbf{q}_{i1}\in\mathbb{R}^{2}$. Extending the decision vector of participant $i$ to $\mathbf{q}_{i} = \begin{bmatrix}\mathbf{q}_{i1}^{\top} \quad z_{i1}\end{bmatrix}^{\top} \in \mathbb{R}^{3}$, we have the parameters $\mathbf{A}_{i} = \begin{bmatrix}\mathbf{C}^{\text{Q}}_{it} \quad \mathbb{0}_{2}\end{bmatrix}$, $\mathbf{d}_{i} = \begin{bmatrix}0 \quad 0 \quad 1\end{bmatrix}^{\top}$, while $\mathbf{b}_{i}=\mathbb{0}_{2}$ and $e_{i}=0$, resulting in a three-dimensional rotated SOC constraint, i.e., $m_{i}=2$.
\end{example}
\medskip
\begin{example}[Electricity network]\label{example:electricity_network}
Known as the branch flow model \citep{Farivar2013}, the SOCP relaxation of power flow equations is exact for the distribution networks that exhibit a radial graph structure under mild conditions. The power flow equations model the nonlinear relation of active and reactive power transported along a line with the current and terminal node voltages. While active power is the actual quantity in MWh of power produced or consumed, reactive power is a necessary component oscillating within the network and regulates voltage across the nodes. Consider a power line $\ell = (n,n') \in \mathcal{L}$ connecting adjacent nodes $n,n'\in\mathcal{N}$. Let ${s_{\ell}^{\text{a}}}\in\mathbb{R}$ and  ${s_{\ell}^{\text{r}}}\in\mathbb{R}$ denote the active power and reactive power; $\theta_{\ell}\in\mathbb{R}_{+}$ and $v_{n}\in\mathbb{R}_{+}$ denote the squared magnitude of current flow in the line $\ell$ and the squared voltage magnitude at the node $n$, respectively. The SOC relaxation for the power flow equations $\forall \ell=(n,n')\in\mathcal{L}$ is
\begin{subequations}
\begin{align}
    &{s_{\ell}^{\text{a}}}^{2} + {s_{\ell}^{\text{r}}}^{2}  \leq \theta_{\ell}v_{n}\label{eq:app_power_def}\\
    &{s_{\ell}^{\text{a}}}^{2} + {s_{\ell}^{\text{r}}}^{2}  \leq {\overline{s}_{\ell}}^{2}, \label{eq:app_power_lims}
\end{align}\label{eq:BFM_relaxation_example}
\end{subequations}
where $\overline{s}_{\ell}\in\mathbb{R}_{+}$ is a parameter denoting the rated power transfer capacity of line $\ell$. Constraints \eqref{eq:BFM_relaxation_example} can be reformulated as SOC constraints admitting the variables $\begin{bmatrix}s_{\ell}^{\text{a}} \quad s_{\ell}^{\text{r}} \quad \theta_{\ell} \quad v_{n} \end{bmatrix}^{\top}$. For instance, \eqref{eq:app_power_def} is equivalent to an SOC constraint of the general form (1) %\eqref{eq:soc_constraint_example} 
with the parameters
\begin{align*}
    \mathbf{A}=\begin{bmatrix*}[l] 2 \quad 0 \quad 0 \quad \phantom{-}0 \\ 0 \quad 2 \quad 0 \quad \phantom{-}0 \\0 \quad 0 \quad 1 \quad -1 \end{bmatrix*}\;,\;\;\mathbf{b} = \mathbb{0}_{3}\;,\;\;\mathbf{d} = \begin{bmatrix}0 \quad 0 \quad 1 \quad 1 \end{bmatrix}^{\top}\;\text{and}\;\;e=0,
\end{align*}which is a four-dimensional SOC constraint. Similarly, \eqref{eq:app_power_lims} is a three-dimensional SOC constraint with a structure similar that in Example 1. %\ref{example:natural_gas_network}.
\end{example}

\section{Market-clearing Problems}\label{app:sec_uncertainty_aware}
For the uncertainty-aware conic market framework proposed in the paper, we first outline the participant models in \cref{ec:subsec_participant_models}, followed by the chance-constrained market-clearing problem and its reformulation as the SOCP problem $\mathcal{M}^{\text{cc}}$ in \cref{ec:subsec_Mcc_formulation}. In \cref{ec:subsec_R1_R2_formulation} we provide formulations for the two LP-based uncertainty-aware benchmark market-clearing problems, $\mathcal{R}1$ and $\mathcal{R}2$. Finally, \cref{ec:subsec_out_of_sample_simulations} formulates the real-time market-clearing problem entailed in the out-of-sample simulation studies discussed in \S4. %\cref{sec:4_numerical_studies}.
\subsection{Modeling of Market Participants} \label{ec:subsec_participant_models}
To enhance the clarity of exposition, we define subsets of market participants $\mathcal{I}$ comprised of flexible power producers $\mathcal{F}\subseteq\mathcal{I}$, energy storage owners $\mathcal{S}\subseteq\mathcal{I}$ and inflexible consumers $\mathcal{D}\subseteq\mathcal{I}$. Let the uncertainty faced by the market-clearing problem arise from a set of weather-dependent renewable power producers $\mathcal{W}\subseteq{\mathcal{I}}$, such that $\mathcal{F}\cup\mathcal{S}\cup\mathcal{D}\cup\mathcal{W} = \mathcal{I}$ and $\mathcal{F}\cap\mathcal{S}\cap\mathcal{D}\cap\mathcal{W}=\emptyset$. We consider consumers to be inflexible. However, the methodology adopted in this paper is extendable to include demand-side flexibility providers.
\subsubsection*{Weather-dependent power producers:}
At the day-ahead stage, the stochastic power production $\widetilde{q}_{it}\in\mathbb{R}$ from weather-dependent power producers $i\in\mathcal{W}=\{1,2,\dots,W\}$ are modeled as
\begin{align}
    \widetilde{q}_{it}(\bm{\xi}_{t}) = \widehat{q}_{it} - \xi_{it},\;\forall i \in \mathcal{W},\forall t, \label{eq:ec_def_windforecast}
\end{align}
where $\widehat{q}_{it}\in\mathbb{R}$ is the nominal production, usually in the form of the best-available day-ahead forecast, while the random variable $\xi_{it} \in \mathbb{R}$ is the forecast error encountered by producer $i$ at hour $t$. The stochastic power production $\widetilde{q}_{it}$ has an upper bound given by the rated power production capacity $\overline{Q}_{i}$, while being bounded below by $\underline{Q}_{i} = 0$. The vector $\bm{\xi} = \begin{bmatrix}\xi_{11} \quad\xi_{21}\quad\dots\quad\xi_{Wt}\quad\dots\quad\xi_{WT}\end{bmatrix}^{\top} \in \mathbb{R}^{WT}$ denotes the overall uncertainty faced by the market-clearing problem, formed by extending $\bm{\xi}_{t}\in\mathbb{R}^{W}$, as defined in Example 2,
%\ref{example:chance_constraints} 
to a multi-period setting. We assume that $\bm{\xi}$ follows an unknown multivariate probability distribution $\mathbb{P}_{\xi}$, characterized by mean $\bm{\mu}\in\mathbb{R}^{WT}$ and covariance $\bm{\Sigma}\in\mathbb{R}^{WT \times WT}$, which are estimated by the system operator having an access to a finite number of historical measurement samples. Without much loss of generality, we assume the distribution $\mathbb{P}_{\xi}$ to have a mean $\bm{\mu} = \mathbb{0}$, as any non-zero elements of the sample mean are used to update the forecast $\widehat{q}_{it}$ in \eqref{eq:ec_def_windforecast}. The structure of ${\bm{\Sigma}}$ is such that its diagonal blocks, comprised of sub-matrices, ${\bm{\Sigma}}_{t} \in \mathbb{R}^{W \times W},\; \forall t$, capture the spatial correlation among the forecast errors at hour $t$, while the off-diagonal blocks contain information about spatio-temporal correlation of uncertainty\footnote{Under probability distribution other than multivariate Gaussian distribution, considered in \S4, %\cref{sec:4_numerical_studies},
the analytical approximation of chance-constrained program as a deterministic SOCP problem relies on independence of the random variables (Nemirovski 2012), %\citep{Nemirovski2012}, 
i.e., the covariance matrix $\bm{\Sigma}$ must be modeled as a diagonal matrix, neglecting any spatio-temporal correlation.}. The net deviation from the day-ahead forecasts realized during real-time operation at hour $t$ is thus given by $\mathbb{1}^{\top}\bm{\xi}_{t}\in\mathbb{R}$. As a sign convention, $\mathbb{1}^{\top}\bm{\xi}_{t} > 0$ implies a deficit of production from renewable energy sources during real-time operation stage as compared to the day-ahead forecast. Weather-dependent power producers are associated with a decision vector ${\bm{q}}_{it} = \begin{bmatrix}\widehat{q}_{it} \quad \xi_{it} \end{bmatrix}^{\top},\;\forall t,\;\forall i\in\mathcal{W}$ such that participation in the two commodities traded in the market is determined as follows. For the commodity energy, i.e., $p=1$, the contribution is given by
\begin{align*}
    {\bm{q}}_{ip} = \begin{bmatrix}\widehat{q}_{i1} \quad \widehat{q}_{i2}\;\dots\;\widehat{q}_{i(T-1)}\quad \widehat{q}_{iT}\end{bmatrix}^{\top},\;\mathbf{G}_{ip} = \text{diag}(\mathbb{1}),\;\forall i\in\mathcal{W},
\end{align*}
where $\mathbb{1}\in\mathbb{R}^{T}$. For $p=2$, i.e., flexibility, the weather-dependent power producers are modeled as the uncertainty sources
\begin{align*}
    &{\bm{q}}_{ip} = \mathbb{1},\;\mathbf{G}_{ip}= -\begin{bmatrix} &\mathbb{1}^{\top}\bm{\xi}_{1} &\cdots &0\\
    &\vdots &\ddots &\vdots\\
    &0 & \cdots &\mathbb{1}^{\top}\bm{\xi}_{T}
    \end{bmatrix},\;\forall i \in \mathcal{W},
\end{align*}
where $\mathbb{1}\in\mathbb{R}^{T}$ and the negative sign arises from the convention adopted in \eqref{eq:ec_def_windforecast}.
\subsubsection*{Flexible power producers:}
For each flexible power producer $i\in\mathcal{F}$, we model the stochastic power production during real-time operation at hour $t$ as
\begin{align}
    \widetilde{q}_{it}(\bm{\xi}_{t}) = \widehat{q}_{it} + h_{it}(\bm{\xi}_{t})\;,\;\forall i \in \mathcal{F},\;\forall t,
\end{align}
where $\widehat{q}_{it}$ denotes the nominal production and the function $h_{it}(\bm{\xi}_{t}) : \mathbb{R}^{W} \mapsto \mathbb{R}$ is the adjustment policy allocated to the market participant $i$, encoding its share in the recourse actions needed to mitigate the net deviation in the electricity system arising from forecast errors realized at hour $t$. Typically, these adjustment policies $h_{it}(\bm{\xi}_{t})$ represent convex decision rules, which may be linear \citep{Georghiou2019} or generalized \citep{Georghiou2015}. In this work, we adopt adjustment policies affinely dependent on the total uncertainty faced by the system operator, such that
\begin{align*}
    h_{it}(\bm{\xi}_{t}) = \mathbb{1}^{\top}\bm{\xi}_{t}\alpha_{it}\;,\;\forall i \in \mathcal{F},\;\forall t,
\end{align*}
where $\alpha_{it}$ is the adjustment policy allocated to participant $i$ at hour $t$. The nominal production quantity $\widehat{q}_{it}$ and the policy $\alpha_{it}$ are optimally decided by the day-ahead market-clearing program. Recalling Example 2, %\ref{example:chance_constraints}, 
we write chance constraints limiting the production from participants to their uppers limit $\overline{Q}_{i}$ as
\begin{subequations}
\begin{align}
    \mathbb{P}_{\xi}\Big(\widehat{q}_{it} + \mathbb{1}^{\top}\bm{\xi}_{t}\alpha_{it} \leq \overline{Q}_{i}\Big) \geq (1 - \widehat{\varepsilon}),\;\forall t,\;\forall i\in\mathcal{F},
\end{align}
which is reformulated as a $W+1$-dimensional SOC constraint
\begin{align}
    r_{\widehat{\varepsilon}}\norm{\mathbf{X}_{t}\mathbb{1}\alpha_{it}} \leq \overline{Q}_{i} - \widehat{q}_{it},\;\forall t,\;\forall i\in\mathcal{F},
\end{align}
\end{subequations}
where $\mathbf{X}_{t}\in\mathbb{R}^{W \times W}$ is obtained by Cholesky decomposition of the submatrix $\bm{\Sigma}_{t}$ of the covariance matrix $\bm{\Sigma}$ such that $\bm{\Sigma}_{t} = \mathbf{X}_{t}\mathbf{X}_{t}^{\top}$. Recall that the parameter $r_{\widehat{\varepsilon}}$ is a safety parameter, related to constraint violations, chosen by the system operator based on the knowledge of distribution $\mathbb{P}_{\xi}$, such that $r_{\widehat{\varepsilon}}$ increases as $\widehat{\varepsilon}$ decreases. Similar reformulation is obtained for the lower bounds on the production, $\underline{Q}_{i}$. Apart from the production bounds, a flexible power producer may have capacity bounds on the flexibility provision, denoted by $\underline{Q}_{i}^{R},\;\overline{Q}_{i}^{R}$, which are reformulated in a similar manner. Lastly, inter-temporal constraints such as the limits on the downward and upward ramping rates $\underline{\Delta}_{i},\;\overline{\Delta}_{i} \in\mathbb{R}$ are modeled as 
\begin{subequations}
\begin{align}
    &\mathbb{P}_{\xi}\Big((\widehat{q}_{it} + \mathbb{1}^{\top}\bm{\xi}_{t}\alpha_{it}) - (\widehat{q}_{i(t-1)} + \mathbb{1}^{\top}\bm{\xi}_{t-1}\alpha_{i(t-1)}) \geq -\underline{\Delta}_{i}\Big) \geq (1 - \widehat{\varepsilon}), \;\forall t>2,\;\forall i\in\mathcal{F}\\
    &\mathbb{P}_{\xi}\Big((\widehat{q}_{it} + \mathbb{1}^{\top}\bm{\xi}_{t}\alpha_{it}) - (\widehat{q}_{i(t-1)} + \mathbb{1}^{\top}\bm{\xi}_{t-1}\alpha_{i(t-1)}) \leq \overline{\Delta}_{i}\Big) \geq (1 - \widehat{\varepsilon}), \;\forall t>2,\;\forall i\in\mathcal{F}
\end{align}
and reformulated as $2W + 1$-dimensional SOC constraints of the form
\begin{align}
    &r_{\widehat{\varepsilon}}\norm{\mathbf{X}_{t-1:t}\begin{bmatrix}\phantom{-}\mathbb{1}^{\top}\alpha_{i(t-1)}\quad&-\mathbb{1}^{\top}\alpha_{it}\end{bmatrix}^{\top}} \leq \underline{\Delta}_{i} - (\widehat{q}_{it-1} - \widehat{q}_{it}), \;\forall t>2,\;\forall i\in\mathcal{F}\\
    &r_{\widehat{\varepsilon}}\norm{\mathbf{X}_{t-1:t}\begin{bmatrix}-\mathbb{1}^{\top}\alpha_{i(t-1)}\quad&\phantom{-}\mathbb{1}^{\top}\alpha_{it}\end{bmatrix}^{\top}} \leq \overline{\Delta}_{i} + (\widehat{q}_{it-1} - \widehat{q}_{it}), \;\forall t>2,\;\forall i\in\mathcal{F},
\end{align}
\end{subequations}
where $\mathbf{X}_{t-1:t}\in\mathbb{R}^{2W \times 2W}$ denotes the factorization of the blocks of covariance matrix $\bm{\Sigma}$ corresponding to the spatio-temporal covariance of forecast errors in two consecutive hours. Flexible power producers participate in the conic market with decision vectors ${\bm{q}}_{it} = \begin{bmatrix}\widehat{q}_{it} \quad \alpha_{it}\end{bmatrix}^{\top}, \forall t,\;\forall i \in \mathcal{F}$. Towards the commodity energy, i.e., $p=1$, flexible generators contribute as
\begin{align*}
    &{\bm{q}}_{ip} = \begin{bmatrix}\widehat{q}_{i1} \quad \widehat{q}_{i2}\;\dots\;\widehat{q}_{i(T-1)}\quad \widehat{q}_{iT}\end{bmatrix}^{\top},\;\mathbf{G}_{ip} = \text{diag}(\mathbb{1}),\;\forall i \in \mathcal{F},
\end{align*}
with $\mathbb{1}\in\mathbb{R}^{T}$. For the commodity flexibility ($p=2$), the contribution is given by
\begin{align*}
    &{\bm{q}}_{ip} = \begin{bmatrix}\alpha_{i1} \quad \alpha_{i2}\;\dots\;\alpha_{i(T-1)}\quad \alpha_{iT}\end{bmatrix}^{\top},\;\mathbf{G}_{ip}= \begin{bmatrix} &\mathbb{1}^{\top}\bm{\xi}_{1} &\cdots &0\\
    &\vdots &\ddots &\vdots\\
    &0 & \cdots &\mathbb{1}^{\top}\bm{\xi}_{T}
    \end{bmatrix},\;\forall i \in \mathcal{F}.
\end{align*}
\subsubsection*{Energy storage operators:} 
%%%----FIGURE-----%%%
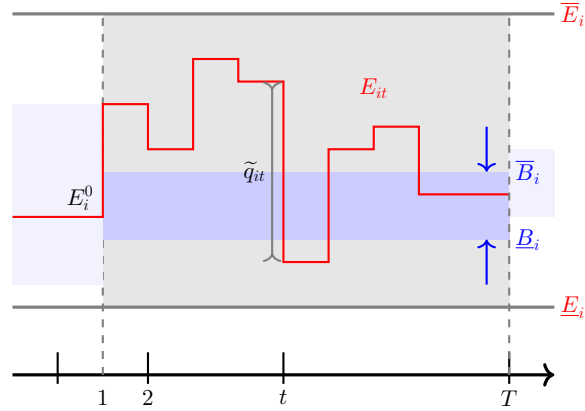
\begin{figure}
\centering
\begin{tikzpicture}[thick,scale=0.6, every node/.style={scale=0.6}]
%DRAW BASE LINE
\draw[very thick,black,->] (0,0) -- (12,0);
\fill[gray!20] (2,1.5) rectangle(11,8);
%TICKS
\draw[thick,black](1,-0.2)--(1,0.5);
\node [below] at(2,-0.2){\large $1$};
\draw[thick,black](3,-0.2)--(3,0.5);
\node [below] at(3,-0.2){\large $2$};
\draw[thick,black](6,-0.2)--(6,0.5);
\node [below] at(6,-0.2){\large $t$};
\draw[thick,black](11,-0.2)--(11,0.5);
\node [below] at(11,-0.2){\large $T$};
%DRAWING VERTICAL LINE
\draw[dashed,thick,gray](2,0)--(2,8);

%Remove this after first screenshot
%Drawing Energy Neutrality Preference Region
\fill[blue!5] (0,2) rectangle(2,6);
\fill[blue!20] (2,3) rectangle(11,4.5);
\fill[blue!5] (11,3.5) rectangle(12,5);
\draw[blue, thick, ->] (10.5,5.5) -- (10.5,4.5);
\node [right, text=blue] at (11,4.5) {\large $\overline{B}_{i}$};
\draw[blue, thick, ->] (10.5,2) -- (10.5,3);
\node [right, text=blue] at (11,3) {\large $\underline{B}_{i}$};

%DRAWING FUNCTION Est(t) 
\draw[red,thick](0,3.5)--(2,3.5)--(2,6)--(3,6)--(3,5)--(4,5)--(4,7)--(5,7)--(5,6.5)--(6,6.5)--(6,2.5)--(7,2.5)--(7,5)--(8,5)--(8,5.5)--(9,5.5) -- (9,4) -- (11,4);
\node [above] at (1.5,3.5){\large $E^{0}_{i}$};
\draw [black!50, >-<](5.75,6.5) -- (5.75,2.5);
\node [left, text=black] at (5.75,4.5){\large $\widetilde{q}_{it}$};
\node [above,text=red] at (8,6){\large $E_{it}$};

%DRAWING E_s_max and E_s_min
\draw[gray, very thick] (0, 8) -- (12,8);
\node[right, text=red] at (12,8){\large $\overline{E}_{i}$};
\draw[gray, very thick] (0, 1.5) -- (12,1.5);
\node[right, text=red] at (12,1.5){\large $\underline{E}_{i}$};

%DRAWING 
%\draw[gray, very thick, dotted] (0, 10.75) -- (12,10.75);
%\node[right] at (12,10.75){$\overline{E}_{i}$};
%\draw[gray, very thick, dotted] (0, 0.8) -- (12,0.8);
%\node[right] at (12,0.8){$E_{i}=0$};

%END OF HORIZON
\draw [dashed, gray](11,0)--(11,8);

\end{tikzpicture}
\caption[Illustration of ESU model]{Illustration of the operation of an energy storage unit for flexibility provision.}
\label{fig:ESR_intertemp_onstraints}
\vspace{-0.25cm}
\end{figure}
For each energy storage unit $s\in\mathcal{S}$, the power produced or consumed during real-time operation at hour $t$ is given by
\begin{align}
    \widetilde{q}_{it}(\bm{\xi}_{t}) = \widehat{q}_{it} +\mathbb{1}^{\top}\bm{\xi}_{t}\gamma_{it}\;,\;\forall i \in \mathcal{S},\;\forall t,
\end{align}
where $\widehat{q}_{it}$ is the nominal production/consumption and $\gamma_{it}$ is the affine adjustment policy allocated. We adopt the sign convention $\widetilde{q}_{it}(\bm{\xi}_{t})> 0$ for hours when the storage is discharging (production), while $\widetilde{q}_{it}(\bm{\xi}_{t})<0$ indicates charging (consumption). In practice, energy storage units are limited by their charging and discharging capacities which are modeled using chance constraints as
\begin{subequations}
\begin{align}
    &\mathbb{P}_{\xi}\Big(\widehat{q}_{it} + \mathbb{1}^{\top}\bm{\xi}_{t}\gamma_{it} \leq \eta_{i}^{\text{D}}\overline{Q}^{\text{D}}_{i}\Big) \geq (1 - \widehat{\varepsilon}),\;\forall t,\;\forall i\in\mathcal{S}\\
    &\mathbb{P}_{\xi}\Big(\widehat{q}_{it} + \mathbb{1}^{\top}\bm{\xi}_{t}\gamma_{it} \geq -\frac{1}{\eta_{i}^{\text{C}}}\overline{Q}^{\text{C}}_{i}\Big) \geq (1 - \widehat{\varepsilon}),\;\forall t,\;\forall i\in\mathcal{S}
\end{align}
and are reformulated as $W+1$-dimensional SOC constraints as
\begin{align}
    &r_{\widehat{\varepsilon}}\norm{\mathbf{X}_{t}\mathbb{1}\gamma_{it}} \leq \eta_{i}^{\text{D}}\overline{Q}^{\text{D}}_{i} - \widehat{q}_{it},\;\forall t,\;\forall i\in\mathcal{S}\\
    &r_{\widehat{\varepsilon}}\norm{\mathbf{X}_{t}\mathbb{1}\gamma_{it}} \leq \frac{1}{\eta_{i}^{\text{C}}}\overline{Q}^{\text{C}}_{i} + \widehat{q}_{it},\;\forall t,\;\forall i\in\mathcal{S},
\end{align}\label{eq:esr_chg_dis_lims}
\end{subequations}
where $\eta_{i}^{\text{D}},\eta_{i}^{\text{C}}\in [0,1]$ are energy conversion factors representing discharging and charging efficiencies, while $\overline{Q}_{i}^{D},\;\overline{Q}_{i}^{C}\in\mathbb{R}$ are, respectively, the maximum discharging and charging capacities of the energy storage unit $i\in\mathcal{S}$. Inter-temporal constraints are critical for energy storage units to ensure that (i) the storage operation trajectory remains within the limits of state of charge, and (ii) the storage unit is neither depleted nor over-charged at the end of market-clearing horizon. While the former ensures safe operation of the storage unit within its operational limits, the latter is relevant in a market-clearing setup where the storage unit is expected to provide flexibility to the grid on an ongoing basis. Figure \ref{fig:ESR_intertemp_onstraints} shows the trajectory (in red) of energy content of a storage unit. At each hour, the changes to energy content due to production (discharging) or consumption (charging) by the storage unit is given by $\widetilde{q}_{it}(\bm{\xi}_{t})$, with the rate of change being limited by the charging and discharging power limits in \eqref{eq:esr_chg_dis_lims}. The energy content of a storage unit evolves as
\begin{subequations}
\begin{align}
    &\mathbb{P}_{\xi}\Big(E_{i}^{0} - \sum_{t'=1}^{t} (\widehat{q}_{it'} + \mathbb{1}^{\top}\bm{\xi}_{t}\gamma_{it'}) \leq \overline{E}_{i}\Big) \geq (1 - \widehat{\varepsilon}),\;\forall t,\;\forall i\in\mathcal{S}\\
    &\mathbb{P}_{\xi}\Big(E_{i}^{0} - \sum_{t'=1}^{t} (\widehat{q}_{it'} + \mathbb{1}^{\top}\bm{\xi}_{t}\gamma_{it'}) \geq \underline{E}_{i}\Big) \geq (1 - \widehat{\varepsilon}),\;\forall t,\;\forall i\in\mathcal{S}, 
\end{align}
which are reformulated as $W\times t + 1,\;\forall t$-dimensional SOC constraints, expressed as
\begin{align}
    &r_{\widehat{\varepsilon}}\norm{\mathbf{X}_{1:t}\begin{bmatrix}\mathbb{1}^{\top}\gamma_{i1} \quad \mathbb{1}^{\top}\gamma_{i2}\;\cdots\;\mathbb{1}^{\top}\gamma_{it}\end{bmatrix}^{\top}} \leq \overline{E}_{i} - E_{i}^{0} + \sum_{t'=1}^{t}\widehat{q}_{it'},\;\forall t,\;\forall i\in\mathcal{S}\\
    &r_{\widehat{\varepsilon}}\norm{\mathbf{X}_{1:t}\begin{bmatrix}\mathbb{1}^{\top}\gamma_{i1} \quad \mathbb{1}^{\top}\gamma_{i2}\;\cdots\;\mathbb{1}^{\top}\gamma_{it}\end{bmatrix}^{\top}} \leq E_{i}^{0} - \underline{E}_{i} - \sum_{t'=1}^{t}\widehat{q}_{it'},\;\forall t,\;\forall i\in\mathcal{S},
\end{align}\label{eq:eod_soc_limits}
\end{subequations}
where $E_{i}^{0} \in \mathbb{R}$ is the energy content at $t=0$ and  $\underline{E}_{i},\;\overline{E}_{i}\in\mathbb{R}$ are the minimum and maximum energy storage capacity of the storage unit, respectively, such that $E_{i}^{0}\in[\underline{E}_{i},\;\overline{E}_{i}]$. Furthermore, to ensure the ongoing market-participation of the storage unit, we utilize the concept of end-of-horizon energy neutrality. We define lower and upper bounds,  $\underline{B}_{i},\;\overline{B}_{i}\in\mathbb{R}$ respectively, around the initial energy stored $E_{i}^{0}$ to reflect the preference of the energy storage owner on the energy content at the end of market-clearing horizon, see Figure \ref{fig:ESR_intertemp_onstraints}. This ensures the end-of-horizon energy content to be within $[E_{i}^{0} -\underline{B}_{i},\;E_{i}^{0} +\overline{B}_{i}]$. These preferences are captured through chance constraints
\begin{subequations}
\begin{align}
    &\mathbb{P}_{\xi}\Big(E_{i}^{0} - \sum_{t'=1}^{T}(\widehat{q}_{it'} + \mathbb{1}^{\top}\bm{\xi}_{t}\gamma_{it'}) \leq E_{i}^{0} + \overline{B}_{i}\Big) \geq (1 - \widehat{\varepsilon}), \;\forall i \in \mathcal{S}\\
    &\mathbb{P}_{\xi}\Big(E_{i}^{0} - \sum_{t'=1}^{T}(\widehat{q}_{it'} + \mathbb{1}^{\top}\bm{\xi}_{t}\gamma_{it'}) \geq E_{i}^{0} - \underline{B}_{i}\Big) \geq (1 - \widehat{\varepsilon}),\;\forall i \in \mathcal{S}
\end{align}
which are reformulated as $W\times T + 1$-dimensional SOC constraints, expressed as
\begin{align}
    &r_{\widehat{\varepsilon}}\norm{\mathbf{X}_{1:T}\begin{bmatrix}\mathbb{1}^{\top}\gamma_{i1} \quad \mathbb{1}^{\top}\gamma_{i2}\;\cdots\;\mathbb{1}^{\top}\gamma_{iT}\end{bmatrix}^{\top}} \leq \overline{B}_{i} + \sum_{t'=1}^{T}\widehat{q}_{it'},\;\forall i \in \mathcal{S}\\
    &r_{\widehat{\varepsilon}}\norm{\mathbf{X}_{1:T}\begin{bmatrix}\mathbb{1}^{\top}\gamma_{i1} \quad \mathbb{1}^{\top}\gamma_{i2}\;\cdots\;\mathbb{1}^{\top}\gamma_{iT}\end{bmatrix}^{\top}} \leq \underline{B}_{i} - \sum_{t'=1}^{T}\widehat{q}_{it'},\;\forall i \in \mathcal{S}.
\end{align}\label{eq:soc_limits}
\end{subequations}
Each energy storage operators participates in the conic market with a decision vector ${\bm{q}}_{it} = \begin{bmatrix}\widehat{q}_{it} \quad \gamma_{it}\end{bmatrix}^{\top}, \forall t,\;\forall i \in \mathcal{S}$. Likewise, towards the commodity energy, i.e., $p=1$, we have 
\begin{align*}
    &{\bm{q}}_{ip} = \begin{bmatrix}\widehat{q}_{i1} \quad \widehat{q}_{i2}\;\dots\;\widehat{q}_{i(T-1)}\quad \widehat{q}_{iT}\end{bmatrix}^{\top},\;\mathbf{G}_{ip} = \text{diag}(\mathbb{1}),\;\forall i \in \mathcal{S},
\end{align*}
with $\mathbb{1}\in\mathbb{R}^{T}$. For the commodity flexibility ($p=2$), the storage unit contributes to the trades as
\begin{align*}
    &{\bm{q}}_{ip} = \begin{bmatrix}\gamma_{i1} \quad \gamma_{i2}\;\dots\;\gamma_{i(T-1)}\quad \gamma_{iT}\end{bmatrix}^{\top},\;\mathbf{G}_{ip}= \begin{bmatrix} &\mathbb{1}^{\top}\bm{\xi}_{1} &\cdots &0\\
    &\vdots &\ddots &\vdots\\
    &0 & \cdots &\mathbb{1}^{\top}\bm{\xi}_{T}
    \end{bmatrix}, \forall i \in \mathcal{S}.
\end{align*}
\subsubsection*{Inflexible consumers:} For each inflexible consumer $d\in\mathcal{D}$, the power consumed during real-time operation at hour $t$ is given by $\widetilde{q}_{it}(\bm{\xi}_{t}) = \widehat{q}_{it},\;\forall i\in\mathcal{D},\;\forall t$, where $\widehat{q}_{it}\in\mathbb{R}_{-}$ is the power demand by consumer $d\in\mathcal{D}$. The above modeling can be extended to include flexible consumers that may respond to the uncertainty by reducing/increasing their consumption based on an affine adjustment policy, similar to flexible generators and energy storage units. However, for clarity of exposition, this paper is restricted to modeling inflexible and perfectly inelastic demand. Each consumer contributes towards the trades of commodity energy, i.e., for $p=1$, as
\begin{align*}
    &{\bm{q}}_{ip} = \begin{bmatrix}\widehat{q}_{i1} \quad \widehat{q}_{i2}\;\dots\;\widehat{q}_{i(T-1)}\quad \widehat{q}_{iT}\end{bmatrix}^{\top},\;\mathbf{G}_{ip} = \text{diag}(\mathbb{1}),\;\forall i \in \mathcal{D},
\end{align*}
where $\mathbb{1}\in\mathbb{R}^{T}$. Lastly, for $p=2$, considering inflexible consumers, we have ${\bm{q}}_{ip} = \mathbb{0},\;\mathbf{G}_{ip} = \mathbf{0},\;\forall i \in \mathcal{D}$. 
\subsection{Chance-constrained Market Clearing}\label{ec:subsec_Mcc_formulation}
In the following, we formulate the chance-constrained market-clearing problem in stochastic optimization variables $\widetilde{q}_{it}(\bm{\xi}_{t}),\;\forall i\in\mathcal{I}$ which are functions of the random variable $\bm{\xi}_{t}$. We drop this dependency for notational convenience, yet it is always implied.
\begin{subequations}\label{prob:Mc_CC}
\begin{align}
    &\minimize{{\widetilde{q}}_{it}}\quad\;\mathbb{E}^{\mathbb{P}_{\xi}}\Big[\sum_{i \in \mathcal{I}\setminus\mathcal{D}}\;\sum_{t\in\mathcal{T}}\;\Big({c}^{\text{Q}}_{it}{\widetilde{q}}_{it}^{2} + {{c}_{it}^{\text{L}}}{\widetilde{q}}_{it}\Big)\Big]\label{objfun_Mc_CC}\\
    &\quad\st\quad\mathbb{P}_{\xi}\left[\begin{aligned}
    &\sum_{i \in \mathcal{I}} \mathbf{G}_{ip}\widetilde{\mathbf{q}}_{ip} = \mathbf{0}_{T}, \; \forall p\in\mathcal{P}, \end{aligned}\right]\overset{\text{a.s.}}{=}1, \label{cc:equalities}\\
    &\phantom{\st}\quad\quad\mathbb{P}_{\xi}\left[\begin{aligned}
    &\underline{Q}_{i} \leq \widetilde{q}_{it} \leq \overline{Q}_{i},\;\forall t,\;\forall i \in \mathcal{F}\cup\mathcal{W}\\
    &-\underline{Q}_{i}^{R} \leq h_{it}(\bm{\xi}_{t}) \leq \overline{Q}_{i}^{R},\;\forall i \in \mathcal{F}\\
    &-\underline{\Delta}_{i} \leq \widetilde{q}_{it} - \widetilde{q}_{i(t-1)} \leq \overline{\Delta}_{i},\;\forall t>2,\;\forall i \in \mathcal{F}\\
    &-\frac{1}{\eta_{i}^{C}}\overline{Q}_{i}^{\text{C}} \leq \widetilde{q}_{it}\leq\eta_{i}^{D}\overline{Q}_{i}^{\text{D}},\;\forall t,\;\forall i \in \mathcal{S}\\
    &\underline{E}_{i} \leq E_{i}^{0} - \sum_{t'=1}^{t} \widetilde{q}_{it'} \leq \overline{E}_{i},\;\forall t,\;\forall i \in \mathcal{S}\\
    &E_{i}^{0} - \underline{B}_{i} \leq E_{i}^{0} - \sum_{t'=1}^{T} \widetilde{q}_{it'} \leq E_{i}^{0} + \overline{B}_{i},\;\forall i \in \mathcal{S} \\
    &-\overline{s}_{\ell} \leq \sum_{n \in \mathcal{N}} [\bm{\Psi}]_{(\ell,n)}\left(\sum_{i \in \mathcal{I}_{n}}\sum_{p\in\mathcal{P}}\;[\mathbf{G}_{ip}\widetilde{\mathbf{q}}_{ip}]_{t}\right) \leq \overline{s}_{\ell},\;\forall t,\;\forall \ell \in \mathcal{L}
    \end{aligned}\right]\geq 1-\varepsilon, \label{cc:inequalities}
\end{align}
\end{subequations}
where the objective function \eqref{objfun_Mc_CC} minimizes the expected cost, which is equivalent to minimizing negative social welfare when facing an inflexible and perfectly inelastic demand. The almost sure (a. s.) constraint \eqref{cc:equalities} ensures the satisfaction of the supply-demand balance constraint for both commodities with a probability of 1, whereas the chance constraint \eqref{cc:inequalities} ensures the inequalities are jointly met with a probability of $(1-\varepsilon)$. The prescribed constraint violation probability $\varepsilon \in (0,1)$ reflects risk tolerance of the system operator towards the violation of technical limits of the system and in our numerical studies discussed in \S4, %\cref{sec:4_numerical_studies},
we set $\varepsilon = 0.05$. The chance-constrained market-clearing problem \eqref{prob:Mc_CC} is computationally intractable since it involves infinitely many constraints arising from the uncertain production from weather-dependent power producers. We gain tractability by expressing the stochastic variables $\widetilde{q}_{it},\;\forall i,\;\forall t$ as affine, finite-dimensional functions of the random variable $\xi$ as discussed in the previous section, resulting in an approximate solution to the infinite-dimensional problem. In the following, we discuss the reformulations to reach the final tractable SOCP-based chance-constrained market-clearing problem.
\subsubsection*{Reformulation of joint chance constraint:}
Observe that \eqref{cc:inequalities} models a joint violation of the constraints, in contrast to the individual chance constraints discussed so far. In this paper, following the Bonferroni approximation of joint chance constraints \citep{Xie2019}, we adopt a consideration of individual chance constraints with the analytical parameterization of risk. Let $N^{\text{ineq}}$ denote the number of individual inequalities forming the joint chance constraint \eqref{cc:inequalities}, such that $\bm{\widehat{\varepsilon}}\in\mathbb{R}_{+}^{{N}^{\text{ineq}}}$ collects all the individual violation probabilities for the inequality constraints. The Bonferroni reformulation of joint chance constraint mandates that the individual constraint violation probabilities be chosen such that $\mathbb{1}^{\top}\bm{\widehat{\varepsilon}} \leq \varepsilon$. Furthermore, this approach provides a joint constraint feasibility guarantee even if the choice of individual probabilities is done trivially, e.g., $\bm{\widehat{\varepsilon}}$ is chosen such that $\widehat{\varepsilon}_{k} = \frac{\varepsilon}{N^{\text{ineq}}},\;\forall k=1,2,\dots,N^{\text{ineq}}.$ We compute the individual constraint violation probabilities in this manner and adopt the SOC reformulation techniques for individual chance constraints discussed so far.
 \subsubsection*{Reformulation of almost sure constraint:} The almost sure constraint \eqref{cc:equalities} must be held at the optimal solution to \eqref{prob:Mc_CC} with a probability of 1. Considering the affine dependency of the stochastic decision variables $\widetilde{q}_{it},\;\forall i,\;\forall t$ with respect to the random variables $\bm{\xi}_{t}$, this amounts to a separation of \eqref{cc:equalities} into nominal and recourse equalities. Corresponding to the two commodities energy and flexibility, these equalities are given by
\begin{subequations}
\begin{align}
    &\sum_{i \in \mathcal{I}} \;\mathbf{G}_{ip}{\mathbf{q}}_{ip} = \mathbb{0},\;p=1 \\
    &\sum_{i \in \mathcal{I}\setminus{(\mathcal{W}\cup\mathcal{D})}}\;\text{diag}(\mathbb{1}){\mathbf{q}}_{ip} = \mathbb{1},\;p=2.
\end{align}
\end{subequations}
\subsubsection*{Power flow limits:} Apart from the participant-specific chance constraints discussed so far, the constraints limiting the power flows in the network are 
\begin{subequations}
\begin{align}
    &\mathbb{P}_{\xi}\Big[\sum_{n \in \mathcal{N}} [\bm{\Psi}]_{(\ell,n)}\left(\sum_{i \in \mathcal{I}_{n}}\sum_{p\in\mathcal{P}}\;[\mathbf{G}_{ip}\widetilde{\mathbf{q}}_{ip}]_{t}\right) \leq \overline{s}_{\ell}\Big] \geq 1-\widehat{\varepsilon},\;\forall t,\;\forall \ell \in \mathcal{L} \label{eq:powerflow_example}\\
    &\mathbb{P}_{\xi}\Big[\sum_{n \in \mathcal{N}} [\bm{\Psi}]_{(\ell,n)}\left(\sum_{i \in \mathcal{I}_{n}}\sum_{p\in\mathcal{P}}\;[\mathbf{G}_{ip}\widetilde{\mathbf{q}}_{ip}]_{t}\right) \geq -\overline{s}_{\ell}\Big] \geq 1-\widehat{\varepsilon},\;\forall t,\;\forall \ell \in \mathcal{L},
\end{align}
\end{subequations}
which require tractable reformulations. In the following, we illustrate the reformulation of \eqref{eq:powerflow_example} as an SOC constraint and adopt a similar approach for the other flow direction. First, we rewrite \eqref{eq:powerflow_example} such that nominal and uncertainty-dependent terms are separable. To that end, we define auxiliary network matrices $\bm{\Psi}^{\text{F}}\in\mathbb{R}^{N\times\abs{\mathcal{F}}}$, $\bm{\Psi}^{\text{S}}\in\mathbb{R}^{N\times\abs{\mathcal{S}}}$, $\bm{\Psi}^{\text{W}}\in\mathbb{R}^{N\times W}$ and $\bm{\Psi}^{\text{D}}\in\mathbb{R}^{N\times\abs{\mathcal{D}}}$ which map the flexible power producers, energy storage units, weather-dependent power producers and consumers, respectively, to the electricity network nodes. We collect all commodity contributions by participant groups at a given hour by defining auxiliary variables $\widetilde{\mathbf{q}}^{\text{F}}_{t}, \;\widehat{\mathbf{q}}^{\text{F}}_{t},\;\bm{\alpha}_{t} \in \mathbb{R}^{|\mathcal{F}|}$ to denote the stochastic production, nominal production and the adjustment policies for the flexible power producers, $\widetilde{\mathbf{q}}^{\text{S}}_{t},\;\widehat{\mathbf{q}}^{\text{S}}_{t},\;\bm{\gamma}_{t} \in \mathbb{R}^{|\mathcal{S}|}$ to denote stochastic production, nominal production and adjustment policies for the storage units, $\widehat{\mathbf{q}}^{\text{W}}_{t} \in \mathbb{R}^{W}$ for the forecasted production from weather-dependent power producers and lastly, $\widehat{\mathbf{q}}^{\text{D}}_{t} \in \mathbb{R}^{|\mathcal{D}|}$ for the inflexible demand from consumers. With that, we rewrite \eqref{eq:powerflow_example} as 
\begin{align}
    \mathbb{P}_{\xi}\Big[[\bm{\Psi}(\bm{\Psi}^{F}\widetilde{\mathbf{q}}^{F}_{t} +  \bm{\Psi}^{S}\widetilde{\mathbf{q}}^{S}_{t} + \bm{\Psi}^{W}(\widetilde{\mathbf{q}}^{W}_{t} - \bm{\xi}_{t}) + \bm{\Psi}^{D}\widehat{\mathbf{q}}^{\text{D}}_{t})]_{\ell} \leq \overline{s}_{\ell}\Big] \geq 1 - \widehat{\varepsilon},\;\forall t,\;\forall \ell \in \mathcal{L},
\end{align}
which we then reformulate as an SOC constraint after separating the nominal and uncertainty-dependent terms. The final SOC reformulation for \eqref{eq:powerflow_example} results in $\forall t,\;\forall \ell \in \mathcal{L}$,
\begin{align}
    r_{\widehat{\varepsilon}}\norm{\mathbf{X}_{t}[\bm{\Psi}(\bm{\Psi}^{F}\bm{\alpha}_{t}\mathbb{1}^{\top} + \bm{\Psi}^{S}\bm{\gamma}_{t}\mathbb{1}^{\top} - \bm{\Psi}^{W})]^{\top}_{(\ell,:)}} \leq \overline{s}_{\ell} - [ \bm{\Psi}(\bm{\Psi}^{F}\widehat{\mathbf{q}}^{F}_{t} + \bm{\Psi}^{S}\widehat{\mathbf{q}}^{S}_{t} + \bm{\Psi}^{W}\widehat{\mathbf{q}}^{W}_{t} + \bm{\Psi}^{D}\widehat{\mathbf{q}}^{D}_{t})]_{\ell}\;\;.
\end{align}
\subsubsection*{Objective function reformulation:} We decompose the objective \eqref{objfun_Mc_CC} among the various participants and expand the stochastic term to its nominal and recourse values to obtain
\begin{align*}
    \mathbb{E}^{\mathbb{P}_{\xi}}\Big[&\underbrace{\sum_{i \in \mathcal{F}}\;\sum_{t\in\mathcal{T}}\;\Big({c}^{\text{Q}}_{it}({\widehat{q}}_{it} + \mathbb{1}^{\top}\bm{\xi}_{t}\alpha_{it})^{2} + {{c}_{it}^{\text{L}}}({\widehat{q}}_{it} + \mathbb{1}^{\top}\bm{\xi}_{t}\alpha_{it})\Big)}_{\text{Term A}} \\
    &\quad + \underbrace{\sum_{i \in \mathcal{S}}\;\sum_{t\in\mathcal{T}}\;\Big({c}^{\text{Q}}_{it}({\widehat{q}}_{it} + \mathbb{1}^{\top}\bm{\xi}_{t}\gamma_{it})^{2} + {{c}_{it}^{\text{L}}}({\widehat{q}}_{it} + \mathbb{1}^{\top}\bm{\xi}_{t}\gamma_{it})\Big)}_{\text{Term B}} + \underbrace{\sum_{i \in \mathcal{W}}\sum_{t\in\mathcal{T}}\;\Big({c}^{\text{Q}}_{it}({\widehat{q}}_{it} - \bm{\xi}_{it})^{2} + {c}_{it}^{\text{L}}({\widehat{q}}_{it} - {\xi}_{it})\Big)}_{\text{Term C}}\Big].
\end{align*}
Linearity of the expectation operator allows us to separate the cost terms and we rewrite the cost of flexible generators, i.e., Term A as
\begin{align*}
    \sum_{i \in \mathcal{F}}\;\sum_{t\in\mathcal{T}}\;\Big({c}^{\text{Q}}_{it}\widehat{q}_{it}^{2} + {c}^{\text{Q}}_{it}\;\mathbb{E}^{\mathbb{P}_{\xi}}[(\mathbb{1}^{\top}\bm{\xi}_{t})^{2}]\alpha_{it}^{2} + 2\widehat{q}_{it}{c}^{\text{Q}}_{it}\;\mathbb{E}^{\mathbb{P}_{\xi}}[\mathbb{1}^{\top}\bm{\xi}_{t}]\alpha_{it} + {{c}_{it}^{\text{L}}}{\widehat{q}}_{it} + c^{\text{L}}_{it}\;\mathbb{E}^{\mathbb{P}_{\xi}}[\mathbb{1}^{\top}\bm{\xi}_{t}]\alpha_{it}\Big).
\end{align*}
The zero-mean assumption made on $\bm{\xi}_{t}$ factors out the terms last two terms under the expectation operator. As discussed in \cite{Bienstock2014}, the first term under expectation operator can be reformulated as the variance of $\bm{\xi}_{t}$, i.e., $\mathbb{E}^{\mathbb{P}_{\xi}}[(\mathbb{1}^{\top}\bm{\xi}_{t})^{2}] = \text{var}(\mathbb{1}^{\top}\bm{\xi}_{t}) = \mathbb{1}^{\top}\bm{\Sigma}_{t}\mathbb{1}$, where $\bm{\Sigma}_{t}$ is the covariance matrix as previously discussed. Therefore, the Term A reduces to 
\begin{align*}
     \sum_{i \in \mathcal{F}}\;\sum_{t\in\mathcal{T}}\;\Big({c}^{\text{Q}}_{it}\widehat{q}_{it}^{2} + {c}^{\text{Q}}_{it}\;\mathbb{1}^{\top}\bm{\Sigma}_{t}\mathbb{1}\alpha_{it}^{2} + {{c}_{it}^{\text{L}}}{\widehat{q}}_{it}\Big).
\end{align*}
Lastly, following Example \ref{example:quad_cost_reformulation}, we reformulate the quadratic cost terms as a rotated SOC constraint in the interest of analytical and computational appeal. Introducing variables $z^{\widehat{q}}_{it}\in\mathbb{R}$ and $z^{\alpha}_{it}\in\mathbb{R}$, for any fixed values of $\widehat{q}_{it}$ and $\alpha_{it}$, Term A of the objective function retrieves the expected cost by solving the following SOCP problem
\begin{subequations}\label{prob:obj_cost_reform}
\begin{align}
    \minimize{z^{\widehat{q}}_{it},z^{\alpha}_{it}}\quad&\sum_{i \in \mathcal{F}}\;\sum_{t\in\mathcal{T}}\;\Big(z^{\widehat{q}}_{it} + z^{\alpha}_{it} +  {{c}_{it}^{\text{L}}}\widehat{q}_{it}\Big)\label{objfun_Mcone_strict_feas}\\
    \st\quad&\norm{({c}^{\text{Q}}_{it})^{\frac{1}{2}}\widehat{q}_{it}}^{2} \le  {z}^{\widehat{q}}_{it},\;\forall t,\;\forall i \in \mathcal{F} \\
    & \norm{\mathbf{X}_{t}\mathbb{1}({c}^{\text{Q}}_{it})^{\frac{1}{2}}\alpha_{it}}^{2} \le  {z}^{\alpha}_{it},\;\forall t,\;\forall i \in \mathcal{F}.
\end{align}
\end{subequations}
Term B of the objective function characterizing the cost of operation of energy storage units follows a similar reformulation. For the set of wind power producers, i.e., the cost in Term C, we use the following equivalence to eliminate the expectation operator
\begin{align*}
    \sum_{i \in \mathcal{W}}\;\sum_{t\in\mathcal{T}}\;\Big({c}^{\text{Q}}_{it}\widehat{q}_{it}^{2} + {c}^{\text{Q}}_{it}\;\mathbb{E}^{\mathbb{P}_{\xi}}[{\xi}_{it}^{2}] + {{c}_{it}^{\text{L}}}{\widehat{q}}_{it}\Big) \Leftrightarrow  \sum_{i \in \mathcal{W}}\;\sum_{t\in\mathcal{T}}\;\Big({c}^{\text{Q}}_{it}\widehat{q}_{it}^{2} + {c}^{\text{Q}}_{it}\;\text{tr}[\bm{\Sigma}_{t}] + {{c}_{it}^{\text{L}}}{\widehat{q}}_{it}\Big).
\end{align*}
The quadratic term in the expression is then reformulated as a rotated SOC constraint as before. The term $c_{it}^{\text{Q}}\text{tr}[\bm{\Sigma}_{t}]$ is a cost term that is constant, depending on the historical forecast error samples.

Adopting the reformulations of the objective function and the chance constraints, we obtain the final tractable chance-constrained market-clearing problem $\mathcal{M}^{\text{cc}}$. Solved centrally by the system operator, the problem $\mathcal{M}^{\text{cc}}$ is an SOCP problem that results in optimal market-clearing quantities and prices for both the commodities traded in the market, i.e., energy and flexibility. 
\begin{subequations}\label{prob:Mc_CC_SOCP}
\begin{align}
    \minimize{\mathcal{V}^{\text{opt}},\mathcal{V}^{\text{aux}}}\quad&\sum_{i \in \mathcal{I}\setminus\mathcal{D}}\;\sum_{t\in\mathcal{T}}\;\Big({z}_{it}^{\widehat{q}_{it}} + c_{it}^{\text{L}}\widehat{q}_{it}\Big) +  \sum_{i \in \mathcal{F}}\;\sum_{t\in\mathcal{T}}\;z_{it}^{\alpha} +  \sum_{i \in \mathcal{S}}\;\sum_{t\in\mathcal{T}}\;z_{it}^{\gamma} +  \sum_{i \in \mathcal{W}}\;\sum_{t\in\mathcal{T}}\;c_{it}^{\text{Q}}\text{tr}[\bm{\Sigma}_{t}]  \label{objfun_Mc_CC_SOCP}\\
    \st\quad&\norm{({c}^{\text{Q}}_{it})^{\frac{1}{2}}\widehat{q}_{it}}^{2} \le  {z}^{\widehat{q}}_{it},\;\forall t,\;\forall i \in \mathcal{I}\setminus\mathcal{D}\\
    & \norm{\mathbf{X}_{t}\mathbb{1}({c}^{\text{Q}}_{it})^{\frac{1}{2}}\alpha_{it}}^{2} \le  {z}^{\alpha}_{it},\;\forall t,\;\forall i \in \mathcal{F}\\
    & \norm{\mathbf{X}_{t}\mathbb{1}({c}^{\text{Q}}_{it})^{\frac{1}{2}}\gamma_{it}}^{2} \le  {z}^{\gamma}_{it},\;\forall t,\;\forall i \in \mathcal{S}\\
    &r_{\widehat{\varepsilon}}\norm{\mathbf{X}_{t}\mathbb{1}\alpha_{it}} \leq \overline{Q}_{i} - \widehat{q}_{it},\;\forall t,\;\forall i \in \mathcal{F}\\
    &r_{\widehat{\varepsilon}}\norm{\mathbf{X}_{t}\mathbb{1}\alpha_{it}} \leq \widehat{q}_{it} - \underline{Q}_{i},\;\forall t,\;\forall i \in \mathcal{F}\\
    &r_{\widehat{\varepsilon}}\norm{\mathbf{X}_{t}\mathbb{1}\alpha_{it}} \leq \overline{Q}^{R}_{i},\;\forall t,\;\forall i \in \mathcal{F}\\
     &r_{\widehat{\varepsilon}}\norm{\mathbf{X}_{t}\mathbb{1}\alpha_{it}} \leq \underline{Q}^{R}_{i},\;\forall t,\;\forall i \in \mathcal{F}\\
    &r_{\widehat{\varepsilon}}\norm{\mathbf{X}_{t-1:t}\begin{bmatrix}-\mathbb{1}^{\top}\alpha_{i(t-1)}l\quad&\phantom{-}\mathbb{1}^{\top}\alpha_{it}\end{bmatrix}^{\top}} \leq \overline{\Delta}_{i} + (\widehat{q}_{it-1} - \widehat{q}_{it}), \;\forall t>2,\;\forall i\in\mathcal{F}\\
    &r_{\widehat{\varepsilon}}\norm{\mathbf{X}_{t-1:t}\begin{bmatrix}\phantom{-}\mathbb{1}^{\top}\alpha_{i(t-1)}\quad&-\mathbb{1}^{\top}\alpha_{it}\end{bmatrix}^{\top}} \leq \underline{\Delta}_{i} - (\widehat{q}_{it-1} - \widehat{q}_{it}), \;\forall t>2,\;\forall i\in\mathcal{F}\\
    &r_{\widehat{\varepsilon}}\norm{\mathbf{X}_{t}\mathbb{1}\gamma_{it}} \leq \eta_{i}^{\text{D}}\overline{Q}^{\text{D}}_{i} - \widehat{q}_{it},\;\forall t,\;\forall i\in\mathcal{S}\\
    &r_{\widehat{\varepsilon}}\norm{\mathbf{X}_{t}\mathbb{1}\gamma_{it}} \leq \frac{1}{\eta_{i}^{\text{C}}}\overline{Q}^{\text{C}}_{i} + \widehat{q}_{it},\;\forall t,\;\forall i\in\mathcal{S}\\
     &r_{\widehat{\varepsilon}}\norm{\mathbf{X}_{1:t}\begin{bmatrix}\mathbb{1}^{\top}\gamma_{i1} \quad \mathbb{1}^{\top}\gamma_{i2}\;\cdots\;\mathbb{1}^{\top}\gamma_{it}\end{bmatrix}^{\top}} \leq \overline{E}_{i} - E_{i}^{0} + \sum_{t'=1}^{t}\widehat{q}_{it'},\;\forall t,\;\forall i\in\mathcal{S}\\
    &r_{\widehat{\varepsilon}}\norm{\mathbf{X}_{1:t}\begin{bmatrix}\mathbb{1}^{\top}\gamma_{i1} \quad \mathbb{1}^{\top}\gamma_{i2}\;\cdots\;\mathbb{1}^{\top}\gamma_{it}\end{bmatrix}^{\top}} \leq E_{i}^{0} - \underline{E}_{i} - \sum_{t'=1}^{t}\widehat{q}_{it'},\;\forall t,\;\forall i\in\mathcal{S}\\
    &r_{\widehat{\varepsilon}}\norm{\mathbf{X}_{1:T}\begin{bmatrix}\mathbb{1}^{\top}\gamma_{i1} \quad \mathbb{1}^{\top}\gamma_{i2}\;\cdots\;\mathbb{1}^{\top}\gamma_{iT}\end{bmatrix}^{\top}} \leq \overline{B}_{i} + \sum_{t'=1}^{T}\widehat{q}_{it'},\;\forall i \in \mathcal{S}\\
    &r_{\widehat{\varepsilon}}\norm{\mathbf{X}_{1:T}\begin{bmatrix}\mathbb{1}^{\top}\gamma_{i1} \quad \mathbb{1}^{\top}\gamma_{i2}\;\cdots\;\mathbb{1}^{\top}\gamma_{iT}\end{bmatrix}^{\top}} \leq \underline{B}_{i} - \sum_{t'=1}^{T}\widehat{q}_{it'},\;\forall i \in \mathcal{S}\\
    &\sum_{i \in \mathcal{I}} \mathbf{G}_{ip}{\mathbf{q}}_{ip} = \mathbb{0},\;p=1 \\
    &\sum_{i \in \mathcal{I}\setminus{(\mathcal{W}\cup\mathcal{D})}}\text{diag}(\mathbb{1}){\mathbf{q}}_{ip} = \mathbb{1}, \; p=2 \\
    &r_{\widehat{\varepsilon}}\norm{\mathbf{X}_{t}[\bm{\Psi}(\bm{\Psi}^{F}\bm{\alpha}_{t}\mathbb{1}^{\top} + \bm{\Psi}^{S}\bm{\gamma}_{t}\mathbb{1}^{\top} - \bm{\Psi}^{W})]^{\top}_{(\ell,:)}} \leq \overline{s}_{\ell} \notag \\
    &\qquad \qquad \qquad \qquad - [ \bm{\Psi}(\bm{\Psi}^{F}\widehat{\mathbf{q}}^{F}_{t} + \bm{\Psi}^{S}\widehat{\mathbf{q}}^{S}_{t} + \bm{\Psi}^{W}\widehat{\mathbf{q}}^{W}_{t} + \bm{\Psi}^{D}\widehat{\mathbf{q}}^{D}_{t})]_{\ell},\;\forall t,\;\forall \ell \in \mathcal{L}\\
    &r_{\widehat{\varepsilon}}\norm{\mathbf{X}_{t}[\bm{\Psi}(\bm{\Psi}^{F}\bm{\alpha}_{t}\mathbb{1}^{\top} + \bm{\Psi}^{S}\bm{\gamma}_{t}\mathbb{1}^{\top} - \bm{\Psi}^{W})]^{\top}_{(\ell,:)}} \leq \overline{s}_{\ell} \notag \\
    &\qquad \qquad \qquad \qquad + [ \bm{\Psi}(\bm{\Psi}^{F}\widehat{\mathbf{q}}^{F}_{t} + \bm{\Psi}^{S}\widehat{\mathbf{q}}^{S}_{t} + \bm{\Psi}^{W}\widehat{\mathbf{q}}^{W}_{t} + \bm{\Psi}^{D}\widehat{\mathbf{q}}^{D}_{t})]_{\ell},\;\forall t,\;\forall \ell \in \mathcal{L},
\end{align}
\end{subequations}
where the set of optimization variables is $\mathcal{V}^{\text{opt}} = \{\widehat{q}_{it},z_{it}^{\widehat{q}_{it}},\alpha_{it},z_{it}^{\widehat{\alpha}_{it}},\gamma_{it},z_{it}^{\widehat{\gamma}_{it}}\}$ and the set of auxiliary variables is $\mathcal{V}^{\text{aux}} = \{\mathbf{q}_{ip},\widehat{\mathbf{q}}^{\text{F}}_{t},\bm{\alpha}_{t},\widehat{\mathbf{q}}^{\text{S}}_{t},\bm{\gamma}_{t},\widehat{\mathbf{q}}^{\text{W}}_{t},\widehat{\mathbf{q}}^{\text{D}}_{t} \}$, formed as previously discussed.
\subsection{LP-based market-clearing benchmarks}\label{ec:subsec_R1_R2_formulation}
In the following, we concisely formulate the two reference uncertainty-aware benchmarks within the LP framework. To account for the quadratic costs of market participants in an LP problem, we perform a piecewise linear approximation of the cost by discretizing the production quantities into a set of bins given by $\mathcal{Y} = \{1,2,\dots,Y\}$.
\subsubsection*{Deterministic two-stage market framework:} In the deterministic market-clearing problem $\mathcal{R}1$, the system operator procures energy and flexibility (in the form of reserve capacity) in the day-ahead market. Thereafter, during real-time operation stage, via another market-clearing mechanism, the reserves are activated based on the allocation capacity bounds cleared during the day-ahead market. This two-stage deterministic market-clearing problem is a natural uncertainty-aware extension of currently-operational electricity markets.

\textbf{Day-ahead market-clearing problem ${\mathcal{R}{1}}_{a}$:}
\begin{subequations}\label{prob:DA_two_stage_LP}
\begin{align}
    \minimize{{q}^{\text{DA}}_{ity}, {q}^{\text{R}}_{it}}\quad&\sum_{i \in \mathcal{I}\setminus{\mathcal{D}}}\;\sum_{t\in\mathcal{T}}\;\sum_{y\in\mathcal{Y}}\;\Big({{c}_{ity}^{\text{Y}}}{q}_{ity}^{\text{DA}} \Big) + \sum_{i\in\mathcal{F}\cup\mathcal{S}}\sum_{t\in\mathcal{T}}c_{it}^{R}q_{it}^{R} \label{objfun_DA_twostage_LP}\\
    \st\quad& 0 \leq {q}^{\text{DA}}_{ity} \leq \frac{1}{Y}(\overline{Q}_{i} - \underline{Q}_{i}),\;\forall y,\;\forall t,\;\forall i \in \mathcal{I}\setminus\mathcal{D}\\
    &\underline{Q}_{i} \leq \sum_{y\in\mathcal{Y}}{q}^{\text{DA}}_{ity} \leq \overline{Q}_{i},\;\forall t,\;\forall i \in \mathcal{F}\cup\mathcal{W}\\
    &\underline{Q}_{i} \leq \sum_{y\in\mathcal{Y}}{q}^{\text{DA}}_{ity} + {q}^{\text{R}}_{it} \leq \overline{Q}_{i},\;\forall t,\;\forall i \in \mathcal{F}\\
    &-\underline{\Delta}_{i} \leq \sum_{y\in\mathcal{Y}}({q}^{\text{DA}}_{ity} - {q}^{\text{DA}}_{i(t-1)y}) \leq \overline{\Delta}_{i},\;\forall t>2,\;\forall i \in \mathcal{F}\\
    &-\frac{1}{\eta_{i}^{\text{C}}}\overline{Q}_{i}^{C} \leq \sum_{y\in\mathcal{Y}}{q}^{\text{DA}}_{ity}\leq\eta_{i}^{\text{D}}\overline{Q}_{i}^{D},\;\forall t,\;\forall i \in \mathcal{S}\\
    &-\frac{1}{\eta_{i}^{\text{C}}}\overline{Q}_{i}^{C} \leq \sum_{y\in\mathcal{Y}}{q}^{\text{DA}}_{ity} + q_{it}^{R}\leq\eta_{i}^{\text{D}}\overline{Q}_{i}^{D},\;\forall t,\;\forall i \in \mathcal{S}\\
    &\underline{E}_{i} \leq E_{i}^{0} - \sum_{t'=1}^{t}( \sum_{y\in\mathcal{Y}}{q}^{\text{DA}}_{it'y} + q_{it}^{R})\leq \overline{E}_{i},\;\forall t,\;\forall i \in \mathcal{S}\\
    &E_{i}^{0} - \underline{B}_{i} \leq E_{i}^{0} - \sum_{t'=1}^{t}( \sum_{y\in\mathcal{Y}}{q}^{\text{DA}}_{it'y} + q_{it'}^{R}) \leq E_{i}^{0} + \overline{B}_{i},\;\forall i \in \mathcal{S} \\
    &0 \leq {q}^{\text{R}}_{it} \leq \overline{Q}^{\text{R}}_{i},\;\forall t,\;\forall i \in \mathcal{F}\cup\mathcal{S}\\
    &\sum_{i\in\mathcal{I}}\sum_{y\in\mathcal{Y}}q_{ity}^{\text{DA}} = 0,\;\forall t\\
    &\sum_{i\in\mathcal{F}\cup\mathcal{S}} {q}^{\text{R}}_{it} \geq M_{\text{R}},\;\forall t\\
    &-\overline{s}_{\ell} \leq \sum_{n \in \mathcal{N}} [\bm{\Psi}]_{(\ell,n)}\left(\sum_{i \in \mathcal{I}_{n}}\;\;\sum_{y\in\mathcal{Y}}{{q}}^{\text{DA}}_{ity}\right) \leq \overline{s}_{\ell},\;\forall t,\;\forall \ell \in \mathcal{L},
\end{align}
\end{subequations}
where the parameter $M_{\text{R}}$ is the exogenously-determined minimum reserve requirement set by the system operator. Next, closer to real-time the following flexibility activation problem is solved for each uncertainty realization, denoted by $\widehat{\bm{\xi}}$.

\textbf{Real-time flexibility activation problem, ${\mathcal{R}{1}}_{b},\;\forall\;\widehat{\bm{\xi}}$:}
\begin{subequations}\label{prob:RT_two_stage_LP}
\begin{align}
    \minimize{{q}^{\text{RT}}_{ity}}\quad&\sum_{i \in \mathcal{I}\setminus{\mathcal{D}}}\;\sum_{t\in\mathcal{T}}\;\sum_{y\in\mathcal{Y}}{c}^{\text{Y}}_{ity}{{{q}}_{ity}^{\text{RT}}} + \sum_{i\in\mathcal{W}}c^{\text{spill}}q^{\text{RT}}_{it} - \sum_{i\in\mathcal{D}}c^{\text{shed}}q^{\text{RT}}_{it}\label{objfun_RT_twostage_LP}\\
    \st\quad& -\frac{{{q}^{\text{R}}_{it} }^{\star}}{Y} \leq {q}^{\text{RT}}_{ity} \leq \frac{{{q}^{\text{R}}_{it} }^{\star}}{Y},\;\forall y,\;\forall t,\;\forall i \in \mathcal{F}\cup\mathcal{S}\\
    & - {{q}^{\text{R}}_{it} }^{\star} \leq \sum_{y\in\mathcal{Y}}q_{ity}^{\text{RT}} \leq {{q}^{\text{R}}_{it} }^{\star},\;\forall t,\;\forall i \in \mathcal{F}\cup\mathcal{S}\\
    & 0 \leq q^{\text{RT}}_{ity} \leq \frac{\widehat{q}_{it} - \widehat{\xi}_{it}}{Y},\;\forall t,\;\forall i \in \mathcal{W}\\
    & -\frac{\widehat{q}_{it}}{Y} \leq q^{\text{RT}}_{ity} \leq 0,\;\forall t,\;\forall i \in \mathcal{D}\\
    & \sum_{i\in\mathcal{I}\setminus\mathcal{W}}\sum_{y\in\mathcal{Y}}q_{ity}^{\text{RT}} = \mathbb{1}^{\top}\widehat{\bm{\xi}}_{t} + \sum_{i\in\mathcal{W}}q^{\text{RT}}_{it},\;\forall t\\
    &-\overline{s}_{\ell} \leq \sum_{n \in \mathcal{N}} [\bm{\Psi}]_{(\ell,n)}\left(\sum_{i \in \mathcal{I}_{n}}\;\sum_{y\in\mathcal{Y}}\Big({{{q}}^{\text{DA}}_{ity}}^{\star} + {q}_{ity}^{\text{RT}}\Big) \right) \leq \overline{s}_{\ell},\;\forall t,\;\forall \ell \in \mathcal{L},
\end{align}
\end{subequations}
where $({q_{ity}^{\text{DA}}}^{\star},{q_{it}^{\text{R}}}^{\star}) = \text{argmin}\;{\mathcal{R}{1}}_{a}$ are day-ahead dispatch and reserve capacity, respectively. 

\subsubsection*{Scenario-based market framework:} The scenario-based stochastic market-clearing problem $\mathcal{R}2$ is a two-stage problem. The first stage involves the day-ahead schedules as a result of the so-called \textit{here-and-now} decisions, whereas the second stage adjusts the day-ahead schedules with real-time adjustments to mitigate the uncertainty from weather-dependent power producers (\textit{wait-and-see}). The adjustment decisions for real-time stage are already included in the day-ahead market-clearing problem by considering a finite number of uncertainty realization scenarios, anticipating that the actual power production is captured in the scenarios considered. Like with $\mathcal{R}1$, we adopt a linear approximation of the quadratic costs to remain within the LP framework. In the following, we provide a concise formulation for this benchmark.
\begin{subequations}\label{prob:Mc_Scen_Sto}
\begin{align}
    \minimize{{q}^{\text{DA}}_{ity}, q_{ityk}^{\text{RT}}}\quad&\sum_{i \in \mathcal{I}\setminus{\mathcal{D}}}\;\sum_{t\in\mathcal{T}}\sum_{y\in\mathcal{Y}}\;\Big({c}^{\text{Y}}_{ity}{{q}}_{ity}^{\text{DA}}\Big) + \sum_{k=1}^{K}\phi_{k} \Big[ \sum_{i\in\mathcal{I}\setminus{\mathcal{D}}}\sum_{t\in\mathcal{T}}\sum_{y\in\mathcal{Y}}\Big({c}^{\text{Y}}_{ity}{{q}}_{ity}^{\text{RT}}\Big)\Big]\label{objfun_Mc_LP}\\
    \st\quad& 0 \leq q_{ity}^{\text{DA}} \leq \frac{1}{Y}(\overline{Q}_{i} - \underline{Q}_{i}),\;\forall y,\;\forall t,\;\forall i\in \mathcal{I}\setminus\mathcal{D}\\
    &\underline{Q}_{i} \leq \sum_{y\in\mathcal{Y}}{q}^{\text{DA}}_{ity} \leq \overline{Q}_{i},\;\forall t,\;\forall i \in \mathcal{F}\cup\mathcal{W}\\
    &-\underline{\Delta}_{i} \leq \sum_{y\in\mathcal{Y}}({q}^{\text{DA}}_{ity} - {q}^{\text{DA}}_{i(t-1)y}) \leq \overline{\Delta}_{i},\;\forall t>2,\;\forall i \in \mathcal{F}\\
    &-\frac{1}{\eta_{i}^{C}}\overline{Q}_{i}^{C} \leq \sum_{y\in\mathcal{Y}}{q}^{\text{DA}}_{ity}\leq\eta_{i}^{D}\overline{Q}_{i}^{D},\;\forall t,\;\forall i \in \mathcal{S}\\
    &\underline{E}_{i} \leq E_{i}^{0} - \sum_{t'=1}^{t}\sum_{y\in\mathcal{Y}} {q}^{\text{DA}}_{it'y} \leq \overline{E}_{i},\;\forall t,\;\forall i \in \mathcal{S}\\
    &E_{i}^{0} - \underline{B}_{i} \leq E_{i}^{0} - \sum_{t'=1}^{T}\sum_{y\in\mathcal{Y}} {q}^{\text{DA}}_{it'y} \leq E_{i}^{0} + \overline{B}_{i},\;\forall i \in \mathcal{S} \\
    &\sum_{i\in\mathcal{I}}\sum_{y\in\mathcal{Y}}q_{ity}^{\text{DA}} = 0,\;\forall t\\
    &-\overline{s}_{\ell} \leq \sum_{n \in \mathcal{N}} [\bm{\Psi}]_{(\ell,n)}\left(\sum_{i \in \mathcal{I}_{n}}\sum_{y\in\mathcal{Y}}\;{{q}}^{\text{DA}}_{ity}\right) \leq \overline{s}_{\ell},\;\forall t,\;\forall \ell \in \mathcal{L}\\
    & -\frac{\underline{Q}_{i}^{R}}{Y} \leq p_{itky}^{\text{RT}} \leq \frac{\overline{Q}_{i}^{R}}{Y}\;\forall y,\;\forall k,\;\forall t,\;\forall i \in \mathcal{F}\cup\mathcal{S} \\
    &-\underline{Q}_{i}^{R} \leq \sum_{y\in\mathcal{Y}}p_{itky}^{\text{RT}} \leq \overline{Q}_{i}^{R},\;\forall k,\;\forall t,\;\forall i \in \mathcal{F}\cup\mathcal{S}\\
    &-\underline{Q}_{i} \leq \sum_{y\in\mathcal{Y}}(p_{ity}^{\text{DA}} + p_{itky}^{\text{RT}}) \leq \overline{Q}_{i},\;\forall k,\;\forall t,\;\forall i \in \mathcal{F} \cup \mathcal{W}\\
    &-\frac{1}{\eta_{i}^{C}}\overline{Q}_{i}^{C} \leq \sum_{y\in\mathcal{Y}}({q}^{\text{DA}}_{ity} + {q}^{\text{RT}}_{itky})\leq\eta_{i}^{D}\overline{Q}_{i}^{D},\;\forall k,\;\forall t,\;\forall i \in \mathcal{S}\\
    &\underline{E}_{i} \leq E_{i}^{0} - (\sum_{t'=1}^{t} \sum_{y\in\mathcal{Y}}({q}^{\text{DA}}_{it'y} + {q}^{\text{RT}}_{it'ky})) \leq \overline{E}_{i},\;\forall t,\;\forall i \in \mathcal{S}\\
    &E_{i}^{0} - \underline{B}_{i} \leq E_{i}^{0} - (\sum_{t'=1}^{t} \sum_{y\in\mathcal{Y}}({q}^{\text{DA}}_{it'y} + {q}^{\text{RT}}_{it'ky})) \leq E_{i}^{0} + \overline{B}_{i},\;\forall k,\;\forall i \in \mathcal{S} \\
    &\sum_{i\in\mathcal{I}}\sum_{y\in\mathcal{Y}}q_{itky}^{\text{RT}} = 0,\;\forall k,\;\forall t\\
    &-\overline{s}_{\ell} \leq \sum_{n \in \mathcal{N}} [\bm{\Psi}]_{(\ell,n)}\left(\sum_{i \in \mathcal{I}_{n}}\sum_{y\in\mathcal{Y}}\;({q}^{\text{DA}}_{ity} + {q}^{\text{RT}}_{itky})\right) \leq \overline{s}_{\ell},\;\forall k,\;\forall t,\;\forall \ell \in \mathcal{L}
\end{align}
\end{subequations}
\subsection{Out-of-sample simulations}\label{ec:subsec_out_of_sample_simulations}
Out-of-sample simulations are performed to evaluate the quality of the day-ahead market-clearing outcomes by fixing the decisions obtained at the day-ahead stage to their optimal values and solving a real-time flexibility activation problem. For the proposed conic market framework, the out-of-sample problem \eqref{prob:out-of-sample} formulated in the following admits fixed day-ahead decisions $({q}_{it}^{\star},\alpha^{\star}_{it},\gamma^{\star}_{it})$ obtained from the solution to $\mathcal{M}^{cc}$. For the LP-based benchmark problems, we use ${q}_{it}^{\star} = \sum_{y\in\mathcal{Y}}{q^{\text{DA}}_{ity}}^{\star}$ from the solution to $\mathcal{R}1_{a}$ for the deterministic benchmark problem and ${q}_{it}^{\star} = \sum_{y\in\mathcal{Y}}{q^{\text{DA}}_{ity}}^{\star}$ from the solution to $\mathcal{R}2$ for the scenario-based stochastic benchmark problem. We have, $\forall \widehat{\bm{\xi}}:$
\begin{subequations}\label{prob:out-of-sample}
\begin{align}
    \minimize{{q}^{\text{RT}}_{it}}\quad&\sum_{i \in \mathcal{I}\setminus{(\mathcal{D}\cup\mathcal{W})}}\;\sum_{t\in\mathcal{T}}\;\Big({c}^{\text{Q}}_{it}({q}_{it}^{\star} + {{{q}}_{it}^{\text{RT}}})^{2} + {{c}_{it}^{\text{L}}}(q_{it}^{\star} + {q}_{it}^{\text{RT}}) \Big) + \sum_{i\in\mathcal{W}}c^{\text{spill}}q^{\text{RT}}_{it} + \sum_{i\in\mathcal{D}}c^{\text{shed}}q^{\text{RT}}_{it}\\
    \st\quad&\underline{Q}_{i} \leq {q}_{it}^{\star} + {q}^{\text{RT}}_{it} \leq \overline{Q}_{i},\;\forall t,\;\forall i \in \mathcal{F}\\
    & -\underline{\Delta}_{i} \leq {q}_{it}^{\star} + {q}^{\text{RT}}_{it} \leq \overline{\Delta}_{i},\;\forall t>1,\;\forall i \in \mathcal{F}\\
    &-\frac{1}{\eta^{\text{C}}}\overline{Q}^{\text{C}}_{i} \leq {q}_{it}^{\star} + {q}^{\text{RT}}_{it} \leq \eta^{\text{D}}\overline{Q}_{i}^{\text{D}},\;\forall t,\;\forall i \in \mathcal{S}\\
    &\underline{E}_{i} \leq E_{i}^{0} - (\sum_{t'=1}^{t} {q}_{it'}^{\star} + {q}^{\text{RT}}_{it'}) \leq \overline{E}_{i},\;\forall t,\;\forall i \in \mathcal{S}\\
    &E_{i}^{0} - \underline{B}_{i} \leq E_{i}^{0} - (\sum_{t'=1}^{t} {q}_{it'}^{\star} + {q}^{\text{RT}}_{it'}) \leq E_{i}^{0} + \overline{B}_{i},\;\forall i \in \mathcal{S} \\
     & 0 \leq q^{\text{RT}}_{it} \leq \widehat{q}_{it} - \widehat{\xi}_{it},\;\forall t,\;\forall i \in \mathcal{W}\\
    & -\widehat{q}_{it} \leq q^{\text{RT}}_{it} \leq 0,\;\forall t,\;\forall i \in \mathcal{D}\\
    &-\underline{Q}_{i}^{R} \leq q_{it}^{RT} \leq \overline{Q}_{i}^{R},\;\forall t,\;\forall i \in \mathcal{F}\cup\mathcal{S} \\
    & \sum_{i\in\mathcal{I}\setminus\mathcal{W}}q_{it}^{\text{RT}} = \mathbb{1}^{\top}\widehat{\bm{\xi}}_{t} + \sum_{i\in\mathcal{W}}q^{\text{RT}}_{it},\;\forall t\\
    &-\overline{s}_{\ell} \leq \sum_{n \in \mathcal{N}} [\bm{\Psi}]_{(\ell,n)}\left(\sum_{i \in \mathcal{I}_{n}}\;{q}_{it}^{\star} + {q}^{\text{RT}}_{it}\right) \leq \overline{s}_{\ell},\;\forall t,\;\forall \ell \in \mathcal{L}\\
    \intertext{For $\mathcal{M}^{\text{cc}}$, these additional constraints restrict the real-time adjustments:}
    &q_{it}^{\text{RT}} = \mathbb{1}^{\top}\widehat{\bm{\xi}}_{t}\alpha^{\star}_{it},\;\forall t,\;\forall i \in \mathcal{F} \; \quad ; \quad  q_{it}^{\text{RT}} = \mathbb{1}^{\top}\widehat{\bm{\xi}}_{t}\gamma^{\star}_{it},\;\forall t,\;\forall i \in \mathcal{S}.
\end{align}
\end{subequations}